\documentclass[pdftex]{aastex631}

\usepackage{amssymb, amsmath}
\usepackage[version=4]{mhchem}

\begin{document}

\title{Physical and Chemical Properties of Galactic Molecular Gas toward QSO J1851+0035}

\author[0009-0000-5913-8555]{Kanako Narita}
\affiliation{Department of Astronomy, Graduate School of Science, The University of Tokyo,  \\
7-3-1 Hongo, Bunkyo-ku, Tokyo 133-0033, Japan}
\affiliation{National Astronomical Observatory of Japan, National Institutes of Natural Sciences,  \\
2-21-1 Osawa, Mitaka, Tokyo 181-8588, Japan}

\author{Seiichi Sakamoto}
\affiliation{Department of Astronomy, Graduate School of Science, The University of Tokyo,  \\
7-3-1 Hongo, Bunkyo-ku, Tokyo 133-0033, Japan}
\affiliation{National Astronomical Observatory of Japan, National Institutes of Natural Sciences, \\
2-21-1 Osawa, Mitaka, Tokyo 181-8588, Japan}
\affiliation{Astronomical Science Program, Graduate Institute for Advanced Studies, SOKENDAI, \\
2-21-1 Osawa, Mitaka, Tokyo 181-8588, Japan}

\author{Jin Koda}
\affiliation{Department of Physics and Astronomy, Stony Brook University,  \\
Stony Brook, NY 11794-3800, USA}

\author{Yuki Yoshimura}
\affiliation{Institute of Astronomy, Graduate School of Science, The University of Tokyo, \\
2-21-1 Osawa, Mitaka, Tokyo 181-0015, Japan}

\author{Kotaro Kohno}
\affiliation{Institute of Astronomy, Graduate School of Science, The University of Tokyo, \\
2-21-1 Osawa, Mitaka, Tokyo 181-0015, Japan}
\affiliation{Research Center for the Early Universe, Graduate School of Science, The University of Tokyo,\\
7-3-1 Hongo, Bunkyo, Tokyo 113-0033, Japan}

%% Note that the \and command from previous CASAs of AASTeX is now
%% depreciated in this CASA as it is no longer necessary. AASTeX 


\begin{abstract}
ALMA data toward QSO J1851+0035 ($l$=$33.498^{\circ}$, $b$=$+0.194^{\circ}$) were used to study absorption lines by Galactic molecular gas. 
We detected 17 species (CO, $^{13}$CO, C$^{18}$O, HCO$^+$, H$^{13}$CO$^+$, HCO, H$_2$CO, C$_2$H, $c$-C$_3$H, $c$-C$_3$H$_2$, CN, HCN, HNC, CS, SO, SiO, and C) and set upper limits to 18 species as reference values for chemical models. 
About 20 independent velocity components at 4.7--10.9 kpc from the Galactic Center were identified. 
Their column density and excitation temperature estimated from the absorption study, as well as the CO intensity distributions obtained from the FUGIN survey, indicate that the components with $\tau$ $\lesssim$ 1 correspond to diffuse clouds or cloud outer edges. 
Simultaneous multiple-Gaussian fitting of CO $J$=1--0 and $J$=2--1 absorption lines shows that these are composed of narrow- and broad-line components. 
The kinetic temperature empirically expected from the high HCN/HNC isomer ratio ($\gtrsim$4) reaches $\gtrsim$40 K and the corresponding thermal width accounts for the line widths of the narrow-line components. 
CN-bearing molecules and hydrocarbons have tight and linear correlations within the groups. 
The CO/HCO$^+$ abundance ratio showed a dispersion as large as 3 orders of magnitude with a smaller ratio in a smaller $N$(HCO$^+$) (or lower $A_{\rm V}$) range. 
Some of the velocity components are detected in single-dish CO emission and ALMA HCO$^+$ absorption but without corresponding ALMA CO absorption. 
This may be explained by the mixture of clumpy CO emitters not resolved with the $\sim$1 pc single-dish beam surrounded by extended components with a very low CO/HCO$^+$ abundance ratio (i.e., CO-poor gas). 
\end{abstract}



\section{Introduction} 
\label{sec:intro}
%Translucent clouds already show widespread filamentary structure. 
%Can we chemically diagnose them?

%Emission line study maps only denser parts where lines are excited, while absorption line study samples everything through a narrow ``borehole''. 
%Bright QSOs behind the Galactic plane can be used as background sources for such study. 

%Although the number of such background sources has been limited to about 30 systems, recent systematic study of ALMA calibration sources significantly increased its total number (\cite{2016PASJ...68....6A}; \cite{yoshimura2020}). 

%QSO J1851+0035 (RA=18:51:46.723, DEC=+00:35:32.364; $l$=$33.498^{\circ}$, $b$=$+0.194^{\circ}$) is one of the sources found in such a manner, and is among the ones with the largest visual extinction. 
%In addition to the observations for calibration purposes, this source was observed in $J$=1--0 and $J$=2--1 transitions of CO, $^{13}$CO and C$^{18}$O by ALMA with high velocity resolution, and also in $J$=1--0 emissions of CO, $^{13}$CO and C$^{18}$O by NRO 45 m telescope (\cite{2019AAS...23325310K}). 
It is known that there is a large amount of interstellar gas, consisting of not only hydrogen and helium but also a wide variety of molecules of more than 200 different species known so far. 
Most of these species have been identified through their radio emission lines. 
The spatial distribution of individual species has also been mapped in line emission. 
However, such emission line studies are biased to denser parts where the lines are excited. In contrast, absorption line studies equally sample dense and diffuse gas through a narrow pencil beam toward background sources. 
Bright QSOs can be used as background sources for such studies with radio interferometers since a 1 Jy source observed at 100 GHz with a $1''$ resolution provides $\sim$120 K background. 

Previous millimetric absorption studies have successfully observed low-excitation molecular gas in the Galaxy \citep[e.g.,][]{1991ApJ...371L..77M, 1993A&A...276L..33L, 1994A&A...282L...5L, 2001A&A...370..576L, 2018A&A...610A..43R}, 
and even in distant QSO host and lensing galaxies \citep[][and references therein]{2008Ap&SS.313..321C,2011A&A...535A.103M}.
Such low-excitation transition regions between fully molecular and fully atomic gas may extend over large areas in the sky and account for a significant ($\sim$30\%) amount of the neutral gas mass in the Galaxy \citep{2013A&A...554A.103P}. 
These gases may provide initial conditions for probing the physical and chemical evolution of molecular clouds, which are parent bodies of star formation. 
Little is known about the spatial extent and mass, internal structure including volume filling factor, detailed excitation conditions, chemical composition, and internal motion of the diffuse molecular gas. 
The study of diffuse molecular gas is thus crucial to understanding the full picture of the evolution of interstellar matter in the Galaxy. 
Although the number of such background sources behind the Galactic plane with known foreground absorption has been limited to about 30 systems, a recent systematic study of ALMA calibration sources significantly increased its total number \citep {2016PASJ...68....6A, yoshimura2020}. 

\begin{figure}[htb!]
\epsscale{0.5}
\plotone{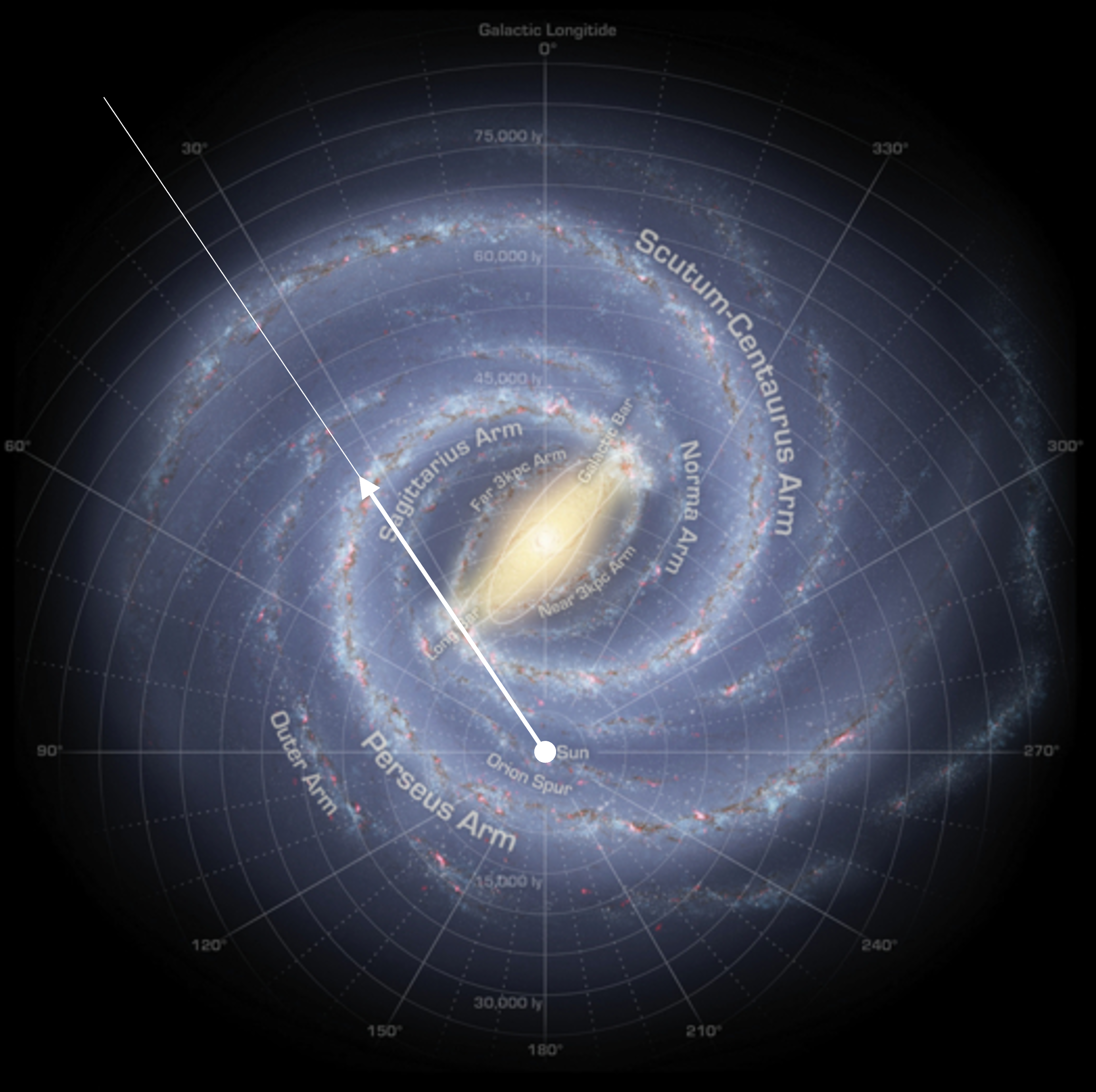}
\caption{The line of sight toward QSO J1851+0035 overlaid on the face-on view of the Galaxy adapted from \citet{2009PASP..121..213C}. 
The tip of the arrow indicates the location along the line of sight where the offset from the Galactic midplane equals the HWHM thickness of molecular gas in the inner Galaxy \citep[45 pc;][]{2015ARA&A..53..583H}, which is 13.3 kpc away from the Sun.
\label{fig:los}}
\end{figure}

The target of this study, QSO J1851+0035 (RA=18:51:46.723, DEC=+00:35:32.364; $l$=$33.498^{\circ}$, $b$=$+0.194^{\circ}$), is one of such sources with the largest visual extinction ($A_{\rm V}$), with $A_{\rm V}$ = 78.59 mag estimated from the $A_{\rm V}$ map of $Planck$ \citep{2014A&A...571A..11P} by \cite{yoshimura2020}. 
In addition to the observations for calibration purposes, this source was observed in $J$=1--0 and $J$=2--1 transition of CO, $^{13}$CO, and C$^{18}$O by ALMA with a high-velocity resolution, and also in $J$=1--0 emissions of CO, $^{13}$CO and C$^{18}$O by the NRO 45 m telescope \citep{2019AAS...23325310K}.
The line of sight toward QSO J1851+0035 ($l$=$33.498^{\circ}$) crosses, from the near to far, the local Aql Rift, near Sagittarius arm, Scutum arm tangent, far Sagittarius arm, Perseus arm, Outer arm, and then Outer Scutum–Centaurus arm as shown in Figure \ref{fig:los}. 

\section{Data and analysis} \label{sec:data}
\subsection{CO absorption data from ALMA}
\label{sec:data_2}
Absorption spectra toward QSO J1851+0035 were observed in Band 3 and 6 in ALMA Cycle 4 (project \#2016.1.00010.S). 
Table \ref{table:archive-1a} shows the three setups of spectral windows that cover the $J$=2--1 and $J$=1--0 transitions of CO, $^{13}$CO, and C$^{18}$O in wide (2 GHz) and narrow ($\sim$59 or 117 MHz) bandwidths. 
The observations were run in three science blocks; each of which started from integrations of bandpass calibrator, J1924-2914, and flux calibrator, J1751+0939, and proceeded to integrations of the target, J1851+0035. 
The targets are bright enough for selfcal, and no separate gain calibrator was observed.
The Common Astronomy Software Applications (CASA) package \citep{2007ASPC..376..127M,cite-key} and ALMA data reduction pipeline were used to carry through the first part of data reduction, including corrections for baseline, system temperature, and water vapor fluctuations. 
After calibrating the visibility data, no deconvolution using the CLEAN method was performed, and a dirty map was created. 
The brightest peak of the QSO was extracted and divided by the continuum intensity to obtain a normalized profile. 
Natural weighting was selected to maximize the continuum sensitivity. 
%We flagged bad data identified by us and also based on information provided by the observatory. Except for the wideband spectral windows, we averaged 20 adjacent channels for bandpass solutions, which removes large-scale variations of bandpass responses. Variations among adjacent channels in bandpass are expected to be negligible. The spectra of flux calibrators were confirmed to be flat after the bandpass calibration. We then derived gain solutions with the targets themselves at a 10 sec interval with the wide spectral windows and applied them to the narrow spectral windows. The continuum spectra of the targets turned out to be slightly polarized (a few $\%$ level). Therefore, the continuum amplitudes of the two polarizations were both adjusted to unity before being averaged.
%One of the advantages of interferometer for absorption studies is that it conveniently resolves out the emission in the MW. We found no noticeable emission in dirty maps with baselines of $>$50 k$\lambda$ ($<$ 4.1$''$), and hence, summed up the data of $>$50 k$\lambda$  to generate final spectra. For a sanity check, we also made spectra with $>$ 100 k$\lambda$  and confirmed that they are the same except elevated noises. We resampled the spectra onto a 0.1 km/s channel width to have the same velocity grid for all CO isotopes and transitions. 
Figure \ref{fig:em_abs} shows the final absorption spectra.
%The RMS noise is listed in Table \ref{table:alma_rms}. A sensitivity with respect to continuum flux $\sigma_l/c$ is used for the absorption line measurement. A typical 1 $\sigma_l/c$ sensitivity in CO($J$=1--0) is 0.03 in a 0.1 km s$^{-1}$ channel. This corresponds to a 1 $\sigma$ sensitivity of $\sim$ 5.3 $\times$ 10$^{12}$ CO/cm$^{2}$ in CO column density under the assumption of Local Thermodynamic Equilibrium (LTE) and an excitation temperature of 5 K (see \ref{sec:velcomp_species}). This translates to a 1$\sigma$ sensitivity of $\sim$$5.3 \times 10^{16}$ H$_{2}$/cm$^{2}$ in H$_{2}$ column density, adopting the CO to H$_{2}$ abundance ratio of $X$(CO) $\sim$ 10$^{-4}$.
%\begin{table}[htb!]
%\caption{RMS Sensitivity of Analyzed Spectra of CO,$^{13}$CO and C$^{18}$O }
% \label{table:alma_rms}
% \centering
% \begin{tabular}{llcl}
%\hline \hline 
%Line &       & Rsolution & RMS                 \\
%     &       & (km/s)   & line/continuum-$\sigma_l/c$ \\
% \hline 
%CO   & $J$=2--1 & 0.1       & 0.03                \\
%     & $J$=1--0 & 0.1       & 0.02                \\
%$^13$CO & $J$=2--1 & 0.1       & 0.028               \\
%     & $J$=1--0 & 0.1       & 0.014               \\
%C$^18$O & $J$=2--1 & 0.1       & 0.018               \\
%     & $J$=1--0 & 0.1       & 0.014                  \\
% \hline
% \end{tabular}
%\end{table}

\subsection{Absorption data from ALMA calibration data}
The absorption study was also performed using archived ALMA data taken for calibration with QSO J1851+0035. 
%For the CO, $^{13}$CO, and C$^{18}$O $J$=1--0 line observations in ALMA Band 3, the angular resolution, frequency resolution, and sensitivity were 0.351$''$, 0.315$''$, 0.315$''$, 61.04, 30.52, 30.52 kHz, and 1.30, 0.70, 0.70 mJy beam$^{-1}$, respectively. 
%For the CO, $^{13}$CO, and C$^{18}$O $J$=2--1 line observations in ALMA Band 6, the angular resolution and frequency resolution was 0.138$''$ and 61.04 kHz, and the sensitivities were 0.72, 0.70, and 0.71 mJy beam$^{-1}$, respectively. 
These data were taken for calibration purposes, and the angular resolution, frequency resolution, and sensitivity were not as good as those of the CO data. 
Details of the data used for this study are shown in Tables \ref{table:archive-1a}--\ref{table:archive-3} of Appendix \ref{sec:archive}. 

The data reduction was conducted using the CASA version 4.7.2, 6.4.3, 6.4.4, 6.5.0, and 6.5.2.

In the archived ALMA calibration data, the raw visibility data for the calibrator were flagged to mask the frequencies with significant absorption lines or contamination by atmospheric lines and interference. 
We thus generated a script to remove the flags in those parts to include them in our analysis.
After calibrating the visibility data, the analysis was performed as described in \ref{sec:data_2}.
%no deconvolution using the CLEAN method was performed, and a dirty map was created. The brightest peak of the QSO was extracted and divided by the intensity of the continuous wave to obtain a normalized profile. Natural weighting was selected to maximize the continuum sensitivity.
Careful removal of the effects of the contamination by other species and other hyperfine components was also done. 
If components are available, the study was done based on these lines without using the possibly contaminated lines. 
Nevertheless, some velocity components were heavily contaminated by the other lines and were not included in the study for CN, HCN, and C$_2$H. 

\subsection{{\rm CO} emission data from NRO 45 m telescope data archive}

To obtain the emissivity of individual components and the lower limit to the CO excitation temperature, the single pointing observations data in CO and $^{13}$CO $J$=1--0 emission line taken with the NRO 45 m telescope on 2019-04-07 UTC (Observation ID: 20190408020229, 20190408031727, 20190408045557, 20190408060934) were downloaded from Nobeyama 45m and ASTE Science Data Archive. 
Frequency resolution and bandwidth were set at 122.07 kHz (0.32 km~s$^{-1}$ at 115 GHz) and 250 MHz (648 km~s$^{-1}$). 
The final spectra of CO and $^{13}$CO after flagging, calibration, and linear baseline subtraction had a 1$\sigma$ sensitivity of 185 and 63 mK, respectively, in the main beam temperature $T_{\rm{mb}}$ at the full velocity resolution. 

\subsection{Additional emission data with NRO 45 m telescope}

To estimate the line ratios and excitation temperatures of dense gas tracers, the emission line data in HCO$^+$, HCN, HNC $J$=1--0 and C$_2$H $N$=1--0 were taken with the NRO 45 m telescope. 
FOREST (FOur-beam REceiver System on the 45-m Telescope) \citep{2016SPIE.9914E..1ZM} was used for observations in a single pointing mode. 
The 2$\times$2 beams were aligned to have a beam spacing of $\sim$50$''$, and the full width of each beam was 18.3$\pm$$0.2''$ (H) 18.1$\pm$$0.2''$ (V) at 86 GHz and was 14.4$\pm$$0.4''$ (H) 14.6$\pm$$0.3''$ (V) at 110 GHz, respectively. 
The LO frequency was set to 95.0 GHz so that four lines (HCO$^+$, HCN, HNC $J$=1--0 and C$_2$H $N$=1--0) could be observed simultaneously. 
The observation was conducted in the mornings of 2023-03-21 and 2023-04-24 JST. 
The system noise temperature, including the sky, ranged from 166 to 358 K at the elevation of the source. 
Off position was set at RA(J2000)=18:59:09.616, DEC(J2000)=$-$00:04:12.52. 
This was confirmed to be emission-free in the observed lines through observations with relatively far different off positions known to be free of CO emission. 
Telescope pointing was established every $\sim$50 minutes with a nearby SiO maser source toward OH31.7$-$0.8 and was accurate at the level of 5$''$. 
The intensity calibration was done using the chopper wheel method. 
W51 (RA(J2000)=19:23:43.9, DEC(J2000)=$+$14:30:30.5) and G34.3$+$0.1 (RA(J2000)=18:53:18.7, DEC(J2000)=$+$01:14:58) were observed as standard sources to establish the absolute line intensity scale of the four beams. 
The scaled values for the HCO$^+$ $J$=1--0 line agree with those taken with the IRAM 30-m telescope by \citep{refId0}. 
The frequency resolution and bandwidth were set at 122.07 kHz (0.32 km~s$^{-1}$ at 115 GHz) and 250 MHz (648 km~s$^{-1}$), respectively. 
After flagging, calibration, and linear baseline subtraction, the spectra had a 1$\sigma$ sensitivity of 2 mK at the 2-binning velocity resolution.

\subsection{{\rm CO} maps from FUGIN survey}

To understand the nature of the molecular gas toward the QSO, the CO and $^{13}$CO $J$=1--0 channel maps of the FUGIN (FOREST unbiased Galactic plane imaging survey with the Nobeyama 45 m telescope) project \citep{2017PASJ...69...78U} were used. 
The original data were sliced to create a subset of $0.5^{\circ}$$\times$$0.5^{\circ}$ area centered at the line of sight toward QSO J1851+0035, and were smoothed in space and velocity domains for better signal-to-noise ratio. 
Specifications of the original and the smoothed sub-sample of the FUGIN data are summarized in Table \ref{table:fugin}. 

\begin{table}[htb!]
 \caption{Specifications of the original and the smoothed FUGIN sub-sample data used for this study}
 \label{table:fugin}
 \centering
  \begin{tabular}{lcc}
   \hline \hline
 & Original & Smoothed \\
   \hline
Spatial resolution & 20$''$ & 40$''$ \\
Spacial area & \multicolumn{2}{c}{$l$=33.25--33.75$^\circ$; $b$=0.04--0.34$^\circ$} \\
Spacial sampling & \multicolumn{2}{c}{$\Delta l$=$\Delta b$=8.5$''$} \\
Velocity range & $-100$--200 km~s$^{-1}$ & $-30$--125 km~s$^{-1}$ \\
Velocity resolution & 0.65 km~s$^{-1}$ & 1.3 km~s$^{-1}$ \\
$1\sigma$ sensitivity & 1.5 K (CO) & 0.5 K (CO) \\
 & 0.9 K ($^{13}$CO) & 0.3 K ($^{13}$CO) \\
   \hline
  \end{tabular}
\end{table}

\section{Results}

\subsection{Identification of species}
\label{sec:line_identification}

\subsubsection{Detected species}
\label{sec:detected_species}

\begin{figure}[htb!]
%\epsscale{0.5}
\plotone{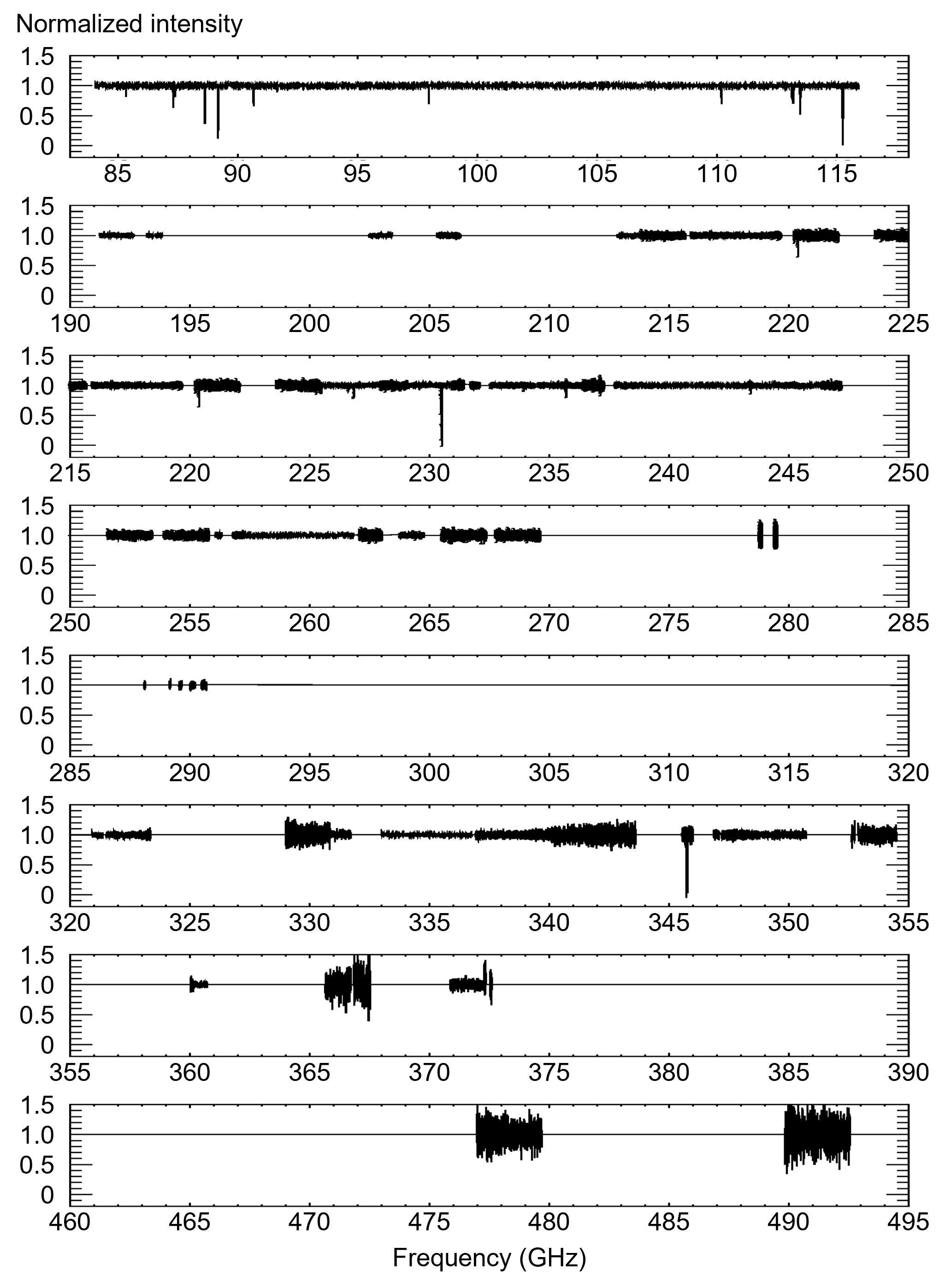}
\caption{Mosaiced absorption line profile toward QSO J1851+0035 for selected frequency ranges in ALMA Band 3 (84--116 GHz), Band 5 (163--211 GHz), Band 6 (211--275 GHz), Band 7 (275--373 GHz) and Band 8 (385--500 GHz). The profile was normalized to the continuum level. 
\label{fig:band35678}}
\end{figure}

Figure \ref{fig:band35678} shows the complete spectrum normalized to the detected continuum level. 
This figure illustrates the frequency coverage and sensitivity of the archive data used for this study, together with major lines and possible interference. 
The continuum level at the $1''$ resolution corresponds to $\sim$0.5 Jy at 100 GHz and $\sim$0.2 Jy at 350 GHz, respectively. 
Changes in the continuum level at this high-temperature range do not significantly affect the results of this study since the excitation temperature of the lines is significantly lower than these values, and the analysis is based on the opacities of the absorption lines. 

Fourteen species (CO, $^{13}$CO, HCO$^+$, H$^{13}$CO$^+$, HCO, C$_2$H, $c$-C$_3$H, $c$-C$_3$H$_2$, CN, HCN, HNC, CS, SO, and SiO) were detected, including five with significant hyperfine structure, i.e., HCO, C$_2$H, $c$-C$_3$H, CN, and HCN. 
The line identification was carried out with the Cologne Database for Molecular Spectroscopy (CDMS) \citep{2001A&A...370L..49M, 2005JMoSt.742..215M}, the Jet Propulsion Laboratory (JPL) Submillimeter, Millimeter, and Microwave Spectral Line Catalog \citep{1998JQSRT..60..883P}, and Spectral Line Atlas of Interstellar Molecules (SLAIM).

Tables \ref{table:mol_data-1} and \ref{table:mol_data-2} show the spectral parameters of all detected or searched transitions. 
For the partition function $Q_{\rm CMB}$, we either referenced value from CDMS or summed up the states with $Q_{\rm CMB}$ = $\sum g_{\rm u} \exp (\frac{-E_{\rm l}}{k_{\rm B} T_{\rm CMB}})$, where $g_{\rm u}$ is the degeneracy of the upper state, $E_{\rm l}$ is the energy level of the lower state, $k_{\rm B}$ is the Boltzmann constant, and $T_{\rm CMB}$ is the cosmic microwave background temperature (2.73 K).

\begin{table}[htb!]
 \caption{Molecular data for detected/searched transitions in ALMA Band 3}
 \label{table:mol_data-1}
 \centering
  \begin{tabular}{lcrlrrrl}
   \hline \hline
  Species & Transitions & \multicolumn{1}{c}{Frequency} & \multicolumn{1}{c}{$\mu$} & \multicolumn{1}{c}{$S_{\rm ul}\mu^2$} & \multicolumn{1}{c}{$Q_{\rm CMB}$} & \multicolumn{1}{c}{$E_{\rm l}/k_{\rm B}$} & Reference \\
    & & \multicolumn{1}{c}{(GHz)} & \multicolumn{1}{c}{(D)} & \multicolumn{1}{c}{(D$^2$)} & & \multicolumn{1}{c}{(K)} \\
   \hline
   HC$^{18}$O$^+$ & $J$=1--0 & 85.162223 & 3.90 & 15.210 & 1.7259 & 0.00 & CDMS \\
   $c$-${\rm C_3H_2}$ & $J_{K_a,K_c}$=$2_{1,2}$--$1_{0,1}$ (ortho) & 85.338893 & 3.43 & 52.945 & 8.8732 & 2.35 &JPL\\
   HCS$^+$ & $J$=2--1 & 85.347890 & 1.86 & 7.667 & 3.0215 & 2.05 & CDMS \\
   SO & $J_K$=$2_2$--$1_1$ & 86.093950 & 1.535 & 3.534 & 4.0346 & 15.18 & CDMS \\
   H$^{13}$CN & $J$=1--0, $F$=2--1 & 86.340163 & 2.9852 & 14.853 & 5.1261 & 0.00 & CDMS \\
   H$^{13}$CO$^+$ & $J$=1--0 & 86.754288 & 3.90 & 15.211 & 1.7028 & 0.00 & CDMS \\
   HCO & $N_K$=$1_{0,1}$--$0_{0,0}$, $J$=$\frac{3}{2}$--$\frac{1}{2}$, $F$=2--1 & 86.670760  & 1.3626 & 3.094 & 6.7785 & 0.02 & JPL \\
   HCO & $N_K$=$1_{0,1}$--$0_{0,0}$, $J$=$\frac{3}{2}$--$\frac{1}{2}$, $F$=1--0 & 86.708360 & 1.3626 & 1.817 & 6.7785 & 0.00 & JPL \\
   HCO & $N_K$=$1_{0,1}$--$0_{0,0}$, $J$=$\frac{1}{2}$--$\frac{1}{2}$, $F$=1--1 & 86.777460 & 1.3626 & 1.817 & 6.7785 & 0.02 & JPL \\
   HCO & $N_K$=$1_{0,1}$--$0_{0,0}$, $J$=$\frac{1}{2}$--$\frac{1}{2}$, $F$=0--1 & 86.805780 & 1.3626 & 0.619 & 6.7785 & 0.02 & JPL \\
   SiO & $J$=2--1 & 86.846960 & 3.098 & 19.197 & 2.98\phantom{00} & 2.08 & CDMS \\
   C$_2$H & $N$=1--0, $J$=$\frac{3}{2}$--$\frac{1}{2}$, $F$=1--1 & 87.284105 & 0.769 & 0.100 & 6.7735 & 0.00 & CDMS \\
   C$_2$H & $N$=1--0, $J$=$\frac{3}{2}$--$\frac{1}{2}$, $F$=2--1 & 87.316898 & 0.769 & 0.986 & 6.7735 & 0.00 & CDMS \\
   C$_2$H & $N$=1--0, $J$=$\frac{3}{2}$--$\frac{1}{2}$, $F$=1--0 & 87.328585 & 0.769 & 0.491 & 6.7735 & 0.00 & CDMS \\
   C$_2$H & $N$=1--0, $J$=$\frac{1}{2}$--$\frac{1}{2}$, $F$=1--1 & 87.401989 & 0.769 & 0.491 & 6.7735 & 0.00 & CDMS \\
   C$_2$H & $N$=1--0, $J$=$\frac{1}{2}$--$\frac{1}{2}$, $F$=0--1 & 87.407165 & 0.769 & 0.197 & 6.7735 & 0.00 & CDMS \\
   C$_2$H & $N$=1--0, $J$=$\frac{1}{2}$--$\frac{1}{2}$, $F$=1--0 & 87.446470 & 0.769 & 0.100 & 6.7735 & 0.00 & CDMS \\
   HCN & $J$=1--0, $F$=1--1 & 88.630416 & 2.984 & 8.912 & 5.03\phantom{00} & 0.00 & CDMS \\
   HCN & $J$=1--0, $F$=2--1 & 88.631848 & 2.984 & 14.852 & 5.03\phantom{00} & 0.00 & CDMS \\
   HCN & $J$=1--0, $F$=0--1 & 88.633936 & 2.984 & 2.971 & 5.03\phantom{00} & 0.00 & CDMS \\
   HCO$^+$ & $J$=1--0 & 89.188525 & 3.90 & 15.210 & 1.6691 & 0.00 & CDMS \\
   HOC$^+$ & $J$=1--0 & 89.487414 & 2.77 & 7.678 & 1.6651 & 0.00 & CDMS \\
   $l$-C$_3$H$^+$ & $J$=4--3 & 89.957625 & 3.0 & 36.000 & 5.3963 & 6.48 & CDMS \\
   HNC & $J$=1--0 & 90.663568 & 3.05 & 9.302 & 1.6497 & 0.00 & CDMS \\
   $c$-C$_3$H &$N_{K_a,K_c}$=$2_{1,2}$--$1_{1,1}$, $J$=$\frac{5}{2}$--$\frac{3}{2}$, $F$=3--2 & 91.494349 & 2.4 & 12.096 & 5.6348 & 2.13 & \cite{1987ApJ...322L..55Y}\\
   $c$-C$_3$H &$N_{K_a,K_c}$=$2_{1,2}$--$1_{1,1}$, $J$=$\frac{5}{2}$--$\frac{3}{2}$, $F$=2--1 & 91.497608 & 2.4 & 7.776 & 5.6348 & 2.13 & \cite{1987ApJ...322L..55Y}\\
   $c$-C$_3$H &$N_{K_a,K_c}$=$2_{1,2}$--$1_{1,1}$, $J$=$\frac{3}{2}$--$\frac{1}{2}$, $F$=2--1 & 91.699471 & 2.4 & 7.720 & 5.6348 & 2.13 & \cite{1987ApJ...322L..55Y}\\
   CH$_3$CN & $J_K$=$5_0$--$4_0$ & 91.987086 & 3.922 & 76.558 & 13.8357 & 8.83 & SLAIM \\
   N$_2$H$^+$ & $J$=1--0, $F1$=2--1, $F$=3--2 & 93.173772 & 3.40 & 26.972 & 14.5633 & 0.00 & CDMS \\
   C$^{34}$S & $J$=2--1 & 96.412950 & 1.958 & 7.650 & 2.7242 & 1.62 & CDMS \\
   CH$_3$OH & $J_{K_a,K_c}$=$2_{0,2}$--$1_{0,1}$, A & 96.741375 & 0.899 & 1.617 & 11.8899 & 2.32 & CDMS \\
   CS & $J$=2--1 & 97.980953 & 1.958 & 7.644 & 2.6827 & 2.35 & CDMS \\
   SO & $J_K$=$3_2$--$2_1$ & 99.299870 & 1.535 & 6.911 & 4.0346 & 4.46 & CDMS \\
   CF$^+$ & $J$=1--0 & 102.587484 & 1.072 & 1.150 & 1.5167 & 0.00 & CDMS \\
   H$_2$CS & $J_{K_a,K_c}$=$3_{0,3}$--$2_{0,2}$ & 103.040452 & 1.649 & 8.158 & 3.7879 & 4.95 & CDMS \\
   SO$_2$ & $J_{K_a,K_c}$=$3_{1,3}$--$2_{0,2}$ & 104.029418 & 1.6331 & 5.373 & 5.3815 & 2.75 & CDMS \\
   $^{13}$CN & $N$=1--0, $J$=$\frac{3}{2}$--$\frac{1}{2}$, $F1$=2--1, $F$=3--2 & 108.780201 & 1.45 & 4.905 & 17.3716 & 0.03 & CDMS \\
   C$^{18}$O & $J$=1--0 & 109.782173 & 0.11049 & 0.012 & 1.4491 & 0.00 & CDMS \\
   $^{13}$CO & $J$=1--0 & 110.201354 & 0.11046 & 0.024 & 2.8912 & 0.00 & CDMS \\
   CN$^-$ & $J$=1--0, $F$=1--1 & 112.263694 & 0.65 & 0.141 & 4.2911 & 0.00 & CDMS \\
   CN$^-$ & $J$=1--0, $F$=2--1 & 112.264993 & 0.65 & 0.235 & 4.2911 & 0.00 & CDMS \\
   CN & $N$=1--0, $J$=$\frac{1}{2}$--$\frac{1}{2}$, $F$=$\frac{1}{2}$--$\frac{3}{2}$ & 113.144157 & 1.45 & 1.249 & 8.5178 & 0.00 & CDMS \\
   CN & $N$=1--0, $J$=$\frac{1}{2}$--$\frac{1}{2}$, $F$=$\frac{3}{2}$--$\frac{1}{2}$ & 113.170492 & 1.45 & 1.220 & 8.5178 & 0.00 & CDMS \\
   CN & $N$=1--0, $J$=$\frac{1}{2}$--$\frac{1}{2}$, $F$=$\frac{3}{2}$--$\frac{3}{2}$ & 113.191278 & 1.45 & 1.584 & 8.5178 & 0.00 & CDMS \\
   CN & $N$=1--0, $J$=$\frac{3}{2}$--$\frac{1}{2}$, $F$=$\frac{3}{2}$--$\frac{1}{2}$ & 113.488120 & 1.45 & 1.584 & 8.5178 & 0.00 & CDMS \\
   CN & $N$=1--0, $J$=$\frac{3}{2}$--$\frac{1}{2}$, $F$=$\frac{5}{2}$--$\frac{3}{2}$ & 113.490970 & 1.45 & 4.205 & 8.5178 & 0.00 & CDMS \\
   CN & $N$=1--0, $J$=$\frac{3}{2}$--$\frac{1}{2}$, $F$=$\frac{1}{2}$--$\frac{1}{2}$ & 113.499644 & 1.45 & 1.249 & 8.5178 & 0.00 & CDMS \\
   CN & $N$=1--0, $J$=$\frac{3}{2}$--$\frac{1}{2}$, $F$=$\frac{3}{2}$--$\frac{3}{2}$ & 113.508907 & 1.45 & 1.220 & 8.5178 & 0.00 & CDMS \\
   CO & $J$=1--0 & 115.271202 & 0.11011 & 0.012 & 1.4053 & 0.00 & CDMS \\
   \hline
  \end{tabular}
\end{table}

\begin{table}[htb!]
 \caption{Molecular data for detected/searched transitions in ALMA Bands 5, 6, 7, 8}
 \label{table:mol_data-2}
 \centering
  \begin{tabular}{lcrlrrrl}
   \hline \hline
  Species & Transitions & \multicolumn{1}{c}{Frequency} & \multicolumn{1}{c}{$\mu$} & \multicolumn{1}{c}{$S_{\rm ul}\mu^2$} & \multicolumn{1}{c}{$Q_{\rm CMB}$} & \multicolumn{1}{c}{$E_{\rm l}/k_{\rm B}$} & Reference \\
    & & \multicolumn{1}{c}{(GHz)} & \multicolumn{1}{c}{(D)} & \multicolumn{1}{c}{(D$^2$)} & & \multicolumn{1}{c}{(K)} \\
   \hline
   C$_2$S & $N$=3--2, $J$=3--3 & 191.940220 & 2.88 & 0.251 & \multicolumn{1}{c}{31.7632} & 3.23 & CDMS \\
   C$_2$S & $N$=3--2, $J$=2--1 & 214.571817 & 2.88 & 0.111 & \multicolumn{1}{c}{31.7632} & 0.53 & CDMS \\
   H$_2$CO & $J_{K_a,K_c}$=$3_{0,3}$--$2_{0,2}$ & 218.222192 & 2.3317 & 16.308 & 2.0165 & 10.48 & CDMS \\
   C$^{18}$O & $J$=2--1 & 219.560354 & 0.11049 & 0.024 & 1.4491 & 5.27 & CDMS \\
   $^{13}$CO & $J$=2--1 & 220.398684 & 0.11046 & 0.049 & 2.8912 & 5.29 & CDMS \\
   CN & $N$=2--1, $J$=$\frac{3}{2}$--$\frac{1}{2}$, $F$=$\frac{3}{2}$--$\frac{3}{2}$ & 226.632190 & 1.45 & 1.258 & 8.5178 & 5.43 & CDMS \\
   CN & $N$=2--1, $J$=$\frac{3}{2}$--$\frac{1}{2}$, $F$=$\frac{5}{2}$--$\frac{3}{2}$ & 226.659558 & 1.45 & 4.191 & 8.5178 & 5.43 & CDMS \\
   CN & $N$=2--1, $J$=$\frac{3}{2}$--$\frac{1}{2}$, $F$=$\frac{1}{2}$--$\frac{1}{2}$ & 226.663693 & 1.45 & 1.249 & 8.5178 & 5.43 & CDMS \\
   CN & $N$=2--1, $J$=$\frac{3}{2}$--$\frac{1}{2}$, $F$=$\frac{3}{2}$--$\frac{1}{2}$ & 226.679311 & 1.45 & 1.554 & 8.5178 & 5.43 & CDMS \\
   CN & $N$=2--1, $J$=$\frac{5}{2}$--$\frac{3}{2}$, $F$=$\frac{5}{2}$--$\frac{3}{2}$ & 226.874191 & 1.45 & 4.247 & 8.5178 & 5.45 & CDMS \\
   CN & $N$=2--1, $J$=$\frac{5}{2}$--$\frac{3}{2}$, $F$=$\frac{7}{2}$--$\frac{5}{2}$ & 226.874781 & 1.45 & 6.728 & 8.5178 & 5.45 & CDMS \\
   CN & $N$=2--1, $J$=$\frac{5}{2}$--$\frac{3}{2}$, $F$=$\frac{3}{2}$--$\frac{1}{2}$ & 226.875896 & 1.45 & 2.527 & 8.5178 & 5.45 & CDMS \\
   CO & $J$=2--1 & 230.538000 & 0.11011 & 0.024 & 1.4053 & 5.53 & CDMS \\
   CO$^+$ & $N$=2--1, $J$=$\frac{3}{2}$--$\frac{1}{2}$ & 235.789641 & 2.61 & 9.046 & 2.7714 & 5.67 & CDMS \\
   CO$^+$ & $N$=2--1, $J$=$\frac{5}{2}$--$\frac{3}{2}$ & 236.062553 & 2.61 & 16.285 & 2.7714 & 5.67 & CDMS \\
   SO & $J_K$=$1_2$--$0_1$ & 329.385477 & 1.535 & 0.103 & 4.0346 & 0.00 & CDMS \\
   C$^{18}$O & $J$=3--2 & 329.330553 & 0.11049 & 0.037 & 1.4491 & 15.81 & CDMS\\
   $^{13}$CO & $J$=3--2 & 330.587965 & 0.11046 & 0.073 & 2.8912 & 15.87 & CDMS\\
   CN & $N$=3--2, $J$=$\frac{7}{2}$--$\frac{5}{2}$, $F$=$\frac{7}{2}$--$\frac{5}{2}$ & 340.247770 & 1.45 & 9.011 & 8.5178 & 16.33 & CDMS \\
   CO & $J$=3--2 & 345.795990 & 0.11011 & 0.036 & 1.4053 & 16.60 & CDMS \\
   SH$^+$ & $N$=0--1, $J$=1--0, $F$=$\frac{3}{2}$--$\frac{1}{2}$ & 345.944350 & 1.285 & 0.954 & 6.0025 & 0.00 & CDMS \\
   ND & $N$=1--0, $J$=0--1, $F1$=1--2, $F$=2--3 & 491.933934 & 1.389 & 1.570 & 26.9872 & 0.00 & CDMS \\
   C & $^3P_1$--$^3P_0$ & 492.160651 & \multicolumn{1}{c}{---} & 0.00017 & 1.0005 & 0.00 & CDMS \\
   \hline
  \end{tabular}
\end{table}

\subsubsection{Marginal and non-detection}
\label{sec:non-detection}

C$^{18}$O, H$_2$CO, and C were marginally detected for some velocity components with the largest $^{13}$CO column density, presumably dark clouds. 
CO$^+$, HC$^{18}$O$^+$, HOC$^+$, CH$_3$OH, $l$-C$_3$H$^+$, $^{13}$CN, CN$^-$, H$^{13}$CN, CH$_3$CN, ND, N$_2$H$^+$, SH$^+$, C$^{34}$S, HCS$^+$, H$_2$CS, C$_2$S, SO$_2$, and CF$^+$ were also searched in the present data sets but were not detected.

The column density $N$ of the absorbing species can be deduced from the opacity $\tau$ of the absorption line integrated over the velocity range $v$ by assuming optically thin lines and the LTE (local thermodynamic equilibrium) conditions, 
\begin{equation}
\label{eq:n_tot}
N = \frac{3hQ(T_{\rm ex})}{8\pi^3\mu^2S_{\rm ul}} \frac{\exp(\frac{E_{\rm l}}{k_{\rm B} T_{\rm ex}})} {[1-\exp(\frac{-h\nu}{k_{\rm B}T_{\rm ex}})]} \int\tau dv, 
\end{equation}
where $h$ is the Planck constant, $\nu$ is the frequency of the transition, $k_{\rm B}$ is the Boltzmann constant, $Q(T_{\rm ex})$ is the partition function at the (assumed) excitation temperature $T_{\rm ex}$, $E_{\rm l}$ is the energy of the lower level, $\mu$ is the dipole moment, and $S_{\rm ul}$ is the line strength. 

The column densities $N$ of these species or their 1$\sigma$ upper limits for Components 7 and 16 deduced in this manner are tabulated in Table \ref{table:non-detection}. 
The abundance was calculated as $X$(M) $\equiv$ $N$(M)/$N$(H$_2$) by assuming $N$(H$_2$) to be $2 \times 10^{21}$ cm$^{-2}$ for Components 7 and 16. 
This $N$(H$_2$) value was determined to be in reasonable agreement with the one estimated from CO emission using the CO-to-H$_2$ conversion factor, the one from optically thin $^{13}$CO emission under the LTE condition, and the ones from HCO$^+$ and $c$-C$_3$H$_2$ absorption column densities. 
The values for TMC-1 Cyanopolyyne Peak (CP), which is often used to test chemical models, and FG0.89SW, which is a $z$ = 0.89 absorber with a large column of absorbing gas ($N$(H$_2$) $>$ $10^{22}$ cm$^{-2}$) toward PKS 1830$-$211, are also included as references. 

\begin{table}[htb!]
 \caption{Column densities, abundances or their 1$\sigma$ upper limits for dark clouds -- Components 7 and 16, TMC-1 Cyanopolyyne Peak, and FG0.89SW.}
 \label{table:non-detection}
  \begin{center}
  \begin{tabular}{lcrcrrcrc}
   \hline \hline
  & \multicolumn{2}{c}{Component 7} & \multicolumn{2}{c}{Component 16} & \multicolumn{2}{c}{TMC-1 (CP)} & \multicolumn{2}{c}{FG0.89SW} \\
  Species & \multicolumn{1}{c}{$N$(M)} & \multicolumn{1}{c}{$\log X$(M)} & \multicolumn{1}{c}{$N$(M)} & \multicolumn{1}{c}{$\log X$(M)} & \multicolumn{1}{c}{$\log X$(M)} & Ref & \multicolumn{1}{c}{$\log X$(M)} & Ref \\
  & \multicolumn{1}{c}{($10^{12}$ cm$^{-2}$)} & & \multicolumn{1}{c}{($10^{12}$ cm$^{-2}$)} & & & & & \\ 
   \hline 
H$_2$ & $2 \times 10^{9}$* & 0\phantom{.00} & $2 \times 10^{9}$* & 0\phantom{.00} & 0\phantom{.00} & & 0\phantom{.00} & \\ 
C & (19.1$\pm$4.17)$\times$$10^4$\dag& $-4.02$ & (6.16$\pm$3.77)$\times$$10^4$\dag & $-4.51$ & & & $-3.93$ & (7) \\ 
CO & (saturated) & ($-4.06$)\ddag & (saturated) & ($-4.52$)\ddag & ($-4.01$) & (5) & $-4.20$ & (7) \\ 
$^{13}$CO & 3461$\pm$47 & $-5.76$ & 1209$\pm$24 & $-6.22$ & $-5.79$ & (5) & $<$$-6.00$ & (7) \\ 
C$^{18}$O & 19.7$\pm$13.5 & $-8.01$ & 30.5$\pm$15.5 & $-7.82$ & $-6.79$ & (5) & $<$$-6.00$ & (7) \\
CO$^+$ & $<$$0.72$ & $<$$-9.44$ & $<$$0.83$ & $<$$-9.38$ & & & & \\
HCO$^+$ & (saturated) & ($-8.76$) & (saturated) & ($-8.64$) & ($-8.00$) & (5) & $-8.06$ & (8) \\ 
H$^{13}$CO$^+$ & 0.07$\pm$0.02 & $-10.46$ & 0.09$\pm$0.02 & $-10.34$ & $-9.78$ & (5) & $-$9.38 & (8) \\ 
HC$^{18}$O$^+$ & $<$$0.03$ & $<$$-10.83$ & 0.05$\pm$0.01 & $-10.64$ & $-10.78$ & (5) & $-$9.74 & (8) \\ 
HOC$^+$ & $<$$0.02$ & $<$$-10.98$ & $<$$0.02$ & $<$$-11.05$ & & & $-9.80$ & (8) \\ 
HCO & 1.85$\pm$0.90 & $-9.03$ & 1.92$\pm$0.65 & $-9.02$ & & & $-9.19$ & (8) \\ 
H$_2$CO & 4.52$\pm$1.70 & $-8.65$ & 2.47$\pm$1.54 & $-8.91$ & $-7.70$ & (2) & $-7.92$ & (8) \\ 
CH$_3$OH & $<$$3.85$ & $<$$-8.72$ & $<$$3.95$ & $<$$-8.70$ & $-8.84$ & (3) & $-8.07$ & (8) \\ 
C$_2$H & 81.36$\pm$5.19 & $-7.39$  & 41.68$\pm$2.59 & $-7.68$ & $-7.12$ & (2) & $-7.20$ & (8) \\ 
$l$-C$_3$H$^+$ & $<$$0.34$ & $<$$-9.77$ & $<$$0.27$ & $<$$-9.87$ & & & & \\ 
$c$-C$_3$H & 0.85$\pm$0.21 & $-9.37$ & 0.45$\pm$0.17 & $-9.65$ & $-8.52$ & (3) & & \\
$c$-C$_3$H$_2$ (o) & 2.99$\pm$0.09 & $-8.83$ & 1.22$\pm$0.09 & $-9.21$ & $-8.73$ & (3) & $-8.71$ & (8) \\
CN & 30.87$\pm$0.37 & $-7.82$ & 22.21$\pm$0.31 & $-7.95$ & $-7.52$ & (2) & & \\ 
$^{13}$CN & $<$$0.09$ & $<$$-10.35$ & $<$$0.09$ & $<$$-10.37$ & & & & \\ 
CN$^-$ & $<$$5.95$ & $<$$-8.53$ & $<$$6.36$ & $<$$-8.50$ & $<$$-9.85$ & (4) & & \\ 
HCN & 9.58$\pm$0.30 & $-8.37$ & 8.42$\pm$0.35 & $-8.38$ & ($-8.15$) & (5) & $-7.82$ & (8) \\ 
H$^{13}$CN & $<$$0.09$ & $<$$-10.35$ & $<$$0.09$ & $<$$-10.37$ & $-9.93$ & (5) & $-9.38$ & (8) \\ 
HNC & 1.97$\pm$0.06 & $-9.00$ & 1.45$\pm$0.05 & $-9.14$ & $-7.70$ & (2) & $-8.29$ & (8) \\ 
CH$_3$CN & $<$$1.62$ & $<$$-9.09$ & $<$$1.35$ & $<$$-9.17$ & $-9.00$ & (2) & $-9.19$ & (8) \\ 
ND & $<$$76.41$ & $<$$-7.42$ & $<$$60.70$ & $<$$-7.52$ & & & & \\ 
N$_2$H$^+$ & $<$$0.49$ & $<$$-9.61$ & $<$$0.42$ & $<$$-9.68$ & $-9.11$ & (5) & $-8.93$ & (8) \\ 
SH$^+$ & $<$$5.04$ & $<$$-8.60$ & $<$$3.66$ & $<$$-8.74$ & & & & \\ 
CS & 7.86$\pm$0.15 & $-8.41$ & 0.87$\pm$0.13 & $-9.36$ & $-7.89$ & (5) & $-7.55$ & (6) \\ 
C$^{34}$S & $<$$0.19$ & $<$$-10.01$ & $<$$0.16$ & $<$$-10.09$ & $-9.30$ & (3) & $-8.58$ & (6) \\
HCS$^+$ & $<$$0.76$ & $<$$-9.42$ & $<$$0.60$ & $<$$-9.52$ & $-9.40$ & (5) & & \\ 
H$_2$CS & $<$$0.69$ & $<$$-9.46$ & $<$$0.71$ & $<$$-9.45$ & $-8.38$ & (3) & $-9.42$ & (8) \\
C$_2$S & $<$$107.54$ & $<$$-7.27$ & $<$$90.92$ & $<$$-7.34$ & $-7.99$ & (3) & $-9.62$ & (8) \\
SO & 2.06$\pm$0.24 & $-9.00$ & $<$$0.21$ & $<$$-9.97$ & $-8.74$ & (5) & $-8.89$ & (8) \\ 
SO$_2$ & $<$$0.94$ & $<$$-9.33$ & $<$$0.79$ & $<$$-9.40$ & $-9.00$ & (2) & $<$$-10.28$ & (8) \\ 
SiO & 0.36$\pm$0.07 & $-9.74$ & $<$$0.06$ & $<$$-10.54$ & $-11.62$ & (1) & $-9.43$ & (8) \\
CF$^+$ & $<$$0.36$ & $<$$-9.75$ & $<$$0.34$ & $<$$-9.77$ & & & & \\ 
   \hline
  \end{tabular}
  \end{center}
\tablenotetext{*}{Assumed to be $N$(H$_2$) = $2 \times 10^{21}$ cm$^{-2}$ for Components 7 and 16 (See Table \ref{table:velcomp}) as the value in reasonable agreement with the one estimated from CO emission using the CO-to-H$_2$ conversion factor, the one from optically thin $^{13}$CO emission under the LTE condition, and the ones from HCO$^+$ and $c$-C$_3$H$_2$ absorption column densities. 
See, Section \ref{sec:molecular_gas_mass} for detailed discussions. }
\tablenotetext{\dag}{Calculated by assuming an excitation temperature of 37.5 K.}
\tablenotetext{\ddag}{Calculated from column density of $^{13}$CO by assuming the CO/$^{13}$CO ratio of 25. 
See, Section \ref{sec:co_isotopic_abundance_ratios} for the justification of this assumption rather than larger values. }
References: (1) \cite{1989ApJ...343..201Z}; (2) \cite{1992IAUS..150..171O}; (3) \cite{2016ApJS..225...25G}; (4) \cite{doi:10.1021/acs.chemrev.6b00480}; (5) \cite{Fuente2019}; (6) \cite{2006A&A...458..417M}; (7) \cite{2009ApJ...690L.130B}; (8) \cite{2011A&A...535A.103M}
\end{table}

\subsubsection{Possible interference}
\label{sec:interference}

Instrumental and atmospheric artifacts are also visible at some frequencies. 
Frequencies of possible interference are listed in Table \ref{table:interference}. 
Most of them were identified to be the atmospheric emission of O$_3$. 
The one at 91.66 GHz is the interference due to the 6th harmonics of the local oscillator for the Water Vapor Radiometer (WVR) installed on the 12 m antennas. 
The one at 213.57 GHz is close to the 14th harmonics but does not perfectly match. 
The possibility of high-$z$ lines will be tested in Section \ref{sec:redshifted_lines}. 

\begin{table}[htb!]
\caption{Possible interference}
\label{table:interference}
\centering
\begin{tabular}{cl}
\hline \hline
Frequency & Identification \\
(GHz) & \\
\hline
\phantom{0}91.66 & WVR LO 6th harmonics \\
\phantom{0}96.23 & atmospheric (O$_3$) \\
101.74 & atmospheric (O$_3$) \\
110.84 & atmospheric (O$_3$) \\
213.57 & unidentified \\
214.80 & atmospheric (O$_3$) \\
231.61 & unidentified \\
235.71 & atmospheric (O$_3$) \\
237.14 & atmospheric (O$_3$) \\
239.07 & atmospheric (O$_3$) \\
242.28 & unidentified \\
243.41 & atmospheric (O$_3$) \\
258.70 & atmospheric (O$_3$) \\
331.06 & unidentified \\
359.60 & atmospheric (O$_3$) \\
478.83 & unidentified \\
\hline
\end{tabular}
\end{table}

\subsection{Identification of velocity components} 
\label{sec:velocity_components}

\subsubsection{Overall characteristics of velocity components and species} 
\label{sec:velcomp_species}

\begin{figure}[htb!]
\epsscale{0.5}
\plotone{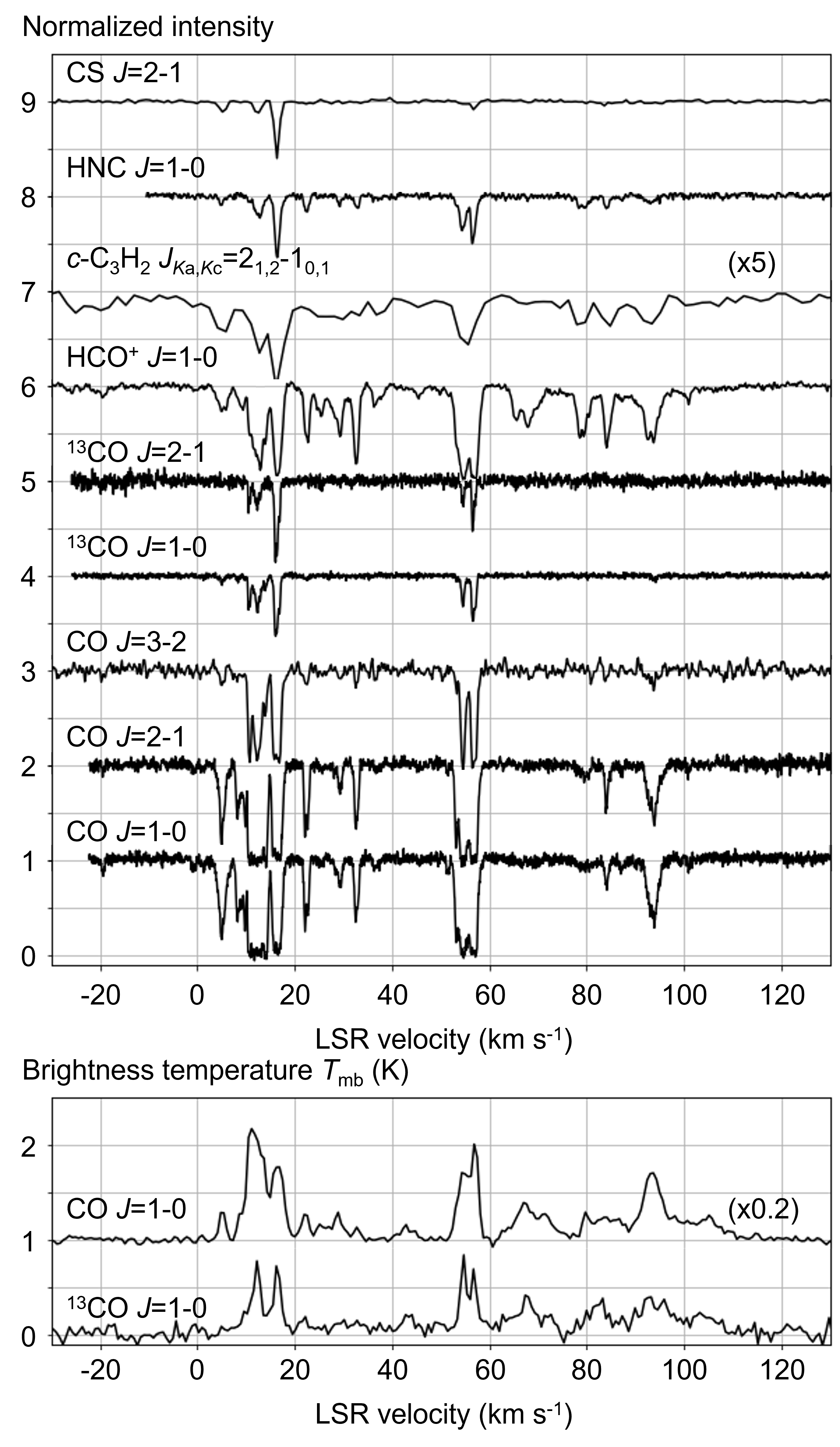}
\caption{Absorption profiles of CS $J$=2--1, HNC $J$=1--0, $c$-${\rm C_3H_2}$ $J_{K_a,K_c}$=$2_{1,2}$--$1_{0,1}$, HCO$^+$ $J$=1--0, $^{13}$CO $J$=2--1, $J$=1--0, and CO $J$=3--2, $J$=2--1, $J$=1--0 obtained with ALMA toward QSO J1851+0035. 
The intensity was normalized to the continuum level. 
Note that the velocity resolution is different. 
The spectra were offset for clarity. 
Emission profiles of CO and $^{13}$CO $J$=1--0 obtained with the NRO 45 m telescope toward the same direction are also shown. 
The intensity of CO $J$=1--0 was divided by 5 so that the components with CO/$^{13}$CO intensity ratio of 5 have the same appearance in these plots. 
See Appendix \ref{sec:absorption_profile} for the full spectrum of the detected/searched lines. \label{fig:em_abs}}
\end{figure}

In Figure \ref{fig:em_abs}, spectra of some species that characterize the absorption profile toward this line of sight are shown. 
HNC is an example of the CN-bearing molecules, $c$-${\rm C_3H_2}$ is that of hydrocarbons, and CS is that of the sulfur-bearing molecules and shock tracers. 
These species were chosen because they are free from hyperfine components, unlike CN, HCN, and C$_2$H. 

We identified velocity components along this line of sight based on the absorption profiles in the CO, $^{13}$CO, and HCO$^+$ $J$=1--0 lines obtained with ALMA, as well as the emission profile in the CO $J$=1--0 line observed with the NRO 45 m telescope as shown in Table \ref{table:velcomp}. 

From the line profiles shown in Figure \ref{fig:em_abs}, we note the following non-trivial characteristics: 
\begin{itemize}
\item[(1)] Both the CO and $^{13}$CO $J$=1--0 emission lines have near continuous emissivity in a broad velocity range from 0 to 120 km~s$^{-1}$ (near the terminal velocity). However, the main beam antenna temperature is usually low ($\sim$1 K). 
Their line profiles mimic each other with an overall intensity ratio of $\sim$6. 
\item[(2)] Both the CO and $^{13}$CO absorption lines are much more sparse than their corresponding emission lines and show discrete velocity components. 
The CO $J$=1--0 absorption is saturated ($e^{-\tau} \sim 0$) only for a few velocity components. 
This means that the common assumption of the CO $J$=1--0 emission being optically thick ($\tau$ $\gg$ 1) and beam-filling ($f$ $\simeq$ 1) is often invalid. 
The $^{13}$CO $J$=1--0 absorption is barely visible except for the saturated CO absorption systems. 
\item[(3)] The CO and $^{13}$CO absorption lines sometimes show (multiple) narrow-line components. 
\item[(4)] The spectra of the CO $J$=1--0 and $J$=2--1 absorption lines mimic each other with significantly smaller $J$=3--2 absorption, suggesting the excitation temperature is close to $E_{J=1}/k_{\rm B}$ (5.53 K for CO). 
\item[(5)] The HCO$^+$ $J$=1--0 absorption lines have near continuous absorption in a broad velocity range from 0 to 120 km~s$^{-1}$. 
The same applies to $c$-${\rm C_3H_2}$ $J_{K_a,K_c}$=$2_{1,2}$--$1_{0,1}$, while the individual features are less prominent in this line due to the poorer velocity resolution. 
\item[(6)] The $^{13}$CO and HNC $J$=1--0, and CS $J$=2--1 absorption lines have similar opacity. 
HNC is more commonly detected than $^{13}$CO in terms of opacity. 
CS is highly concentrated in one of the velocity components with the largest $^{13}$CO absorption column density. 
\item[(7)] There are a few velocity components where the CO emission and HCO$^+$ $J$=1--0 absorption are observed without significant CO absorption. 
The most prominent case is found near 68 km~s$^{-1}$. 
\end{itemize}

These findings will be discussed in the following sections. 

\subsubsection{Characteristics of the parent clouds} 
\label{sec:parent_clouds}

The types of the parent clouds of individual velocity components were identified in Table \ref{table:velcomp} based on their column density in Tables \ref{table:col_co_abs} and \ref{table:col_co_em} and their appearances in the CO and $^{13}$CO $J$=1--0 channel maps in Appendix \ref{sec:channel_map}, which includes locations of HII regions and dense clumps. 

We classified the molecular clouds based on the following criteria: 
\renewcommand{\labelitemi}{-}
\begin{itemize}
\item Diffuse clouds have a hydrogen column density $N$(H$_2$) $\lesssim 1 \times 10^{21}$ cm$^{-2}$. 
They are difficult to visualize in emission line mapping because their excitation temperature is close to the CMB.

\item Dark clouds are denser than diffuse clouds and have $N$(H$_2$) $\gtrsim 1 \times 10^{21}$ cm$^{-2}$, though their edges may have less column density. 
Most of the velocity components identified in this study belong to this group. 

\item GMCs have large peak brightness due to the heating by nearby/embedded sources or accompanied HII regions and/or ATLASGAL GCSC sources in the neighborhood ($\sim$10 pc). 
The association with HII regions was studied based on the HII region catalog in Table \ref{table:hrds} of Appendix \ref{sec:hrds}. 
Similarly, the association with dense clumps was studied based on the ATLASGAL GCSC source catalog in Table \ref{table:atlasgal} of Appendix \ref{sec:atlasgal}. 

\item Globules are isolated small clouds with a discontinuous brightness distribution and sizes up to a few pc when viewed in the channel maps and the position-velocity diagrams. 
They are much smaller in size than dark clouds and GMCs that extend several tens of parsecs. 
\end{itemize}

The Galactocentric distance ($R_{\rm G}$) was calculated from the center of the velocity range of the component by assuming the distance to the Galactic Center ($R_{\rm 0}$) to be 8.5 kpc and the flat rotation curve at 220 km~s$^{-1}$. 

\begin{figure}[htb!]
\plotone{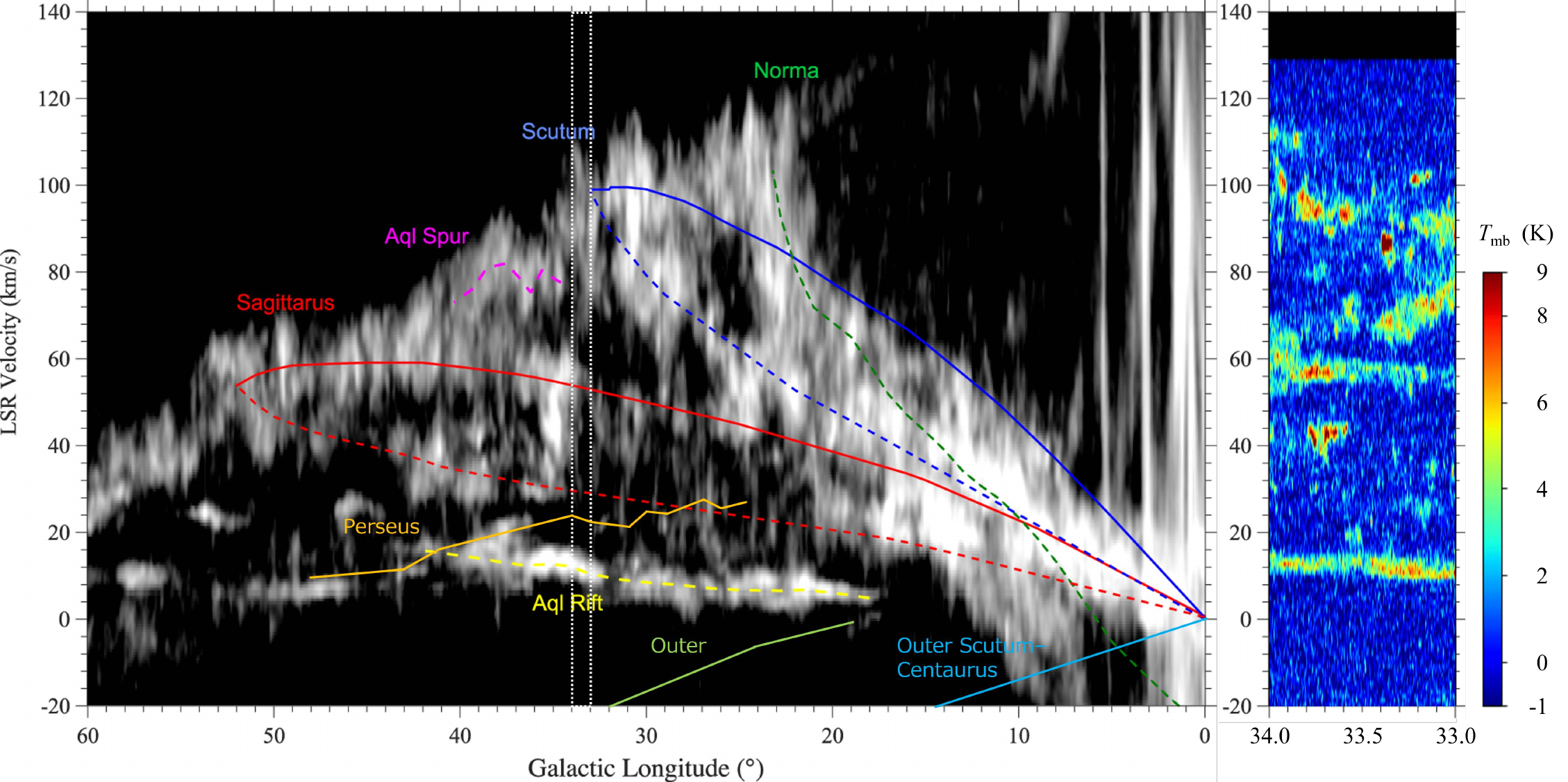}
\caption{({\it left}) CO longitude-velocity ($l$--$v$) diagram and traces of the Norma, Scutum, and Sagittarius arms as well as Aquila Rift in the first Galactic quadrant, with the near and far sides of the arms as dotted and solid lines, respectively \citep{2016ApJ...823...77R}. 
Traces of the Perseus, Outer, and Outer Scutum-Centaurus arms were also added. 
({\it right}) The $l$--$v$ diagram of CO $J$=1--0 emission obtained from the FUGIN survey \citep{2017PASJ...69...78U}, centered at the line of sight toward the QSO with $\Delta l$=1.0$^\circ$. 
\label{fig:arm}}
\end{figure}

The association with the spiral arms and some other features is given based on the identification of spiral arms by \cite{2016ApJ...823...77R}. 
These arms can be traced in the longitude-velocity ($l$--$v$) diagram (Figure \ref{fig:arm}), and possible association of the individual velocity components was judged based on this information and the apparent angular extent of the surrounding emission features. 

\begin{table}[htb!]
 \caption{Identification of velocity components.}
 \label{table:velcomp}
\begin{center}
\begin{tabular}{crrcllcl}
\hline \hline
& \multicolumn{2}{c}{Velocity} & & & & & \\
Comp & $v_{\rm min}$ & $v_{\rm max}$ & $R_{\rm G}$ & \multicolumn{1}{c}{Spiral arm?} & Cloud type/ & Case & \multicolumn{1}{c}{Note} \\
ID & \multicolumn{2}{c}{(km~s$^{-1}$)} & (kpc) & \multicolumn{1}{c}{(near/far)} & location & & \\
\hline 
\phantom{0}1 &$-$30.0 &$-$23.0 & 10.87\phantom{0} & interarm  & diffuse cloud & 3 & no CO emission \\
\phantom{0}2 &$-$23.0 &$-$17.0 & 10.18\phantom{0} & Outer Arm & diffuse cloud & 2 & weak CO emission \\
\phantom{0}3 &$-$17.0 &$-$6.0 & 9.39 & --- & --- & & no absorption/emission \\
\phantom{0}4 & $-$6.0 &  0.0 & 8.72 & interarm        & diffuse cloud & 2 & no CO emission \\
\phantom{0}5 &  0.0 &  7.0 & 8.26 & Per arm?          & diffuse cloud & 1 & \\
\phantom{0}6 &  7.0 & 14.5 & 7.81 & Aql Rift          & dark cloud   & 1 & saturated CO absorption \\
\phantom{0}7 & 14.5 & 20.0 & 7.44 & interarm (far)    & dark cloud   & 1 & saturated CO absorption; shock tracers \\
\phantom{0}8 & 20.0 & 23.5 & 7.21 & interarm          & globule edge & 2 & \\
\phantom{0}9 & 23.5 & 26.0 & 7.06 & Sgr arm (near)    & globule edge & 3 & \\
\phantom{}10 & 26.0 & 30.5 & 6.90 & Sgr arm (near)    & globule edge & 2 & \\
\phantom{}11 & 30.5 & 34.5 & 6.71 & Sgr arm (near)    & globule edge & 2 & \\
\phantom{}12 & 34.5 & 40.0 & 6.50 & interarm          & GMC edge     & 2 & ATLASGAL GCSC source \\
\phantom{}13 & 40.0 & 47.0 & 6.26 & interarm (far)    & GMC edge     & 3 & ATLASGAL GCSC source \\
\phantom{}14 & 47.0 & 51.5 & 6.05 & interarm (far)    & dark cloud edge & 3 & \\
\phantom{}15 & 51.5 & 55.5 & 5.90 & Sgr arm (far)     & dark cloud   & 1 & saturated CO absorption \\
\phantom{}16 & 55.5 & 60.0 & 5.76 & Sgr arm (far)     & dark cloud   & 1 & saturated CO absorption \\
\phantom{}17 & 60.0 & 76.0 & 5.45 & interarm (far)    & GMC edge     & 3 & adjacent to HCO$^+$/HCN clump \\
\phantom{}18 & 76.0 & 81.0 & 5.16 & Scu arm (tangent) & GMC edge     & 2 & ATLASGAL GCSC source \\
\phantom{}19 & 81.0 & 88.0 & 5.01 & Scu arm (tangent) & GMC edge     & 2 & \\
\phantom{}20 & 88.0 & 97.0 & 4.82 & Scu arm (tangent) & dark cloud edge & 1 & \\
\phantom{}21 & 97.0 &102.0 & 4.69 & Scu arm (tangent) & GMC edge     & 3 & ATLASGAL GCSC source \\
\phantom{}22 &102.0 &110.0 & 4.69 & Scu arm (tangent) & dark cloud edge & 3 & \\
\phantom{}23 &110.0 &120.0 & 4.69 & Scu arm (tangent) & dark cloud edge & 3 & \\
\phantom{}24 &120.0 &130.0 & 4.69 & --- & --- & & no absorption/emission \\
\hline
\end{tabular}
\end{center}
\end{table}

Column densities of CO, $^{13}$CO, HCO$^+$, HNC, C$_2$H, $c$-C$_3$H, and $c$-C$_3$H$_2$ are shown in Tables \ref{table:col_co_abs}--\ref{table:col_c2h_c3h_c3h2} for the individual velocity components. 
Components 6, 7, 15, and 16 are saturated CO absorbers and may be classified as dark clouds. 
Among these, Component 7 may be the densest cloud as judged by the excitation temperature of $^{13}$CO. 
For H$^{13}$CO$^+$, HCO, and H$_2$CO, their column densities are shown in Table \ref{table:col_h13co+_hco_and C} for Components 6, 7, 15, 16, and 17. 
Column densities of CS, SO, and SiO are shown in Table \ref{table:col_cs_so_sio} for Components 5, 6, 7, 15, and 16. 

\begin{table}[htb!]
 \caption{Column densities of CO and $^{13}$CO in absorption lines.}
 \label{table:col_co_abs}
\begin{center}
\begin{tabular}{crrDcDDcD}
\hline \hline
& \multicolumn{2}{c}{Velocity} & 
\multicolumn{5}{c}{CO absorption} & 
\multicolumn{5}{c}{$^{13}$CO absorption} \\
Comp & $v_{\rm min}$ & $v_{\rm max}$ & 
\multicolumn{2}{c}{$\int\tau_{1-0}dv$} & $T_{\rm ex}$ & \multicolumn{2}{c}{$N$(CO)} & 
\multicolumn{2}{c}{$\int\tau_{1-0}dv$} & $T_{\rm ex}$ & \multicolumn{2}{c}{$N$($^{13}$CO)} \\
ID & \multicolumn{2}{c}{(km~s$^{-1}$)} & 
\multicolumn{2}{c}{(km~s$^{-1}$)} & (K) & \multicolumn{2}{c}{($10^{14}$ cm$^{-2}$)} & 
\multicolumn{2}{c}{(km~s$^{-1}$)} & (K) & \multicolumn{2}{c}{($10^{14}$ cm$^{-2}$)} \\
\hline 
\phantom{0}1 & $-$30.0 & $-$23.0 &$-$0.04 &$\pm$0.05 & & $-$0.45&$\pm$0.58 & 0.01 &$\pm$0.02 & & 0.12&$\pm$0.17 \\
\phantom{0}2 & $-$23.0 & $-$17.0 &0.11 &$\pm$0.05 &2.27$\pm$10.63 & 1.18&$\pm$0.54 & $-$0.01 &$\pm$0.01 & & $-$0.08&$\pm$0.15 \\
\phantom{0}3 & $-$17.0 & $-$6.0 &$-$0.15 &$\pm$0.07 & & $-$1.59&$\pm$0.72 & 0.04 &$\pm$0.02 & & 0.42&$\pm$0.21 \\
\phantom{0}4 & $-$6.0 & 0.0 &0.01 &$\pm$0.05 & & 0.11&$\pm$0.54 &$-$0.01 &$\pm$0.01 & & $-$0.06&$\pm$0.15 \\
\phantom{0}5 & 0.0 & 7.0 & 2.17 &$\pm$0.05 & 4.30$\pm$0.18 & 38.75&$\pm$0.71 & 0.11&$\pm$0.01 & 2.74$\pm$4.52 & 1.27&$\pm$0.17 \\
\phantom{0}6 & 7.0 & 14.5 & \multicolumn{3}{c}{(saturated)} & (393&$\pm$6)\dag & 0.97 &$\pm$0.02 & 3.80$\pm$0.28 & 15.71&$\pm$0.25 \\
\phantom{0}7 & 14.5 & 20.0 & \multicolumn{3}{c}{(saturated)} & (865&$\pm$12)\dag & 0.96 &$\pm$0.01 & 6.75$\pm$0.11 & 34.61&$\pm$0.47 \\
\phantom{0}8 & 20.0 & 23.5 &0.96 &$\pm$0.04 & 5.69$\pm$0.22 & 24.98&$\pm$0.81 & 0.04 &$\pm$0.01 & & 0.40 &$\pm$0.12 \\
\phantom{0}9 & 23.5 & 26.0 &0.05 &$\pm$0.03 &4.67$\pm$4.58 & 0.92&$\pm$0.76 & $-$0.00 &$\pm$0.01 & & $-$0.04&$\pm$0.10 \\
\phantom{}10 & 26.0 & 30.5 &0.34 &$\pm$0.04 &6.00$\pm$0.69 & 9.57&$\pm$1.01 & 0.01 &$\pm$0.01 & & 0.11&$\pm$0.13 \\
\phantom{}11 & 30.5 & 34.5 &0.80 &$\pm$0.04 &5.28$\pm$0.30 & 18.85&$\pm$1.08 & 0.02 &$\pm$0.01 & & 0.22&$\pm$0.13 \\
\phantom{}12 & 34.5 & 40.0 &0.04 &$\pm$0.05 &4.75$\pm$7.50 & 0.84&$\pm$1.08 &$-$0.01 &$\pm$0.01 & & $-$0.15&$\pm$0.15 \\
\phantom{}13 & 40.0 & 47.0 &$-$0.07 &$\pm$0.05 & & $-$0.78&$\pm$0.57 &$-$0.01 &$\pm$0.01 & & $-$0.07&$\pm$0.16 \\
\phantom{}14 & 47.0 & 51.5 &0.02 &$\pm$0.04 & & 0.17&$\pm$0.57 &$-$0.01 &$\pm$0.01 & & $-$0.12&$\pm$0.13 \\
\phantom{}15 & 51.5 & 55.5 & \multicolumn{3}{c}{(saturated)} & (114&$\pm$4)\dag & 0.30 &$\pm$0.01 & 3.60$\pm$0.72 & 4.58&$\pm$0.17 \\
\phantom{}16 & 55.5 & 60.0 & \multicolumn{3}{c}{(saturated)} & (302&$\pm$6)\dag & 0.61 &$\pm$0.01 & 4.45$\pm$0.26 & 12.09&$\pm$0.24 \\
\phantom{}17 & 60.0 & 76.0 & 0.04&$\pm$0.08 & & 0.37&$\pm$0.88 &$-$0.03 &$\pm$0.02 & & $-$0.35&$\pm$0.25 \\
\phantom{}18 & 76.0 & 81.0 & 0.22&$\pm$0.05 &6.98$\pm$1.06 & 7.67&$\pm$1.18 & $-$0.01 &$\pm$0.01 & & $-$0.10&$\pm$0.14 \\
\phantom{}19 & 81.0 & 88.0 &0.48 &$\pm$0.05 &6.39$\pm$0.60 & 14.71&$\pm$1.82 & $-$0.01 &$\pm$0.01 & & $-$0.15&$\pm$0.16 \\
\phantom{}20 & 88.0 & 97.0 &2.20 &$\pm$0.06 &4.64$\pm$0.18 & 43.33&$\pm$1.59 & 0.03 &$\pm$0.02 & & 0.32 &$\pm$0.19 \\
\phantom{}21 & 97.0 & 102.0 &0.00 &$\pm$0.05 & & $-$0.04&$\pm$0.49 & 0.01 &$\pm$0.01 & & 0.07&$\pm$0.14 \\
\phantom{}22 & 102.0 & 110.0 &$-$0.14 &$\pm$0.06 & & $-$1.52&$\pm$0.61 &$-$0.04 &$\pm$0.02 & & $-$0.41&$\pm$0.18 \\
\phantom{}23 & 110.0 & 120.0 &$-$0.24 &$\pm$0.07 & & $-$2.61&$\pm$0.67 &$-$0.03 &$\pm$0.02 & & $-$0.36&$\pm$0.20 \\
\phantom{}24 & 120.0 & 130.0 &$-$0.17 &$\pm$0.07 & & $-$1.78&$\pm$0.68 & 0.01 &$\pm$0.02 & & 0.07&$\pm$0.20 \\
\hline
$\Sigma$ & $-$30.0 & 130.0 & & & & 1827& & 2.94 & & & 68.10 & \\
\hline
\end{tabular}
\end{center}
\tablenotetext{\dag}{Calculated from column density of $^{13}$CO by assuming the CO/$^{13}$CO ratio of 25. 
See, Section \ref{sec:co_isotopic_abundance_ratios} for the justification of this assumption rather than larger values. }
\end{table}

\begin{table}[htb!]
\caption{Column densities of HCO$^+$ and HNC in absorption lines.}
\label{table:col_hco+_hnc}
\begin{center}
\begin{tabular}{crrDDDDDD}
\hline \hline
& \multicolumn{2}{c}{Velocity} & 
\multicolumn{4}{c}{HCO$^+$ absorption} & 
\multicolumn{4}{c}{HNC absorption} \\
Comp & $v_{\rm min}$ & $v_{\rm max}$ & 
\multicolumn{2}{c}{$\int\tau dv$} & \multicolumn{2}{c}{$N$(HCO$^+$)} & 
\multicolumn{2}{c}{$\int\tau dv$} & \multicolumn{2}{c}{$N$(HNC)} \\
ID & \multicolumn{2}{c}{(km~s$^{-1}$)} & 
\multicolumn{2}{c}{(km~s$^{-1}$)} & \multicolumn{2}{c}{($10^{12}$ cm$^{-2}$)} & 
\multicolumn{2}{c}{(km~s$^{-1}$)} & \multicolumn{2}{c}{($10^{12}$ cm$^{-2}$)} \\
\hline 
\phantom{0}1 & $-$30.0 & $-$23.0 & 0.13 &$\pm$0.03 & 0.15 &$\pm$0.04 & 0.03&$\pm$0.01 & 0.06&$\pm$0.02 \\
\phantom{0}2 & $-$23.0 & $-$17.0 & 0.30 &$\pm$0.03 & 0.33 &$\pm$0.03 & 0.00&$\pm$0.01 & 0.00&$\pm$0.02 \\
\phantom{0}3 & $-$17.0 & $-$6.0 & 0.01 &$\pm$0.04 & 0.02 &$\pm$0.04 & 0.01&$\pm$0.01 & 0.02&$\pm$0.03 \\
\phantom{0}4 & $-$6.0 & 0.0 & 0.05 &$\pm$0.03 & 0.05 &$\pm$0.03 & $-$0.01&$\pm$0.03 & $-$0.02&$\pm$0.05 \\
\phantom{0}5 & 0.0 & 7.0 & 0.75 &$\pm$0.04 & 0.83 &$\pm$0.04 & 0.12&$\pm$0.03 & 0.21&$\pm$0.05 \\
\phantom{0}6 & 7.0 & 14.5 & 4.80 &$\pm$0.04 & 5.34 &$\pm$0.04 & 0.49&$\pm$0.03 & 0.87&$\pm$0.05 \\
\phantom{0}7 & 14.5 & 20.0 & \multicolumn{2}{c}{(saturated)} & (3.49&$\pm$1.18)\dag & 1.11&$\pm$0.03 & 1.97&$\pm$0.06 \\
\phantom{0}8 & 20.0 & 23.5 & 1.01 &$\pm$0.03 & 1.12 &$\pm$0.04 & 0.19&$\pm$0.02 & 0.35&$\pm$0.04 \\
\phantom{0}9 & 23.5 & 26.0 & 0.58 &$\pm$0.02 & 0.64 &$\pm$0.03 & $-$0.01&$\pm$0.02 & $-$0.01&$\pm$0.03 \\
\phantom{}10 & 26.0 & 30.5 & 1.56 &$\pm$0.04 & 1.73 &$\pm$0.04 & 0.09&$\pm$0.02 & 0.16&$\pm$0.04 \\
\phantom{}11 & 30.5 & 34.5 & 1.99 &$\pm$0.05 & 2.21 &$\pm$0.06 & 0.15&$\pm$0.02 & 0.28&$\pm$0.04 \\
\phantom{}12 & 34.5 & 40.0 & 0.44 &$\pm$0.03 & 0.49 &$\pm$0.03 & 0.06&$\pm$0.02 & 0.11&$\pm$0.04 \\
\phantom{}13 & 40.0 & 47.0 & 0.30 &$\pm$0.04 & 0.33 &$\pm$0.04 & $-$0.04&$\pm$0.03 & $-$0.07&$\pm$0.05 \\
\phantom{}14 & 47.0 & 51.5 & 0.16 &$\pm$0.03 & 0.18 &$\pm$0.03 & 0.00&$\pm$0.02 & 0.00&$\pm$0.04 \\
\phantom{}15 & 51.5 & 55.5 & \multicolumn{2}{c}{(saturated)} & (3.48&$\pm$1.01)\dag & 0.63&$\pm$0.02 & 1.12&$\pm$0.04 \\
\phantom{}16 & 55.5 & 60.0 & \multicolumn{2}{c}{(saturated)} & (4.58&$\pm$1.08)\dag & 0.81&$\pm$0.03 & 1.45&$\pm$0.05 \\
\phantom{}17 & 60.0 & 76.0 & 2.68 &$\pm$0.06 & 2.98 &$\pm$0.06 & 0.05&$\pm$0.04 & 0.09&$\pm$0.07 \\
\phantom{}18 & 76.0 & 81.0 & 1.88 &$\pm$0.04 &2.10 &$\pm$0.05 & 0.32&$\pm$0.02 & 0.57&$\pm$0.04 \\
\phantom{}19 & 81.0 & 88.0 & 2.00 &$\pm$0.05 & 2.22 &$\pm$0.05 & 0.26&$\pm$0.03 & 0.47&$\pm$0.05 \\
\phantom{}20 & 88.0 & 97.0 & 3.17 &$\pm$0.05 &3.52 &$\pm$0.06 & 0.16&$\pm$0.03 & 0.28&$\pm$0.06 \\
\phantom{}21 & 97.0 & 102.0 & 0.46 &$\pm$0.03 & 0.51 &$\pm$0.03 & $-$0.03&$\pm$0.02 & $-$0.06&$\pm$0.04 \\
\phantom{}22 & 102.0 & 110.0 & 0.28 &$\pm$0.04 & 0.31 &$\pm$0.04 & $-$0.03&$\pm$0.03 & $-$0.05&$\pm$0.05 \\
\phantom{}23 & 110.0 & 120.0 & 0.09 &$\pm$0.04 & 0.10 &$\pm$0.04 & $-$0.10&$\pm$0.03 & $-$0.17&$\pm$0.06 \\
\phantom{}24 & 120.0 & 130.0 & $-$0.06 &$\pm$0.04 & $-$0.07 &$\pm$0.04 & $-$0.06&$\pm$0.03 & $-$0.10&$\pm$0.06 \\
\hline
$\Sigma$ & $-$30.0 & 130.0 & & & 36.67 & & 4.22 & & 7.53 & \\
\hline
\end{tabular}
\end{center}
\tablenotetext{\dag}{Calculated from column density of H$^{13}$CO$^+$ by assuming the HCO$^+$/H$^{13}$CO$^+$ isotopologue ratio of 50. }
\end{table}

\begin{table}[htb!]
\caption{Column densities of C$_2$H, $c$-C$_3$H, and $c$-C$_3$H$_2$ in absorption lines.}
\label{table:col_c2h_c3h_c3h2}
\begin{center}
\begin{tabular}{crrDDDDDD}
\hline \hline
& \multicolumn{2}{c}{Velocity} & 
\multicolumn{4}{c}{C$_2$H absorption} & 
\multicolumn{4}{c}{$c$-C$_3$H absorption} & 
\multicolumn{4}{c}{$c$-C$_3$H$_2$ absorption} \\
Comp & $v_{\rm min}$ & $v_{\rm max}$ & 
\multicolumn{2}{c}{$\int\tau dv$} & \multicolumn{2}{c}{$N$(C$_2$H)} & 
\multicolumn{2}{c}{$\int\tau dv$} & \multicolumn{2}{c}{$N$($c$-C$_3$H)} & 
\multicolumn{2}{c}{$\int\tau dv$} & \multicolumn{2}{c}{$N$($c$-C$_3$H$_2$)} \\
ID & \multicolumn{2}{c}{(km~s$^{-1}$)} & 
\multicolumn{2}{c}{(km~s$^{-1}$)} & \multicolumn{2}{c}{($10^{12}$ cm$^{-2}$)} & 
\multicolumn{2}{c}{(km~s$^{-1}$)} & \multicolumn{2}{c}{($10^{12}$ cm$^{-2}$)} & 
\multicolumn{2}{c}{(km~s$^{-1}$)} & \multicolumn{2}{c}{($10^{12}$ cm$^{-2}$)} \\
\hline 
\phantom{0}1 & $-$30.0 & $-$23.0 & $-$0.05&$\pm$0.04 & $-$7.71&$\pm$6.00 & $-$0.05&$\pm$0.02 & $-$0.51&$\pm$0.25 & 0.02&$\pm$0.04 & 0.09&$\pm$0.20 \\
\phantom{0}2 & $-$23.0 & $-$17.0 & $-$0.03&$\pm$0.04 & $-$3.68&$\pm$5.61 & $-$0.04&$\pm$0.02 & $-$0.45&$\pm$0.25 & 0.12&$\pm$0.02 & 0.55&$\pm$0.11 \\
\phantom{0}3 & $-$17.0 & $-$6.0 & $-$0.09&$\pm$0.05 & $-$12.76&$\pm$7.65 & $-$0.02&$\pm$0.03 & $-$0.17&$\pm$0.32 & $-$0.01&$\pm$0.03 & $-$0.07&$\pm$0.13 \\
\phantom{0}4 & $-$6.0 & 0.0 & $-$0.06&$\pm$0.04 & $-$8.72&$\pm$5.61 & $-$0.03&$\pm$0.02 & $-$0.32&$\pm$0.25 & 0.02&$\pm$0.02 & 0.07&$\pm$0.11 \\
\phantom{0}5 & 0.0 & 7.0 & 0.19&$\pm$0.05 & 27.10&$\pm$6.36 & $-$0.02&$\pm$0.02 & $-$0.17&$\pm$0.25 & 0.28&$\pm$0.02 & 1.25&$\pm$0.11 \\
\phantom{0}6 & 7.0 & 14.5 & 0.31&$\pm$0.05 & 43.99&$\pm$6.36 & $-$0.01&$\pm$0.03 & $-$0.09&$\pm$0.27 & 0.38&$\pm$0.02 & 1.72&$\pm$0.11 \\
\phantom{0}7 & 14.5 & 20.0 & 0.58&$\pm$0.04 & 81.36&$\pm$5.19 & 0.08&$\pm$0.02 & 0.85&$\pm$0.21 & 0.66&$\pm$0.02 & 2.99&$\pm$0.09 \\
\phantom{0}8 & 20.0 & 23.5 & 0.06&$\pm$0.03 & 4.05&$\pm$2.11 & 0.02&$\pm$0.02 & 0.25&$\pm$0.21 & 0.07&$\pm$0.02 & 0.31&$\pm$0.08 \\
\phantom{0}9 & 23.5 & 26.0 & 0.11&$\pm$0.03 & 7.89&$\pm$1.83 & 0.01&$\pm$0.01 & 0.09&$\pm$0.12 & 0.10&$\pm$0.02 & 0.46&$\pm$0.08 \\
\phantom{}10 & 26.0 & 30.5 & 0.37&$\pm$0.04 & 26.07&$\pm$2.59 & 0.04&$\pm$0.02 & 0.36&$\pm$0.21  & 0.13&$\pm$0.02 & 0.58&$\pm$0.08 \\
\phantom{}11 & 30.5 & 34.5 & 0.21&$\pm$0.03 & 15.10&$\pm$2.36 & 0.03&$\pm$0.02 & 0.27&$\pm$0.17 & 0.18&$\pm$0.02 & 0.82&$\pm$0.09 \\
\phantom{}12 & 34.5 & 40.0 & 0.21&$\pm$0.04 & 14.46&$\pm$2.59 & 0.00&$\pm$0.02 & 0.00&$\pm$0.25 & 0.11&$\pm$0.02 & 0.50&$\pm$0.09 \\
\phantom{}13 & 40.0 & 47.0 & 0.23&$\pm$0.04 & 16.44&$\pm$2.99 &  0.05&$\pm$0.02 & 0.46&$\pm$0.25 & 0.04&$\pm$0.02 & 0.18&$\pm$0.11 \\
\phantom{}14 & 47.0 & 51.5 & 0.08&$\pm$0.04 & 5.59&$\pm$2.59 & 0.00&$\pm$0.02 & 0.04&$\pm$0.21 & 0.03&$\pm$0.02 & 0.15&$\pm$0.09 \\
\phantom{}15 & 51.5 & 55.5 & 0.40&$\pm$0.03 & 28.06&$\pm$2.11 & 0.05&$\pm$0.02 & 0.55&$\pm$0.21 & 0.17&$\pm$0.02 & 0.76&$\pm$0.08 \\
\phantom{}16 & 55.5 & 60.0 & 0.59&$\pm$0.04 & 41.68&$\pm$2.59 & 0.04&$\pm$0.02 & 0.45&$\pm$0.17 & 0.27&$\pm$0.02 & 1.22&$\pm$0.09\\
\phantom{}17 & 60.0 & 76.0 & 0.15&$\pm$0.07 & 10.35&$\pm$4.60 & $-$0.05&$\pm$0.04 & $-$0.55&$\pm$0.39 & 0.06&$\pm$0.04 & 0.28&$\pm$0.16 \\
\phantom{}18 & 76.0 & 81.0 & 0.27&$\pm$0.04 & 19.15&$\pm$2.59 & $-$0.05&$\pm$0.02 & $-$0.55&$\pm$0.25 & 0.19&$\pm$0.02 & 0.83&$\pm$0.09 \\
\phantom{}19 & 81.0 & 88.0 & 0.28&$\pm$0.04 & 19.97&$\pm$2.99 & 0.00&$\pm$0.02 & $-$0.01&$\pm$0.25 & 0.22&$\pm$0.02 & 0.98&$\pm$0.11 \\
\phantom{}20 & 88.0 & 97.0 & 0.43&$\pm$0.05 & 30.01&$\pm$3.50 & 0.03&$\pm$0.03 & 0.34&$\pm$0.27 & 0.27&$\pm$0.03 & 1.23&$\pm$0.12 \\
\phantom{}21 & 97.0 & 102.0 & 0.03&$\pm$0.04 & 2.13&$\pm$2.59 & 0.02&$\pm$0.02 & 0.19&$\pm$0.25 & 0.00&$\pm$0.02 & 0.01&$\pm$0.09 \\
\phantom{}22 & 102.0 & 110.0 & 0.01&$\pm$0.05 & 0.38&$\pm$3.34 &$-$0.03&$\pm$0.03 & $-$0.30&$\pm$0.27 & 0.06&$\pm$0.03 & 0.27&$\pm$0.12 \\
\phantom{}23 & 110.0 & 120.0 & $-$0.02&$\pm$0.05 & $-$1.44&$\pm$3.50 & 0.00&$\pm$0.03 & 0.01&$\pm$0.30 & $-$0.02&$\pm$0.03 & $-$0.09&$\pm$0.13 \\
\phantom{}24 & 120.0 & 130.0 & 0.10&$\pm$0.05 & 7.17&$\pm$3.66 & $-$0.05&$\pm$0.03 & $-$0.56&$\pm$0.30 & $-$0.02&$\pm$0.03 & $-$0.09&$\pm$0.12 \\
\hline
$\Sigma$ & $-$30.0 & 130.0 & 4.36 & & 366.64 & & 0.02 & & 0.18 & & 3.34 & & 15.02 \\
\hline
\end{tabular}
\end{center}
\end{table}

\begin{table}[htb!]
\caption{Column densities of H$^{13}$CO$^+$, HCO, H$_2$CO, and $c$-C$_3$H in absorption lines toward selected clouds.}
\label{table:col_h13co+_hco_and C}
\centering
\begin{tabular}{crrDDDDDD}
\hline \hline
& \multicolumn{2}{c}{Velocity} & 
\multicolumn{4}{c}{H$^{13}$CO$^+$ absorption} & 
\multicolumn{4}{c}{HCO absorption} & 
\multicolumn{4}{c}{H$_2$CO absorption} \\
Comp & $v_{\rm min}$ & $v_{\rm max}$ & 
\multicolumn{2}{c}{$\int\tau dv$} & \multicolumn{2}{c}{$N$(H$^{13}$CO$^+$)} & 
\multicolumn{2}{c}{$\int\tau dv$} & \multicolumn{2}{c}{$N$(HCO)} & 
\multicolumn{2}{c}{$\int\tau dv$} & \multicolumn{2}{c}{$N$(H$_2$CO)} \\
ID & \multicolumn{2}{c}{(km~s$^{-1}$)} & 
\multicolumn{2}{c}{(km~s$^{-1}$)} & \multicolumn{2}{c}{($10^{12}$ cm$^{-2}$)} & 
\multicolumn{2}{c}{(km~s$^{-1}$)} & \multicolumn{2}{c}{($10^{12}$ cm$^{-2}$)} & 
\multicolumn{2}{c}{(km~s$^{-1}$)} & \multicolumn{2}{c}{($10^{12}$ cm$^{-2}$)} \\
\hline 
\phantom{0}6 &  7.0 & 14.5 & 0.02&$\pm$0.02 & 0.03&$\pm$0.03 & 0.01&$\pm$0.03 & 0.41&$\pm$1.06 & 0.03&$\pm$0.04 & 1.51&$\pm$1.93 \\
\phantom{0}7 & 14.5 & 20.0 & 0.06&$\pm$0.02 & 0.07&$\pm$0.02 & 0.05&$\pm$0.02 & 1.85&$\pm$0.90 & 0.10&$\pm$0.04 & 4.52&$\pm$1.70 \\
\phantom{}15 & 51.5 & 55.5 & 0.06&$\pm$0.02 & 0.07&$\pm$0.02 & 0.04&$\pm$0.02 & 1.73&$\pm$0.62 & $-$0.02&$\pm$0.03 & $-$0.80&$\pm$1.42 \\
\phantom{}16 & 55.5 & 60.0 & 0.08&$\pm$0.02 & 0.09&$\pm$0.02 & 0.05&$\pm$0.02 & 1.92&$\pm$0.65 & 0.05&$\pm$0.03 & 2.47&$\pm$1.54 \\
\phantom{}17 & 60.0 & 76.0 & $-$0.00&$\pm$0.03 & $-$0.00&$\pm$0.04 & $-$0.02&$\pm$0.03 & $-$0.65&$\pm$1.23 & $-$0.08&$\pm$0.06 & $-$3.61&$\pm$2.85 \\
\hline
\end{tabular}
\end{table}

\begin{table}[htb!]
 \caption{Column densities of CS, SO, and SiO in absorption lines toward selected clouds.}
 \label{table:col_cs_so_sio}
 \centering
\begin{tabular}{crrDDDDDD}
\hline \hline
& \multicolumn{2}{c}{Velocity} & 
\multicolumn{4}{c}{CS absorption} & 
\multicolumn{4}{c}{SO absorption} & 
\multicolumn{4}{c}{SiO absorption} \\
Comp & $v_{\rm min}$ & $v_{\rm max}$ & 
\multicolumn{2}{c}{$\int\tau dv$} & \multicolumn{2}{c}{$N$(CS)} & 
\multicolumn{2}{c}{$\int\tau dv$} & \multicolumn{2}{c}{$N$(SO)} & 
\multicolumn{2}{c}{$\int\tau dv$} & \multicolumn{2}{c}{$N$(SiO)} \\
ID & \multicolumn{2}{c}{(km~s$^{-1}$)} & 
\multicolumn{2}{c}{(km~s$^{-1}$)} & \multicolumn{2}{c}{($10^{12}$ cm$^{-2}$)} & 
\multicolumn{2}{c}{(km~s$^{-1}$)} & \multicolumn{2}{c}{($10^{12}$ cm$^{-2}$)} & 
\multicolumn{2}{c}{(km~s$^{-1}$)} & \multicolumn{2}{c}{($10^{12}$ cm$^{-2}$)} \\
\hline 
\phantom{0}5 &  0.0 &  7.0 & 0.22&$\pm$0.02 & 1.80&$\pm$0.17 & $-$0.04&$\pm$0.02 & $-$1.05&$\pm$0.58 & $-$0.00&$\pm$0.02 & $-$0.01&$\pm$0.07 \\
\phantom{0}6 &  7.0 & 14.5 & 0.22&$\pm$0.02 & 1.81&$\pm$0.17 & $-$0.00&$\pm$0.02 & $-$0.13&$\pm$0.60 & 0.04&$\pm$0.02 & 0.15&$\pm$0.08 \\
\phantom{0}7 & 14.5 & 20.0 & 0.97&$\pm$0.02 & 7.85&$\pm$0.15 & 0.16&$\pm$0.02 & 4.54&$\pm$0.52 & 0.11&$\pm$0.02 & 0.36&$\pm$0.06 \\
\phantom{}15 & 51.5 & 55.5 & 0.02&$\pm$0.02 & 0.20&$\pm$0.13 & 0.02&$\pm$0.02 & 0.66&$\pm$0.44 & 0.01&$\pm$0.02 & 0.05&$\pm$0.06 \\
\phantom{}16 & 55.5 & 60.0 & 0.11&$\pm$0.02 & 0.87&$\pm$0.13 & 0.00&$\pm$0.02 & 0.12&$\pm$0.47 & 0.01&$\pm$0.02 & 0.04&$\pm$0.06 \\
\hline
\end{tabular}
\end{table}

The H$_2$ column density was also deduced through CO and $^{13}$CO $J$=1--0 emission lines, as shown in Table \ref{table:col_co_em}. 
The standard Galactic CO-to-H$_2$ conversion factor \citep[$2 \times 10^{20}$ $\rm cm^{-2}~(K~km~s^{-1})^{-1}$;][]{2013ARA&A..51..207B} was used for the calculation from the CO integrated intensity to $N$(H$_2$). 
We also obtained the $^{13}$CO column density with its $J$=1--0 emission in a manner described in Appendix \ref{sec:reliability}. 
The H$_2$ column density was then deduced by using the $^{13}$CO/H$_2$ abundance ratio ($X$($^{13}$CO) = $2 \times 10^{-6}$) as a canonical value for Galactic disk clouds \citep[][and references therein]{2008ApJ...679..481P}. 

\begin{table}[htb!]
 \caption{Column density of H$_2$ through CO and $^{13}$CO $J$=1--0 emission lines}
 \label{table:col_co_em}
 \centering
\begin{tabular}{crrcDDDD}
\hline \hline
& \multicolumn{2}{c}{Velocity} & 
\multicolumn{5}{c}{CO emission} & 
\multicolumn{4}{c}{$^{13}$CO emission} \\
Comp & $v_{\rm min}$ & $v_{\rm max}$ & $T_{\rm peak}$ & 
\multicolumn{2}{c}{$\int T_{\rm mb}$(CO)$dv$} & \multicolumn{2}{c}{$N$(H$_2$)} & 
\multicolumn{2}{c}{$\int T_{\rm mb}$($^{13}$CO)$dv$} & \multicolumn{2}{c}{$N$(H$_2$)} \\
ID & \multicolumn{2}{c}{(km~s$^{-1}$)} & (K) & 
\multicolumn{2}{c}{($\rm K~km~s^{-1}$)} & \multicolumn{2}{c}{($10^{20} \rm cm^{-2}$)} & 
\multicolumn{2}{c}{($\rm K~km~s^{-1}$)} & \multicolumn{2}{c}{($10^{20} \rm cm^{-2}$)} \\
\hline 
\phantom{0}1 &$-$30.0&$-$23.0&2.93 & 0.45 &$\pm$0.83 &  1 &$\pm$2 & 0.12 &$\pm$0.19 &  0.57 &$\pm$0.89 \\
\phantom{0}2 &$-$23.0&$-$17.0&2.83 & 0.22 &$\pm$0.77 &  0 &$\pm$2 & 0.26 &$\pm$0.17 &  1.22 &$\pm$0.83 \\
\phantom{0}3 &$-$17.0&$-$6.0&2.81 &$-$1.04&$\pm$1.04 &$-$2&$\pm$2&$-$0.17&$\pm$0.24 &$-$0.80&$\pm$1.12 \\
\phantom{0}4 &$-$6.0&  0.0 & 2.82 &$-$0.53&$\pm$0.77 &$-$1&$\pm$2 & 0.08 &$\pm$0.17 &  0.38 &$\pm$0.83 \\
\phantom{0}5 &  0.0 &  7.0 & 4.15 &  1.85 &$\pm$0.83 &  4 &$\pm$2 & 0.21 &$\pm$0.17 &  0.95 &$\pm$0.79 \\
\phantom{0}6 &  7.0 & 14.5 & 8.50 & 22.60 &$\pm$0.86 & 45 &$\pm$2 & 0.21 &$\pm$0.19 &  1.05 &$\pm$0.96 \\
\phantom{0}7 & 14.5 & 20.0 & 6.43 & 12.26 &$\pm$0.73 & 25 &$\pm$1 & 1.61 &$\pm$0.17 & 7.46 &$\pm$0.77 \\
\phantom{0}8 & 20.0 & 23.5 & 3.90 &  2.13 &$\pm$0.59 &  4 &$\pm$1 & 0.40 &$\pm$0.13 &  1.80 &$\pm$0.60 \\
\phantom{0}9 & 23.5 & 26.0 & 3.34 &  0.92 &$\pm$0.50 &  2 &$\pm$1 & 0.17 &$\pm$0.11 &  0.79 &$\pm$0.52 \\
\phantom{}10 & 26.0 & 30.5 & 3.95 &  2.97 &$\pm$0.66 &  6 &$\pm$1 & 0.41 &$\pm$0.15 &  1.87 &$\pm$0.68 \\
\phantom{}11 & 30.5 & 34.5 & 3.15 &  0.79 &$\pm$0.63 &  2 &$\pm$1 & 0.43 &$\pm$0.14 &  2.00&$\pm$0.66\\
\phantom{}12 & 34.5 & 40.0 & 2.65 &$-$0.93&$\pm$0.73 &$-$2&$\pm$1 & 0.38 &$\pm$0.17 &  1.82 &$\pm$0.81 \\
\phantom{}13 & 40.0 & 47.0 & 3.28 &  1.01 &$\pm$0.83 &  2 &$\pm$2 & 0.77 &$\pm$0.19 &  3.55 &$\pm$0.87 \\
\phantom{}14 & 47.0 & 51.5 & 2.67 &$-$0.77&$\pm$0.66 &$-$2&$\pm$1 & 0.40 &$\pm$0.15 &  1.94 &$\pm$0.73 \\
\phantom{}15 & 51.5 & 55.5 & 6.10 &  6.51 &$\pm$0.63 & 13 &$\pm$1 & 1.22 &$\pm$0.14 & 5.58 &$\pm$0.65 \\
\phantom{}16 & 55.5 & 60.0 & 7.60 & 10.48 &$\pm$0.66 & 21 &$\pm$1 & 1.48 &$\pm$0.15 & 7.05 &$\pm$0.72 \\
\phantom{}17 & 60.0 & 76.0 & 4.48 & 10.34 &$\pm$1.25 & 21 &$\pm$3 & 2.67 &$\pm$0.28 & 12.01 &$\pm$1.28 \\
\phantom{}18 & 76.0 & 81.0 & 3.92 &  1.93 &$\pm$0.70 &  4 &$\pm$1 & 0.52 &$\pm$0.16 &  2.35 &$\pm$0.72 \\
\phantom{}19 & 81.0 & 88.0 & 3.75 &  5.80 &$\pm$0.83 & 12 &$\pm$2 & 1.53 &$\pm$0.19 & 6.96 &$\pm$0.85
\\
\phantom{}20 & 88.0 & 97.0 & 6.10 & 16.02 &$\pm$0.94 & 32 &$\pm$2 & 2.37 &$\pm$0.21 & 10.87 &$\pm$0.98\\
\phantom{}21 & 97.0 &102.0 & 3.83 &  4.30 &$\pm$0.70 &  9 &$\pm$1 & 0.80 &$\pm$0.16 &  3.64 &$\pm$0.72 \\
\phantom{}22 &102.0 &110.0 & 3.84 &  5.42 &$\pm$0.89 & 11 &$\pm$2 & 1.17 &$\pm$0.20 &  5.29 &$\pm$0.91 \\
\phantom{}23 &110.0 &120.0 & 3.05 &  0.31 &$\pm$0.99 &  1 &$\pm$2 & 0.22 &$\pm$0.22 &  1.01 &$\pm$1.05 \\
\phantom{}24 &120.0 &130.0 & 2.83 &$-$0.61&$\pm$0.99 &$-$1&$\pm$2 & 0.27 &$\pm$0.22 &  1.26 &$\pm$1.07 \\
\hline
$\Sigma$ & $-$30.0 & 130.0 & & 102.41 & & 205 & & 17.52 & & 80.62 & \\
\hline
\end{tabular}
\end{table}

\subsection{Excitation analysis and physical properties} 
\label{sec:excitation_analysis}

\subsubsection{Excitation temperature through multi-transition study of {\rm CO} absorption} \label{sec:Tex_CO}

An excitation analysis can be done if some species are detected in several transitions.
Under the Rayleigh-Jeans approximation ($h\nu \ll k_{\rm B}T$), Equation \ref{eq:n_tot} can be rewritten as, 
\begin{equation}
\label{eq:equation-3}
\log \left(\frac{3 k_{\rm B} \int \tau dv}{8 \pi^3 \nu \mu^2 S_{\rm ul}}\right)=\log \left(\frac{N_{\rm tot}}{Q\left(T_{\rm ex}\right) T_{\rm ex}}\right)-\frac{E_{\rm l} \log (e)}{k_{\rm B} T_{\rm ex}}.
\end{equation}
%Excitation analysis can be performed for CO and $^{13}$CO because multiple transitions have been detected. 
%We first conducted excitation analysis of CO and $^{13}$CO for individual velocity components. 
The excitation temperature for each component can be calculated as an inverse of the slope of the rotation diagram.

%\begin{figure}[htb!]
%\epsscale{0.75}
%\plotone{rot_dia.png}
%\caption{Rotation diagrams of CO and $^{13}$CO (and C$^{18}$O as a red dot) for the individual velocity components. 
%\label{fig:rot_dia}}
%\end{figure}
However, the excitation temperature is quite small for the current absorption study and the assumption for the Rayleigh-Jeans approximation ($h\nu \ll k_{\rm B}T_{\rm ex}$) is not met. 
We thus need to deduce $T_{\rm ex}$ from $\tau_{1-0}$ and $\tau_{2-1}$ without using this approximation. 

For $J$=1--0 and $J$=2--1 absorption, Equation \ref{eq:n_tot} can be rewritten as, 
\begin{equation}
\label{eq:n_tot_1}
N = \frac{3hQ(T_{\rm ex})}{8\pi^3\mu^2S_{1,0}} \frac{\exp(\frac{E_{J=0}}{k_{\rm B} T_{\rm ex}})}{[1-\exp(\frac{-h\nu_{1-0}}{k_{\rm B}T_{\rm ex}})]} \int\tau_{1-0} dv, 
\end{equation}
\begin{equation}
\label{eq:n_tot_2}
N = \frac{3hQ(T_{\rm ex})}{8\pi^3\mu^2S_{2,1}} \frac{\exp(\frac{E_{J=1}}{k_{\rm B} T_{\rm ex}})}{[1-\exp(\frac{-h\nu_{2-1}}{k_{\rm B}T_{\rm ex}})]} \int\tau_{2-1} dv. 
\end{equation}
Under the LTE condition, $T_{\rm ex}$ remains constant for all transitions, and thus $T_{\rm ex}$ shall meet 
\begin{equation}
\label{eq:tex_deduction1}
\frac{1}{S_{1,0}} \frac{\exp(\frac{E_{J=0}}{k_{\rm B} T_{\rm ex}})} {[1-\exp(\frac{-h\nu_{1-0}}{k_{\rm B}T_{\rm ex}})]} \int\tau_{1-0} dv = 
\frac{1}{S_{2,1}} \frac{\exp(\frac{E_{J=1}}{k_{\rm B} T_{\rm ex}})} {[1-\exp(\frac{-h\nu_{2-1}}{k_{\rm B}T_{\rm ex}})]} \int\tau_{2-1} dv. 
\end{equation}
Here, $E_{J=0}/k_{\rm B}$ = 0 K, $E_{J=1}/k_{\rm B}$ = 5.53 K (CO) or 5.29 K ($^{13}$CO), $h\nu_{1-0}$ = $E_{J=1}$, $\nu_{2-1}$ $\simeq$ 2$\nu_{1-0}$, and $S_{2,1}$ $\simeq$ $2S_{1,0}$, and thus, 
\begin{equation}
\label{eq:tex_deduction2}
\frac{\int\tau_{2-1} dv}{ \int\tau_{1-0} dv} = 
\frac{2\exp(\frac{-E_{J=1}}{k_{\rm B} T_{\rm ex}})[1-\exp(\frac{-2E_{J=1}}{k_{\rm B}T_{\rm ex}})]}{[1-\exp(\frac{-E_{J=1}}{k_{\rm B}T_{\rm ex}})]} = 
2\exp\left(\frac{-E_{J=1}}{k_{\rm B} T_{\rm ex}}\right)\left[1+\exp\left(\frac{-E_{J=1}}{k_{\rm B}T_{\rm ex}}\right)\right].
\end{equation}
By defining $x \equiv \exp\left(\frac{-E_{J=1}}{k_{\rm B} T_{\rm ex}}\right)$, the formula for the solution of the quadratic equation gives, 
\begin{equation}
\label{eq:tex_deduction3}
x = -\frac{1}{2} \pm \frac{1}{2}\sqrt{2\frac{\int\tau_{2-1} dv}{\int\tau_{1-0} dv}+ 1}. 
\end{equation}
Since $x$ shall be positive, we finally get the analytical solution of $T_{\rm ex}$ as, 
\begin{equation}
\label{eq:tex_deduction4}
T_{\rm ex} = -\frac{E_{J=1}}{k_{\rm B}} \left[\ln\left(-\frac{1}{2} + \frac{1}{2}\sqrt{2\frac{\int\tau_{2-1} dv}{\int\tau_{1-0} dv}+ 1}\right)\right]^{-1}.  
\end{equation}
The obtained excitation temperatures for CO and its rarer isotopologue were listed in Table \ref{table:col_co_abs}.
%In this manner, the excitation temperature of $^{13}$CO is sometimes significantly lower than that of CO estimated from $T_{\rm mb}$ and optically thick CO emission with a filling factor near unity.
%Such differences in excitation temperatures of CO, $^{13}$CO, and C$^{18}$O are reported for the translucent cloud along the line of sight toward HD 62542 \citep{2020ApJ...897...36W} and for Orion B giant molecular cloud \citep{2021A&A...645A..26R}. 
%The difference in the excitation temperatures among the CO isotopologues may be explained by more effective radiative trapping in the optically thicker lines or by the presence of kinetic temperature gradients along the line of sight, especially near photodissociation regions, possibly combined with density gradients \citep {2021A&A...645A..26R}.

\subsubsection{Validity of CO excitation temperature estimated from CO main beam antenna temperature}
\label{sec:Tex_Tb}

\begin{figure}[htb!]
\epsscale{0.5}
\plotone{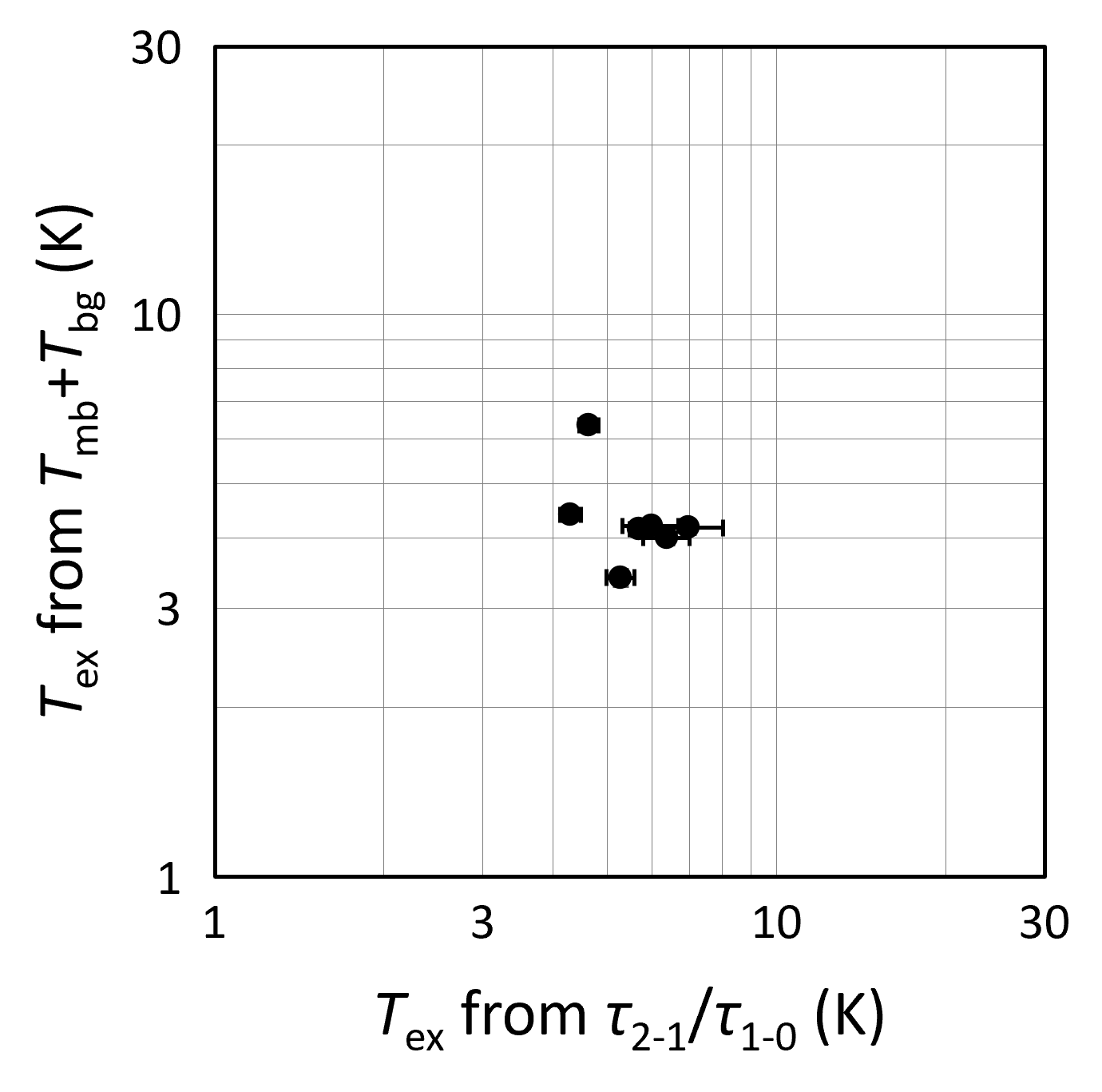}
\caption{Correlation plot of the CO excitation temperatures obtained through excitation analysis of the absorption lines and those estimated from the CO brightness temperatures by assuming the optically thick ($\tau \gg 1$) and beam-filling ($f = 1$) CO emission. }
\label{fig:tex_tex}
\end{figure}

The alternative used for the estimation of the CO excitation temperature is the CO brightness temperature since both will agree if the emission is optically thick ($\tau \gg 1$) and the filling factor ($f$) is close to unity. 
To check the validity of this method toward general Galactic disk clouds, the CO excitation temperatures obtained through the excitation analysis of the absorption lines were compared with the CO brightness temperatures. 
The CO brightness temperature toward the QSO was estimated from the baseline-subtracted profile by adding the 2.73 K cosmic background and the QSO continuum diluted by the single-dish beam (0.25 K). 
Figure \ref{fig:tex_tex} shows the correlation plot of these CO excitation temperatures taken from Tables \ref{table:col_co_abs} and \ref{table:col_co_em}. 

There is a trend that the CO excitation temperatures obtained through the excitation analysis of the absorption lines tend to be higher than those estimated from the CO brightness temperatures by assuming the optically thick and beam-filling CO emission. 
The opacity explains this discrepancy because the excitation analysis of the CO absorption lines can only be done when the lines are not saturated ($\tau \lesssim 1$). 
Given that nearly half of the molecular gas mass along this line of sight is attributed to the velocity components not fully saturated in the CO absorption (Table \ref{table:col_co_em}), nearly half of the Galactic molecular gas mass does not satisfy the assumption of the optically thick and beam-filling CO emission and their excitation temperature is underestimated. 

\subsubsection{Single-dish emission/absorption study of {\rm HCO$^+$}, {\rm HCN}, {\rm HNC} and {\rm C$_2$H}} 
\label{sec:Tex_HCO+}

The obtained excitation temperature for CO or its rarer isotopologue was less than 6 K in most cases, as described in the previous subsection. 
The excitation temperature for molecules with larger dipole moments shall be much less than this value and is close to that of the cosmic microwave background. 
To test the validity of the above assumption, the emission and absorption studies of HCO$^+$, HCN, HNC, and C$_2$H lines were done with the NRO 45 m telescope. 

\begin{figure}[htb!]
\epsscale{0.5}
\plotone{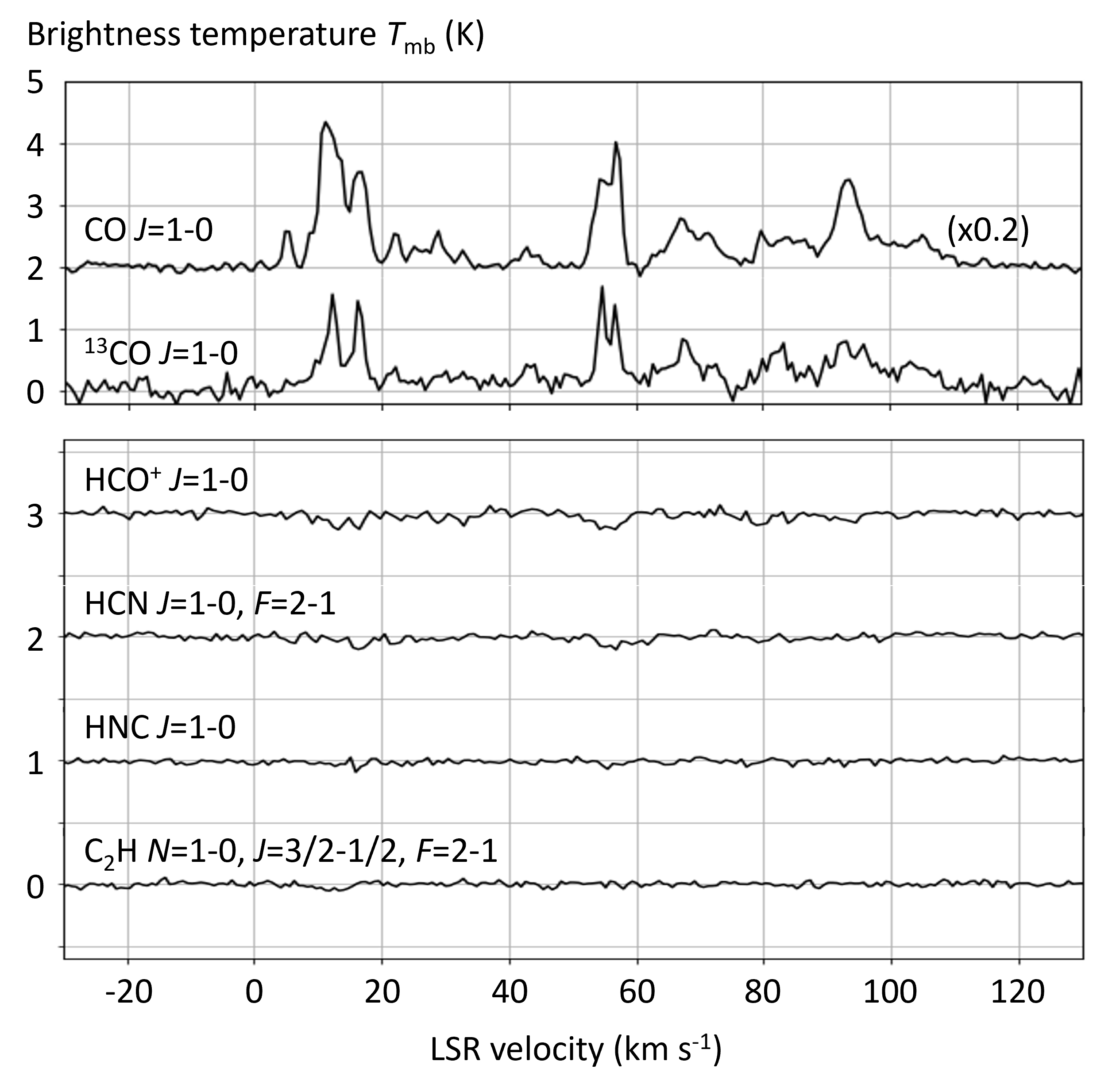}
\caption{Baseline-subtracted line profiles toward the QSO in CO, $^{13}$CO, HCO$^+$, HCN, HNC, and C$_2$H obtained with the NRO 45 m telescope. 
The subtracted continuum level of the QSO is estimated to be $\sim$0.25 K. 
The spectra were offset for clarity. 
Note that the HCN and C$_2$H lines have hyperfine components. } 
\label{fig:45m_abs}
\end{figure}

The line profiles obtained toward the QSO with the NRO 45 m telescope are shown in Figure \ref{fig:45m_abs}. 
The subtracted continuum level of the QSO was diluted to $\sim$0.25 K due to the relatively large beam of the telescope. 
No significant emission lines were detected in the QSO direction for the four molecular species of HCO$^+$, HCN, HNC, and C$_2$H in the QSO direction. 
%For the four-velocity components with the largest $^{13}$CO column densities, absorption lines were observed at a level of $\sim$0.1 K against the 0.25 K QSO continuum averaged over the NRO 45 m telescope beam. 
This suggests that these species' excitation temperatures are very low. 

The four beams of the FOREST receiver separated by $50 ''$ (1.2 pc at 5 kpc) in RA and DEC directions provided us with three additional data sets as a bonus. 
The analysis of these emission-line data from adjacent areas will be presented in Appendix \ref{sec:emission_line_ratio}. 

\subsection{Total amount of molecular gas along the line of sight}
\label{sec:molecular_gas_mass}

We estimated the total molecular gas column density toward this line of sight using several different means. 
Here, we verify the validity of these estimations. 

\begin{itemize}
\item[(1)] Visual extinction -- 
$A_{\rm V}$ = 78.59 mag was estimated by \cite{yoshimura2020} from the all-sky $A_{\rm V}$ map taken with {\it Planck} \citep{2014A&A...571A..11P}. 
With the standard gas-to-extinction ratio ($N$(H)+2$N$(H$_2$))/$A_{\rm V}$ = $2 \times 10^{21}$ cm$^{-2}$ mag$^{-1}$ \citep[][and references therein]{2017MNRAS.471.3494Z} and by introducing the molecular mass fraction $f_{\rm mol}$ $\equiv$ 2$N$(H$_2$)/($N$(H)+2$N$(H$_2$)), this reads $N$(H$_2$) $\sim 8 \times 10^{22}$ $f_{\rm mol}$ cm$^{-2}$. 
\item[(2)] The CO $J$=1--0 integrated intensity -- 
$N$(H$_2$) in the corresponding column of Table \ref{table:col_co_em} was estimated from $\int T_{\rm mb}({\rm CO}) dv$ and the standard Galactic CO-to-H$_2$ conversion factor \citep[$2 \times 10^{20}$ $\rm cm^{-2}~(K~km~s^{-1})^{-1}$;][]{2013ARA&A..51..207B}. 
The observed values is $N$(H$_2$) $\sim 2 \times 10^{22}$ cm$^{-2}$. 
\item[(3)] The optically thin LTE $^{13}$CO $J$=1--0 emission -- 
$N$(H$_2$) in the corresponding column of Table \ref{table:col_co_em} was estimated by assuming the LTE conditions, optically thin $^{13}$CO emission, the $^{13}$CO excitation temperature equal to the CO peak brightness temperature, and the canonical $^{13}$CO abundance as described in detail in Appendix \ref{sec:reliability}. 
This yields $N$(H$_2$) $\sim 0.8 \times 10^{22}$ cm$^{-2}$ with large uncertainty. 
%When the dilution factor is less than 1, the excitation temperature may be underestimated and $N$(H$_2$) may be overestimated. In contrast, $N$(H$_2$) may be underestimated if the $^{13}$CO excitation temperature is lower than the excitation temperature estimated by the CO peak brightness temperature, as is the case for Component 5 in Table \ref{table:col_co_abs}. As described above, deviations from the true excitation temperature affect the estimation of the column density.
%If the estimated excitation temperature value is 10$\%$ above or below the true one, the column density is about 10 $\%$ overestimated or underestimated.
\item[(4)] The CO absorption -- 
The total CO column density in Table \ref{table:col_co_abs} can be converted to $N$(H$_2$) by assuming the CO/H$_2$ abundance ratio \citep[$X$(CO) = $1 \times 10^{-4}$ as a canonical value for Galactic disk clouds;][and references therein]{2008ApJ...679..481P}. 
The extracted value of $N$(H$_2$) = $1.8 \times$ $10^{21}$ cm$^{-2}$ is about one order of magnitude smaller than the others. 
The total $^{13}$CO column density in Table \ref{table:col_co_abs} and the canonical $X$($^{13}$CO) = $2 \times 10^{-6}$ also yields small value ($N$(H$_2$) = $3.4 \times 10^{21}$ cm$^{-2}$). 
\item[(5)] The HCO$^+$ column density -- 
The total HCO$^+$ column density in Table \ref{table:col_co_abs} can be converted to $N$(H$_2$) by assuming the universal HCO$^+$ abundance \citep[$X$(HCO$^+$) = $3 \times 10^{-9}$;][]{2023ApJ...943..172L}. 
$N$(HCO$^+$) = $5.3 \times 10^{13}$ cm$^{-2}$ yields $N$(H$_2$) $\sim 2 \times 10^{22}$ cm$^{-2}$
\item[(6)] The total $c$-C$_3$H$_2$ column density -- 
The total $c$-C$_3$H$_2$ column density in Table \ref{table:col_c2h_c3h_c3h2} can be converted to $N$(H$_2$) by assuming the universal $c$-C$_3$H$_2$ abundance \citep[$X$($c$-C$_3$H$_2$) = (2--3) $\times 10^{-9}$;][]{2012ApJ...753L..28L}. 
$N$($c$-C$_3$H$_2$) = $1.5 \times 10^{13}$ cm$^{-2}$ yields $N$(H$_2$) $\sim 0.6 \times 10^{22}$ cm$^{-2}$. 
\end{itemize}

All these estimations of total molecular column density are in reasonable agreement with $N$(H$_2$) $\sim$ (0.6--2) $\times 10^{22}$ cm$^{-2}$ except the ones estimated from $A_{\rm V}$ and CO absorption. 
The deviation of the first one may be attributed at least in part to the molecular fraction, which is $\sim$0.5 near the solar circle and even smaller out of the Galactic plane \citep{2016ApJ...823...76K,2016PASJ...68...63S}. 
The extremely low $N$(H$_2$) obtained through the conversion from the CO absorption column density to $N$(H$_2$) may be due to the ubiquity of the ``CO-poor'' molecular gas due to the differential photo-dissociation. 
We will touch on this in Section \ref{sec:co_hco+}. 

\subsection{Identification of velocity substructures}
\label{sec:internal_motion_of_substructures}

\subsubsection{Multiple-Gaussian line fit of {\rm CO} absorption} 
\label{sec:line_fit}

\begin{figure}[htb!]
%\epsscale{0.836}
\plotone{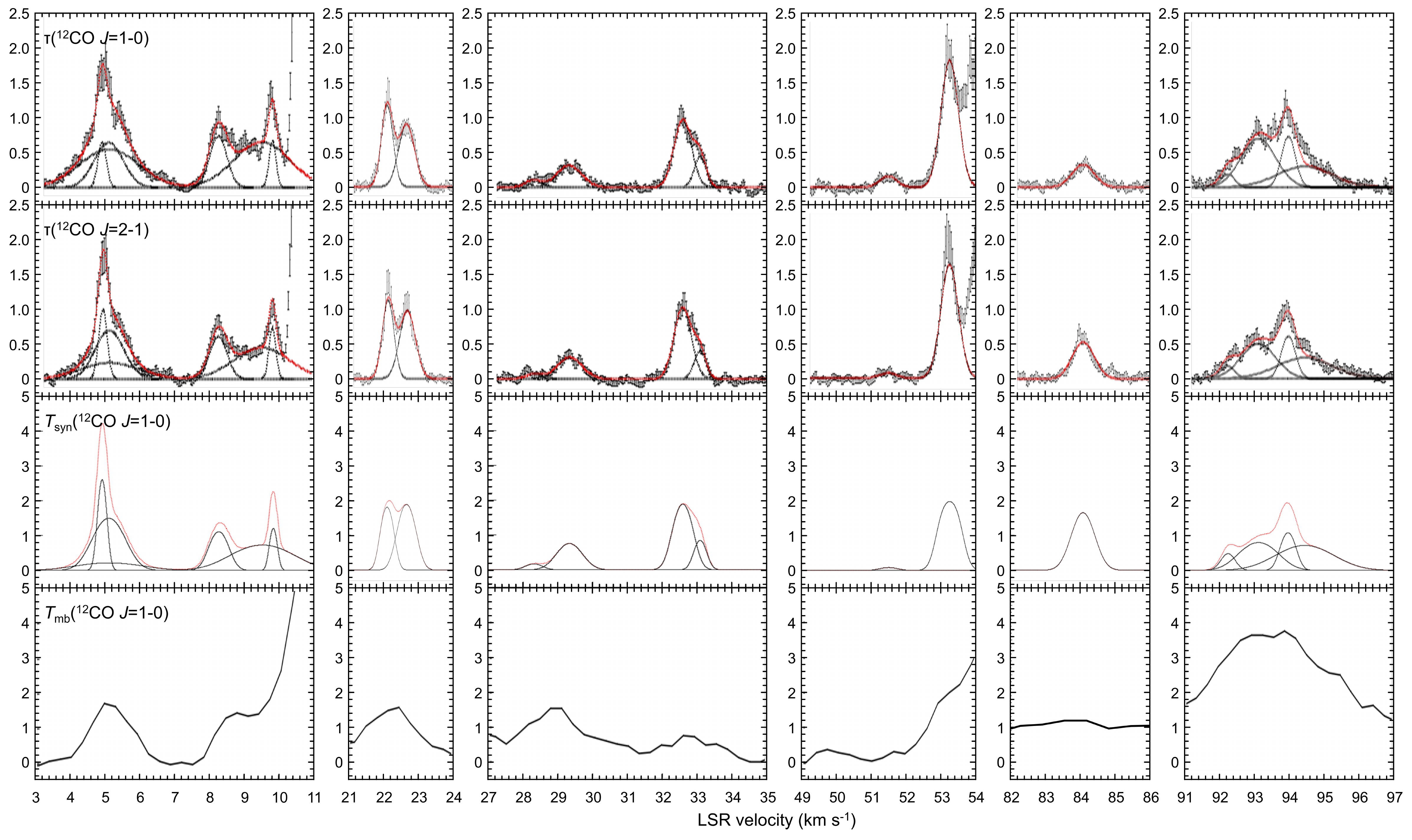}
\caption{Results of simultaneous multiple-Gaussian fitting of CO $J$=1--0 and $J$=2--1 absorption profiles in the velocity ranges $+3.0$--$+11.0$, $+21.0$--$+24.0$, $+27.0$--$+35.0$, $+49.0$--$+54.0$, $+82.0$--$+86.0$, and $+91.0$--$+97.0$ km~s$^{-1}$. 
Synthesized main beam antenna temperature ($T_{\rm syn} = \sum f (1 - e^{-\tau}) (T_{\rm ex} - T_{\rm CMB})$) with $T_{\rm ex}$ estimated from $\tau_{2-1}/\tau_{1-0}$ ratio and by assuming $f$ = 1. 
Single-dish CO $J$=1--0 emission spectra in the same velocity ranges are also plotted at the bottom.}
\label{fig:gaussfit12}
\end{figure}

\begin{figure}[htb!]
%\epsscale{0.5}
\plotone{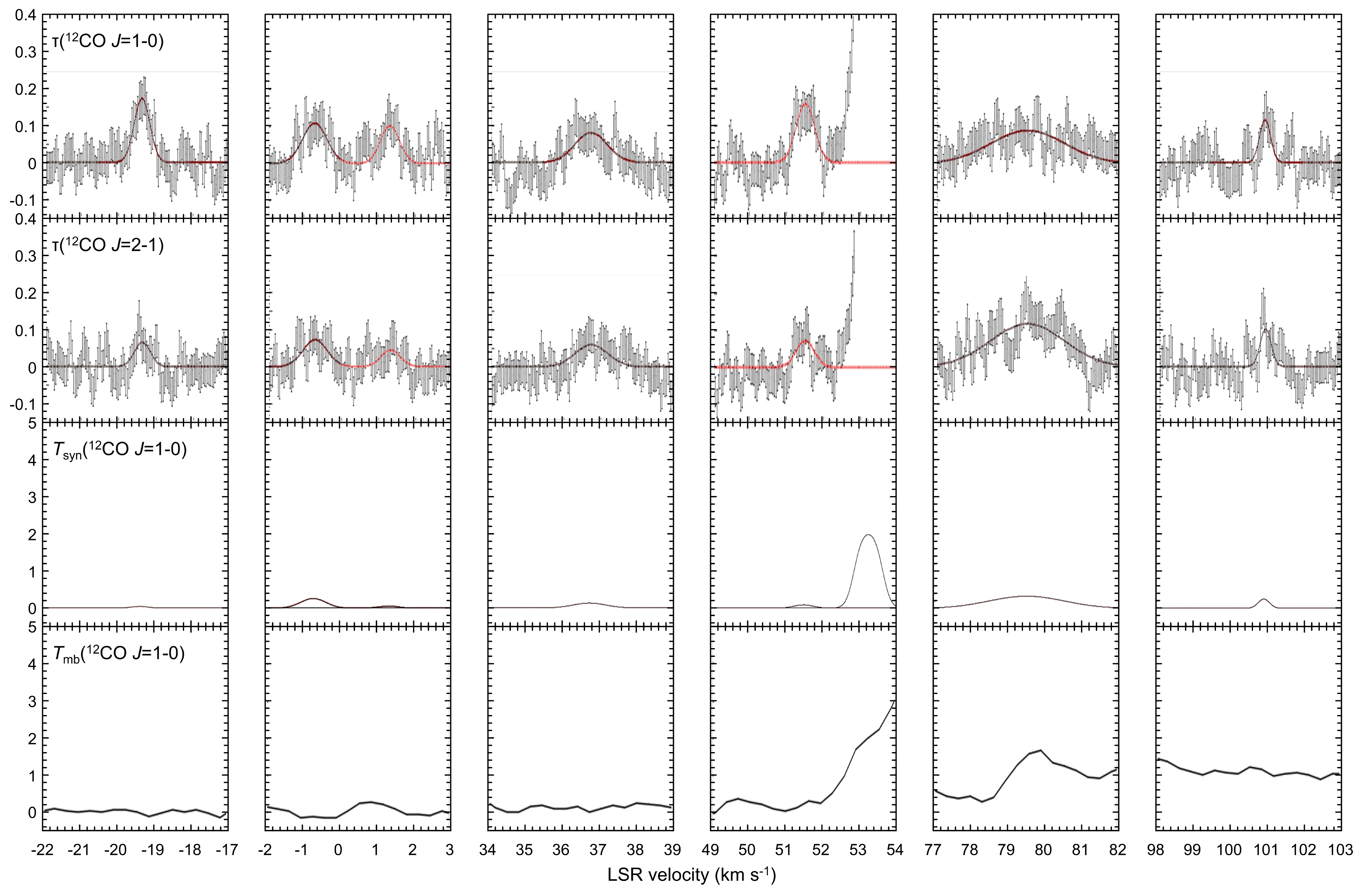}
\caption{Same as Figure \ref{fig:gaussfit12} but with magnified vertical scale of the opacity for faint components in the velocity ranges $-22.0$--$-17.0$, $-2.0$--$+3.0$, $+34.0$--$+39.0$, $+49.0$--$+54.0$, $+77.0$--$+82.0$, and $+98.0$--$+103.0$ km~s$^{-1}$. }
\label{fig:gaussfit12b}
\end{figure}

\begin{figure}[htb!]
\epsscale{0.452}
\plotone{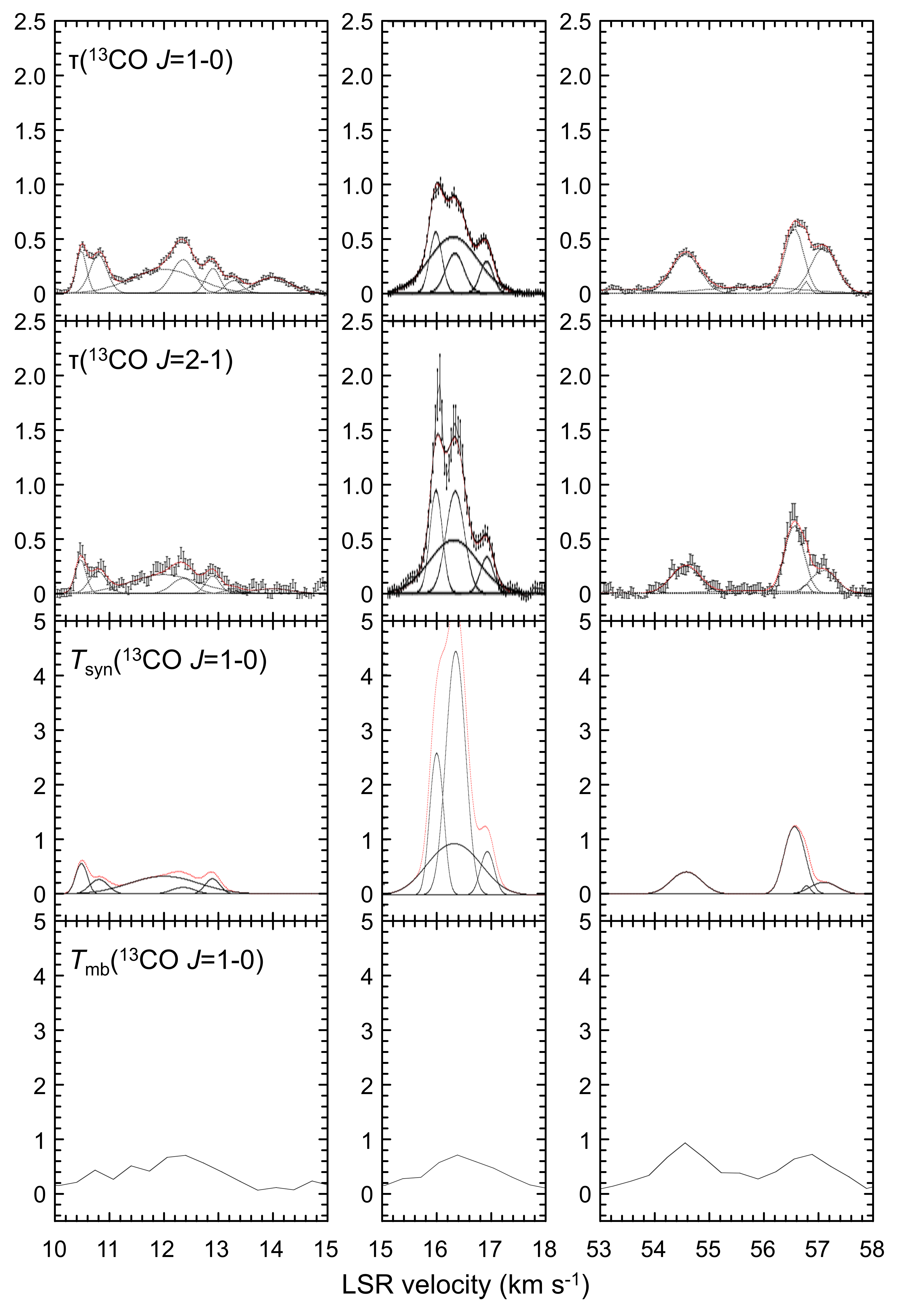}
\caption{Results of simultaneous multiple-Gaussian fitting of $^{13}$CO $J$=1--0 and $J$=2--1 absorption profiles in the velocity ranges $+10.0$--$+15.0$, $+15.0$--$+18.0$, and $+53.0$--$+58.0$ km~s$^{-1}$. 
Synthesized main beam antenna temperature ($T_{\rm syn} = \sum f (1 - e^{-\tau}) (T_{\rm ex} - T_{\rm CMB})$) with $T_{\rm ex}$ estimated from $\tau_{2-1}/\tau_{1-0}$ ratio and by assuming $f$ = 1. 
Single-dish $^{13}$CO $J$=1--0 emission spectra in the same velocity ranges are also plotted at the bottom.}
\label{fig:gaussfit13}
\end{figure}

To understand the internal motion of velocity components and to resolve the velocity substructures, simultaneous multiple-Gaussian fitting was done with the CO $J$=1--0 and $J$=2--1 absorption profiles by assuming the same center velocity ($v_{\rm c}$) and half velocity width at half maximum ($\Delta v_{1/2}$) to both transitions. 
For those velocity components with saturated CO absorption, the $^{13}$CO $J$=1--0 and $J$=2--1 absorption profiles were used instead. 
The fitting results are summarized in Figures \ref{fig:gaussfit12}--\ref{fig:gaussfit13} and Tables \ref{table:gaussfit12} and  \ref{table:gaussfit13}. 
The corresponding CO emission line profiles were also shown. 
The overall fitting was quite satisfactory, and the corresponding reduced $\chi^2$ ranged from 0.54 to 1.17. 
Although there are some hints of even narrower line components, the limited signal-to-noise ratio prevented us from further identification. 
More sensitive observations are needed to pursue this possibility further.

\begin{table}[htb!]
 \caption{Results of simultaneous multiple-Gaussian decomposition of CO $J$=1--0 and $J$=2--1 for some of the velocity components.}
 \label{table:gaussfit12}
 \centering
  \begin{tabular}{cccDDDDcc}
   \hline \hline
 Sub-comp & \multicolumn{1}{c}{$v_{\rm c}$} & $\Delta v_{1/2}$ & \multicolumn{2}{c}{$\tau_{\rm max,1-0}$} & \multicolumn{2}{c}{$\tau_{\rm max,2-1}$} & \multicolumn{2}{c}{$T_{\rm ex}$} & \multicolumn{2}{c}{$N$(CO)} & $n$(H$_2$) & $X$(CO)/($dv/dr$) \\
 ID & \multicolumn{1}{c}{(km~s$^{-1}$)} & (km~s$^{-1}$) & & & & & \multicolumn{2}{c}{(K)} & \multicolumn{2}{c}{($10^{14}$ cm$^{-2}$)} & (cm$^{-3}$) & ($\rm (km~s^{-1}~pc^{-1})^{-1}$) \\
\hline 
\phantom{0}2-1 & \phantom{}$-$19.37 & 0.25 & 0.17&$\pm$0.02 & 0.07&$\pm$0.02 & 3.07&$\pm$1.50 & 1.11&$\pm$0.12 & $3.1 \times 10^1$ & $2.3 \times 10^{-6}$ \\
\phantom{0}4-1 & \phantom{0}$-$0.71 & 0.39 & 0.11&$\pm$0.01 & 0.07&$\pm$0.01 & 4.15&$\pm$1.04 & 1.53&$\pm$0.19 & $1.4 \times 10^2$ & $4.8 \times 10^{-7}$ \\
\phantom{0}4-2 & \phantom{0$-$}1.32 & 0.30 & 0.10&$\pm$0.02 & 0.05&$\pm$0.02 & 3.29&$\pm$2.38 & 0.86&$\pm$0.20 & $5.4 \times 10^1$ & $8.3 \times 10^{-7}$ \\
\phantom{0}5-1 & \phantom{0$-$}4.93 & 0.14 & 0.65&$\pm$0.11 & 1.00&$\pm$0.11 & 8.22&$\pm$2.40 &9.04&$\pm$1.58 & $3.4 \times 10^2$ & $2.6 \times 10^{-6}$ \\
\phantom{0}5-2 & \phantom{0$-$}5.12 & 0.49 & 0.65&$\pm$0.14 & 0.70&$\pm$0.07 & 5.88&$\pm$1.03 &18.42&$\pm$4.12 & $2.1 \times 10^2$ & $2.7 \times 10^{-6}$ \\
\phantom{0}5-3 & \phantom{0$-$}5.13 & 1.00 & 0.55&$\pm$0.12 & 0.24&$\pm$0.07 & 3.24&$\pm$1.55 &15.02&$\pm$3.36 & $3.8 \times 10^1$ & $6.4 \times 10^{-6}$ \\
\phantom{0}6-1 & \phantom{0$-$}8.27 & 0.29 & 0.74&$\pm$0.04 & 0.62&$\pm$0.03 & 4.83&$\pm$0.42 & 9.58&$\pm$0.49 & $1.4 \times 10^2$ & $3.7 \times 10^{-6}$ \\
\phantom{0}6-2 & \phantom{0$-$}9.52 & 0.95 & 0.65&$\pm$0.02 & 0.45&$\pm$0.01 & 4.25&$\pm$0.16 &23.20&$\pm$0.65 & $1.1 \times 10^2$ & $3.6 \times 10^{-6}$ \\
\phantom{0}6-3 & \phantom{0$-$}9.83 & 0.12 & 0.66&$\pm$0.06 & 0.73&$\pm$0.08 & 6.00&$\pm$1.56 & 4.66&$\pm$0.41 & $2.2 \times 10^2$ & $2.8 \times 10^{-6}$ \\
\phantom{0}8-1 & \phantom{$-$}22.12 & 0.20 & 1.18&$\pm$0.05 & 1.13&$\pm$0.05 & 5.32&$\pm$0.51 &11.79&$\pm$0.53 & $1.4 \times 10^2$ & $6.7 \times 10^{-6}$ \\
\phantom{0}8-2 & \phantom{$-$}22.68 & 0.27 & 0.91&$\pm$0.03 & 0.98&$\pm$0.04 & 5.85&$\pm$0.32 &13.35&$\pm$0.49 & $1.9 \times 10^2$ & $4.4 \times 10^{-6}$ \\
\phantom{}10-1 & \phantom{$-$}28.31 & 0.29 & 0.11&$\pm$0.02 & 0.08&$\pm$0.02 & 4.44&$\pm$1.49 & 1.20&$\pm$0.19 & $1.6 \times 10^2$ & $4.3 \times 10^{-7}$ \\
\phantom{}10-2 & \phantom{$-$}29.33 & 0.41 & 0.32&$\pm$0.02 & 0.32&$\pm$0.02 & 5.45&$\pm$0.32 & 6.60&$\pm$0.31 & $2.2 \times 10^2$ & $1.2 \times 10^{-6}$ \\
\phantom{}11-1 & \phantom{$-$}32.59 & 0.30 & 0.97&$\pm$0.03 & 1.03&$\pm$0.04 & 5.75&$\pm$0.29 &16.21&$\pm$0.56 & $1.8 \times 10^2$ & $4.8 \times 10^{-6}$ \\
\phantom{}11-2 & \phantom{$-$}33.09 & 0.20 & 0.44&$\pm$0.05 & 0.40&$\pm$0.06 & 5.16&$\pm$1.41 & 4.48&$\pm$0.54 & $1.9 \times 10^2$ & $1.8 \times 10^{-6}$ \\
\phantom{}12-1 & \phantom{$-$}36.74 & 0.51 & 0.08&$\pm$0.01 & 0.06&$\pm$0.01 & 4.42&$\pm$1.05 & 1.78&$\pm$0.27 & $1.6 \times 10^2$ & $3.1 \times 10^{-7}$ \\
\phantom{}14-1 & \phantom{$-$}51.50 & 0.28 & 0.16&$\pm$0.02 & 0.07&$\pm$0.01 & 3.35&$\pm$1.26 & 1.28&$\pm$0.13 & $5.7 \times 10^1$ & $1.3 \times 10^{-6}$ \\
\phantom{}15-1 & \phantom{$-$}53.26 & 0.29 & 1.82&$\pm$0.06 & 1.64&$\pm$0.05 & 5.08&$\pm$0.26 &25.42&$\pm$0.81 & $9.6 \times 10^1$ & $1.4 \times 10^{-5}$ \\
\phantom{}18-1 & \phantom{$-$}79.50 & 1.18 & 0.09&$\pm$0.01 & 0.12&$\pm$0.01 & 6.96&$\pm$0.28 & 7.81&$\pm$0.61 & $3.7 \times 10^2$ & $2.7 \times 10^{-7}$ \\
\phantom{}19-1 & \phantom{$-$}84.09 & 0.37 & 0.33&$\pm$0.02 & 0.53&$\pm$0.02 & 8.73&$\pm$0.32 &12.98&$\pm$0.65 & $4.3 \times 10^2$ & $1.1 \times 10^{-6}$ \\
\phantom{}20-1 & \phantom{$-$}92.24 & 0.24 & 0.19&$\pm$0.03 & 0.19&$\pm$0.02 & 5.50&$\pm$1.35 & 2.38&$\pm$0.34 & $2.4 \times 10^2$ & $6.4 \times 10^{-7}$ \\
\phantom{}20-2 & \phantom{$-$}93.12 & 0.59 & 0.70&$\pm$0.02 & 0.50&$\pm$0.33 & 4.32&$\pm$3.00 &17.71&$\pm$0.61 & $1.1 \times 10^2$ & $3.9 \times 10^{-6}$ \\
\phantom{}20-3 & \phantom{$-$}93.98 & 0.24 & 0.74&$\pm$0.04 & 0.61&$\pm$0.04 & 4.81&$\pm$0.61 & 7.63&$\pm$0.46 & $1.4 \times 10^2$ & $3.7 \times 10^{-6}$ \\
\phantom{}20-4 & \phantom{$-$}94.45 & 0.93 & 0.30&$\pm$0.00 & 0.30&$\pm$0.00 & 5.50&$\pm$0.07 &14.85&$\pm$0.07 & $2.3 \times 10^2$ & $1.1 \times 10^{-6}$ \\
\phantom{}21-1 & \phantom{$-$}100.90\phantom{0} & 0.18 & 0.12&$\pm$0.02 & 0.10&$\pm$0.02 & 4.98&$\pm$2.44 & 0.95&$\pm$0.18 & $2.1 \times 10^2$ & $ 4.2 \times 10^{-7}$ \\
\hline
\end{tabular}
\end{table}

\begin{table}[htb!]
 \caption{Results of simultaneous multiple-Gaussian decomposition of $^{13}$CO $J$=1--0 and $J$=2--1 for some of the velocity components.}
 \label{table:gaussfit13}
 \centering
  \begin{tabular}{cccDDDDcc}
   \hline \hline
 Sub-comp & $v_{\rm c}$ & $\Delta v_{1/2}$ & \multicolumn{2}{c}{$\tau_{\rm max,1-0}$} & \multicolumn{2}{c}{$\tau_{\rm max,2-1}$} & \multicolumn{2}{c}{$T_{\rm ex}$} & \multicolumn{2}{c}{$N$($^{13}$CO)} & $n$(H$_2$) & $X$($^{13}$CO)/($dv/dr$) \\
 ID & (km~s$^{-1}$) & (km~s$^{-1}$) & & & & & \multicolumn{2}{c}{(K)} & \multicolumn{2}{c}{($10^{14}$ cm$^{-2}$)} & (cm$^{-3}$) & ($\rm (km~s^{-1}~pc^{-1})^{-1}$) \\
\hline 
\phantom{0}6-4 & 10.49 & 0.11 & 0.39&$\pm$0.02 & 0.31&$\pm$0.03 & 4.45&$\pm$0.47 & 1.88&$\pm$0.12 & $1.3 \times 10^2$ & $2.1 \times 10^{-6}$ \\
\phantom{0}6-5 & 10.81 & 0.18 & 0.35&$\pm$0.01 & 0.20&$\pm$0.02 & 3.63&$\pm$0.47 & 2.11&$\pm$0.08 & $6.8 \times 10^1$ & $2.7 \times 10^{-6}$ \\
\phantom{0}6-6 & 12.01 & 0.72 & 0.22&$\pm$0.02 & 0.17&$\pm$0.02 & 4.42&$\pm$0.67 & 6.57&$\pm$0.71 & $1.4 \times 10^2$ & $1.1 \times 10^{-6}$ \\
\phantom{0}6-7 & 12.36 & 0.24 & 0.31&$\pm$0.05 & 0.20&$\pm$0.02 & 3.93&$\pm$0.89 & 2.63&$\pm$0.45 & $9.2 \times 10^1$ & $1.9 \times 10^{-6}$ \\
\phantom{0}6-8 & 12.90 & 0.15 & 0.23&$\pm$0.05 & 0.16&$\pm$0.05 & 4.10&$\pm$1.57 & 1.31&$\pm$0.28 & $1.1 \times 10^2$ & $1.3 \times 10^{-6}$ \\
\phantom{0}6-9 & 13.28 & 0.17 & 0.12&$\pm$0.03 & 0.04&$\pm$0.03 & 2.81&$\pm$3.75 & 0.50&$\pm$0.13 & $ < 3.0 \times 10^2$ & -- \\
\phantom{}6-10 & 14.00 & 0.37 & 0.14&$\pm$0.01 & 0.04&$\pm$0.01 & 2.61&$\pm$1.22& 1.29&$\pm$0.06 & $ < 9.3 \times 10^1$ & --\\
\phantom{0}7-1 & 15.99 & 0.13 & 0.56&$\pm$0.03 & 0.94&$\pm$0.08 & 8.74&$\pm$0.38 & 8.61&$\pm$0.43 & $3.4 \times 10^2$ & $2.6 \times 10^{-6}$ \\
\phantom{0}7-2 & 16.31 & 0.52 & 0.52&$\pm$0.06 & 0.49&$\pm$0.08 & 5.05&$\pm$0.87 &13.49&$\pm$1.59 & $1.5 \times 10^2$ & $2.6 \times 10^{-6}$ \\
\phantom{0}7-3 & 16.34 & 0.20 & 0.37&$\pm$0.06 & 0.93&$\pm$0.09 &17.16&$\pm$0.74 &26.44&$\pm$4.65 & $7.2 \times 10^2$ & $2.0 \times 10^{-6}$ \\
\phantom{0}7-4 & 16.92 & 0.14 & 0.29&$\pm$0.02 & 0.33&$\pm$0.04 & 5.85&$\pm$0.58 & 2.55&$\pm$0.21 & $2.3 \times 10^2$ & $1.2 \times 10^{-6}$ \\
\phantom{}15-2 & 54.56 & 0.29 & 0.34&$\pm$0.00 & 0.24&$\pm$0.02 & 4.10&$\pm$0.31 & 3.80&$\pm$0.02 & $1.0 \times 10^2$ & $2.0 \times 10^{-6}$ \\
\phantom{}16-1 & 55.04 & 1.05 & 0.06&$\pm$0.00 & 0.03&$\pm$0.01 & 3.17&$\pm$1.36 & 1.82&$\pm$0.09 & $3.9 \times 10^1$ & $6.9 \times 10^{-7}$ \\
\phantom{}16-2 & 56.55 & 0.21 & 0.61&$\pm$0.02 & 0.62&$\pm$0.03 & 5.32&$\pm$0.23 & 6.88&$\pm$0.20 & $1.6 \times 10^2$ & $3.1 \times 10^{-6}$ \\
\phantom{}16-3 & 56.77 & 0.09 & 0.16&$\pm$0.03 & 0.12&$\pm$0.04 & 4.34&$\pm$1.70 & 0.56&$\pm$0.11 & $1.3 \times 10^2$ & $7.8 \times 10^{-7}$ \\
\phantom{}16-4 & 57.10 & 0.28 & 0.43&$\pm$0.01 & 0.22&$\pm$0.02 & 3.39&$\pm$0.34 & 3.60&$\pm$0.08 & $4.8 \times 10^1$ & $4.4 \times 10^{-6}$ \\
\hline
\end{tabular}
\end{table}

\subsubsection{Narrow- and broad-line components} 
\label{sec:narrow_broad}

The CO absorption profile often comprises narrow-line components ($\Delta v_{1/2}$ $\simeq$ 0.1 km~s$^{-1}$) and broad-line components with $\Delta v_{1/2}$ $\gtrsim$ 0.3 km~s$^{-1}$. 
The shape of the emission lines tends to be smoother than that of the absorption lines, though the velocity resolution for the emission study is sufficiently high. 
Narrow-line components seen in ALMA absorption spectra are less prominent in the single-dish emission spectra. 

At $T_{\rm k}$ = 40 K, the thermal line width of CO is 0.16 km~s$^{-1}$, which is comparable with the line widths of narrow-line components. 
This implies that the non-thermal motion does not dominate the internal motion of these components. 
On the other hand, the internal motion of the broad-line components is apparently dominated by non-thermal motion. 
In Subsections \ref{sec:substructure-1} and \ref{sec:substructure-2}, we will estimate the physical conditions and sizes of substructures and test if the broad-line components can be explained as ensembles of narrow-line components. 

\subsubsection{Link between line width and excitation temperature} 
\label{sec:linewidth-tex}

\begin{figure}[htb!]
\epsscale{0.5}
\plotone{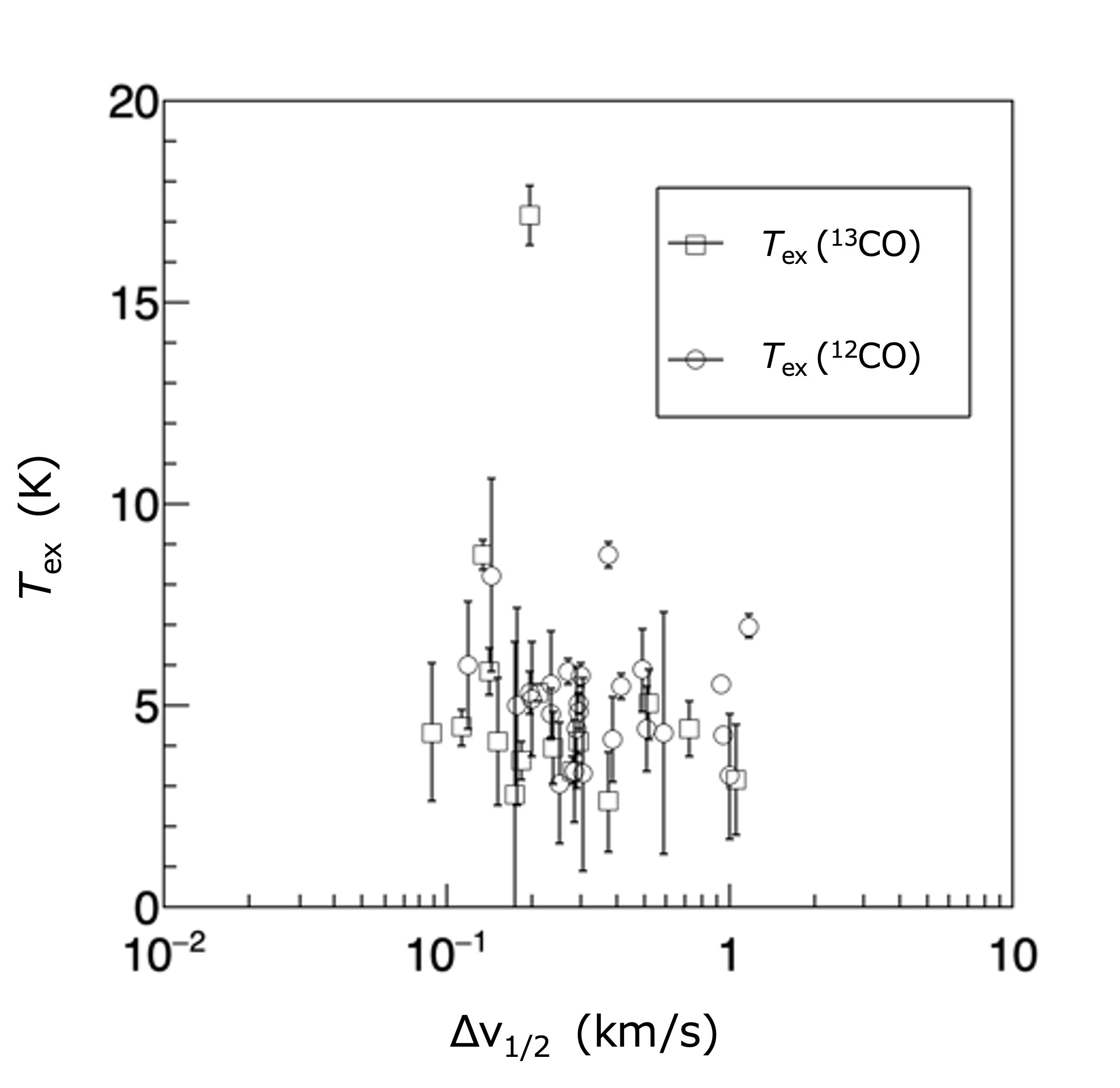}
\caption{Correlation plot of the line width and the excitation temperature of individual Gaussian components. 
Note that individual components appear only for CO or $^{13}$CO data. 
\label{fig:linewidth-tex}}
\end{figure}

The correlation plot of the line width and the excitation temperature of individual Gaussian components is shown in Figure \ref{fig:linewidth-tex}. 
The data were taken from Tables \ref{table:gaussfit12} and \ref{table:gaussfit13}. 
The histogram of line widths is not bimodal and is scattered over the wide velocity range from 0.1 to 1 km~s$^{-1}$. 
This means that the partition between the narrow- and broad-line components is unclear. 
All components with high excitation temperature ($>$ 8 K) have half maximum half widths narrower than 0.4 km~s$^{-1}$. 
This may be understood as the kinematic and physical co-evolution of clumps or the result of confusion of multiple components with similar center velocities and excitation temperatures. 

\subsection{Chemical compositions}
\label{sec:chemical_compositions}

Here, we deduce the chemical composition of individual velocity components. 
The subcomponents resolved by multiple Gaussian fitting of CO and $^{13}$CO are not separated in this analysis because of the poorer sensitivity and velocity resolution for most of the subcomponents.
For CO and $^{13}$CO, multiple transitions were detected, and the column density was obtained by using the results of the excitation analysis in Section \ref{sec:excitation_analysis}. 
For most cases, the column density was obtained by assuming LTE for optically thin lines and that the excitation temperature is the CMB temperature. 
We also detected multiple transitions of CN, but excitation analysis was not done because of the severe hyperfine contamination. 

In the following subsections, the column density of HCO$^+$, $N$(HCO$^+$), is used as an indicator of $N$(H$_2$) because we cannot directly measure it for individual Galactic disk clouds that overlap along the line of sight toward the QSO behind the Galactic plane. 
Note that the reliability of $N$(HCO$^+$) as a proxy of $N$(H$_2$) in diffuse clouds with a constant $N$(HCO$^+$)/$N$(H$_2$) = $3 \times 10^{-9}$ is well established as described in \cite{2023ApJ...943..172L}. 
This value is consistent with the results for low-density ($<$$10^3$ cm$^{-3}$) regions in star-forming clouds \citep{2021A&A...648A.120R}. 
Very rough conversion to $A_{\rm V}$ may also be possible by assuming the standard gas-to-extinction ratio ($N$(H)+2$N$(H$_2$))/$A_{\rm V}$ = $2 \times 10^{21}$ cm$^{-2}$ mag$^{-1}$ \citep[][and references therein]{2017MNRAS.471.3494Z} and that the molecular fraction is close to unity. 
In the horizontal axis of the following figures, $N$(HCO$^+$) = $10^{12}$ cm$^{-2}$ roughly corresponds to $A_{\rm V}$ = 0.3 mag. 

\subsubsection{{\rm CN}-bearing molecules}
\label{sec:cn-bearing_molecules}

\begin{figure}[htb!]
\epsscale{0.5}
\plotone{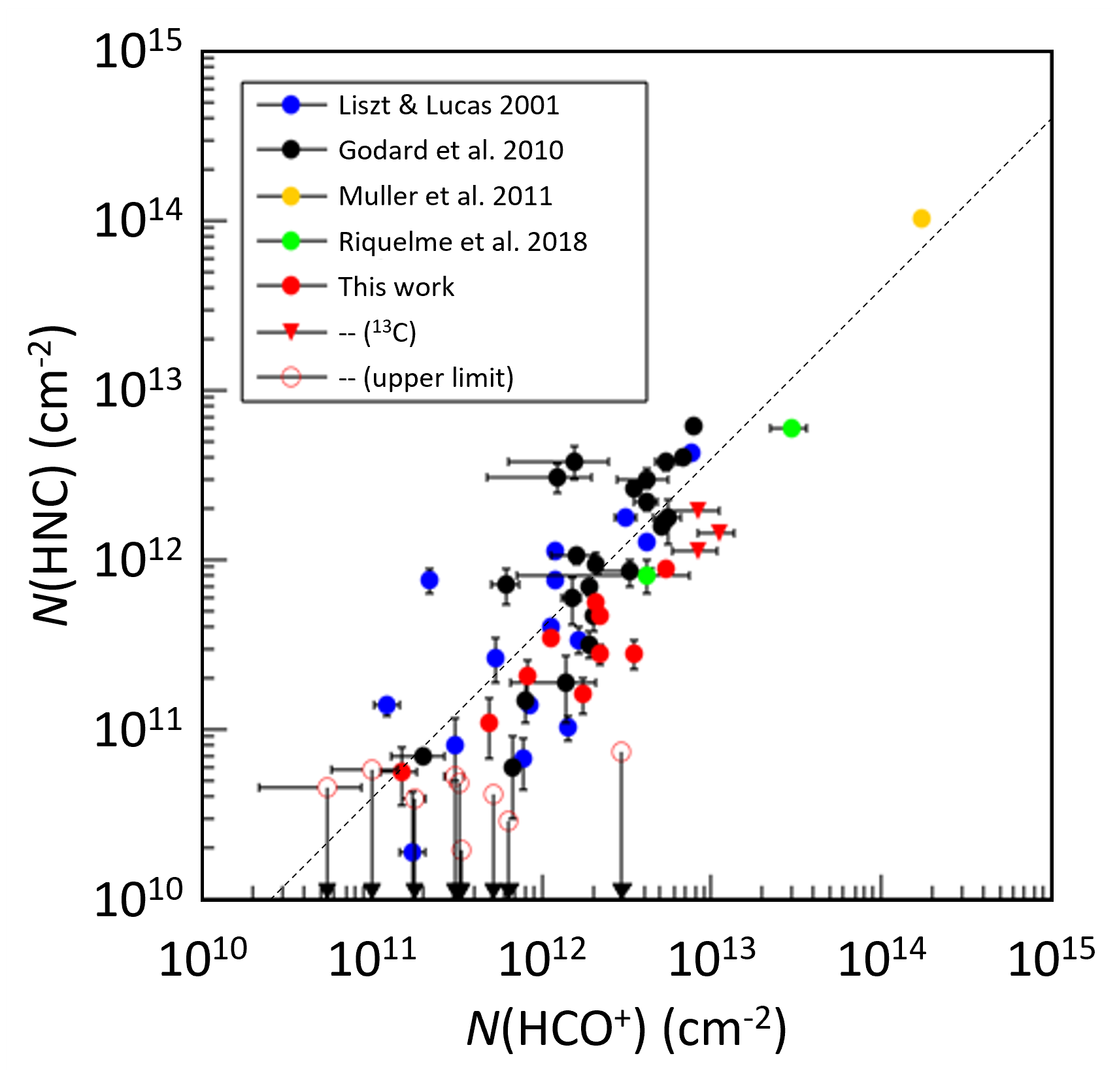}
\caption{Comparison of the column densities of HCO$^+$ and HNC. 
The column density of HCO$^+$ marked with red triangles was calculated from that of H$^{13}$CO$^+$ by assuming the HCO$^+$/H$^{13}$CO$^+$ isotopologue ratio of 50. 
The data for Galactic absorption systems \citep{2001A&A...370..576L,2010A&A...520A..20G,2016PASJ...68....6A,2018A&A...610A..43R} and $z$=0.89 QSO absorption system \citep{2011A&A...535A.103M} are also plotted. 
The diagonal line indicates the constant $N$(HNC)/$N$(HCO$^+$) ratio of 0.4. 
\label{fig:hco+_hnc}}
\end{figure}

\paragraph{{\rm HNC} and {\rm HCO$^+$}}
A correlation plot of HNC and HCO$^+$ column densities from this study is shown in Figure \ref{fig:hco+_hnc} together with the results of the previous studies \citep{2001A&A...370..576L,2010A&A...520A..20G, 2011A&A...535A.103M,2016PASJ...68....6A}. 
The overall correlation is good, as we see in Figure \ref{fig:hco+_hnc}. 
The behavior of our data points is quite similar to that of \cite{2018A&A...610A..43R}, who observed the molecular clouds in the inner Galaxy. 
The results of \cite{2001A&A...370..576L} also show a similar trend except for one outlier (29.10 km~s$^{-1}$ component of G10.62$-$0.38) in the anti-center direction. 

Our data also show that the abundance of HNC over HCO$^+$ significantly decreases at $N$(HCO$^+$) $< 5 \times 10^{12}$ cm$^{-2}$ ($A_{\rm V}$ $<$ 1.5 mag).
This may be attributed to the selective dissociation of HNC in diffuse molecular gas. 

\paragraph{{\rm CN}, {\rm HCN}, and {\rm HNC}}

To understand the correlation of column densities of CN-bearing molecules, narrower velocity ranges were set to minimize the blending by the other hyperfine components of CN and HCN. 
The column densities of HCN, HNC, and CN obtained in this manner are shown in Table \ref{table:col_cn}. 
Note that the opacity was obtained for the clean component. Thus, the conversion from the integrated opacity to the column density varies by the hyperfine component used (See, Equation \ref{eq:n_tot} and Tables \ref{table:mol_data-1} and \ref{table:mol_data-2}). 
Due to the velocity range being narrower than Table \ref{table:col_hco+_hnc}, the column density of HNC is slightly lower by about 10\%. 
The same shall apply to CN and HCN. 
This slight difference does not affect the overall conclusion of this work. 

\begin{table}[htb!]
 \caption{Column densities of HCN, HNC, and CN in absorption lines in selected narrower velocity ranges.}
 \label{table:col_cn}
 \centering
\begin{tabular}{crrDDDDDD}
\hline \hline
& \multicolumn{2}{c}{Velocity} & 
\multicolumn{4}{c}{HCN absorption} & 
\multicolumn{4}{c}{HNC absorption} & 
\multicolumn{4}{c}{CN absorption} \\
Comp & $v_{\rm min}$ & $v_{\rm max}$ & 
\multicolumn{2}{c}{$\int\tau dv$} & \multicolumn{2}{c}{$N$(HCN)} & 
\multicolumn{2}{c}{$\int\tau dv$} & \multicolumn{2}{c}{$N$(HNC)} & 
\multicolumn{2}{c}{$\int\tau dv$} & \multicolumn{2}{c}{$N$(CN)} \\
ID & \multicolumn{2}{c}{(km~s$^{-1}$)} & 
\multicolumn{2}{c}{(km~s$^{-1}$)} & \multicolumn{2}{c}{($10^{12}$ cm$^{-2}$)} & 
\multicolumn{2}{c}{(km~s$^{-1}$)} & \multicolumn{2}{c}{($10^{12}$ cm$^{-2}$)} & 
\multicolumn{2}{c}{(km~s$^{-1}$)} & \multicolumn{2}{c}{($10^{12}$ cm$^{-2}$)} \\
\hline 
\phantom{0}1a & $-$28.0 & $-$23.0 & $-$0.13  &$\pm$0.03 & $-$22.2  &$\pm$0.48 & 0.027  &$\pm$0.004 & 0.05  &$\pm$0.01 & 0.07  &$\pm$0.06 & 1.29  &$\pm$1.07 \\
\phantom{0}2a & $-$20.0 &  $-$19.0 & $-$0.02  &$\pm$0.01 & $-$0.27  &$\pm$0.22 & $-$0.001  &$\pm$0.002 & 0.00  &$\pm$0.00 & $-$0.02  &$\pm$0.03 & $-$0.43  &$\pm$0.51 \\
\phantom{0}4a & $-$1.4 & 0.0 & $-$0.01  &$\pm$0.02 & $-$0.19  &$\pm$0.26 & 0.009  &$\pm$0.008 & 0.02  &$\pm$0.01 & $-$0.06  &$\pm$0.03 & $-$4.17  &$\pm$1.98 \\
\phantom{0}5a & 0.0 & 2.8 & $-$0.05  &$\pm$0.02 & $-$0.82  &$\pm$0.36 & $-$0.003  &$\pm$0.011 & $-$0.01  &$\pm$0.02 & $-$0.09  &$\pm$0.04 & $-$1.77  &$\pm$0.81 \\
\phantom{0}5b & 2.9 & 7.0 & 0.03  &$\pm$0.03 & 0.60  &$\pm$0.43 & 0.111  &$\pm$0.013 & 0.20  &$\pm$0.02 & $-$0.08  &$\pm$0.05 & $-$5.13  &$\pm$3.30 \\
\phantom{0}6a & 7.3 & 10.0 & $-$0.04  &$\pm$0.02 & $-$0.72 &$\pm$0.35 & $-$0.018  &$\pm$0.011 & $-$0.03  &$\pm$0.02 & $-$0.08  &$\pm$0.04 & $-$5.37  &$\pm$2.72 \\
\phantom{0}6b & 10.2 & 14.5 & 1.46 &$\pm$0.03 & 5.03  &$\pm$0.09 & 0.503  &$\pm$0.015 & 0.90  &$\pm$0.03 & 0.95  &$\pm$0.05 & 17.95  &$\pm$0.98 \\
\phantom{0}7a & 15.2 & 17.3 & \multicolumn{4}{c}{(contaminated)} & 1.034  &$\pm$0.017 & 1.85  &$\pm$0.03 & 0.49  &$\pm$0.04 & 31.01  &$\pm$2.32 \\
\phantom{0}8a & 21.5 & 23.3 & 0.40  &$\pm$0.02 & 2.31 &$\pm$0.10 & 0.189  &$\pm$0.010 & 0.34  &$\pm$0.02 & 0.37  &$\pm$0.03 & 6.99  &$\pm$0.64 \\
\phantom{0}9a & 24.0 & 26.0 & \multicolumn{4}{c}{(contaminated)} & $-$0.002  &$\pm$0.009 & 0.00  &$\pm$0.02 & $-$0.02  &$\pm$0.04 & $-$1.52  &$\pm$2.38 \\
\phantom{}10a & 27.8 & 30.2 & 0.39  &$\pm$0.02 & 1.33  &$\pm$0.07 & 0.087  &$\pm$0.010 & 0.16  &$\pm$0.02 & 0.22  &$\pm$0.04 & 4.23  &$\pm$0.74 \\
\phantom{}11a & 31.9 & 33.6 & 0.48  &$\pm$0.02 & 1.67  &$\pm$0.06 & 0.141  &$\pm$0.009 & 0.25  &$\pm$0.02 & 0.06  &$\pm$0.03 & 3.03  &$\pm$1.61 \\
\phantom{}12a & 36.0 & 37.5 & 0.03  &$\pm$0.02 & 0.18  &$\pm$0.09 & 0.042  &$\pm$0.008 & 0.08  &$\pm$0.01 & $-$0.03  &$\pm$0.03 & $-$1.64  &$\pm$1.61 \\
\phantom{}13a & 44.3 & 47.0 & \multicolumn{4}{c}{(contaminated)} & $-$0.024  &$\pm$0.011 & $-$0.04  &$\pm$0.02 & 0.01  &$\pm$0.04 & 0.28  &$\pm$0.79 \\
\phantom{}14a & 48.8 & 50.7 & 0.00  &$\pm$0.02 & $-$0.01  &$\pm$0.30 & $-$0.001 &$\pm$0.009 & 0.00  &$\pm$0.02 & $-$0.07  &$\pm$0.04 & $-$1.28  &$\pm$0.66 \\
\phantom{}14b & 51.0 & 52.0 & $-$0.02  &$\pm$0.01 & $-$0.27  &$\pm$0.22 & 0.000  &$\pm$0.006 & 0.00  &$\pm$0.01 & 0.00  &$\pm$0.02 & 0.00  &$\pm$0.47 \\
\phantom{}15a & 54.0 & 55.4 & 1.07  &$\pm$0.01 & 6.12  &$\pm$0.08 & 0.498  &$\pm$0.0011 & 0.89  &$\pm$0.02 & 0.93 &$\pm$0.03 & 17.52  &$\pm$0.57 \\
\phantom{}16a & 56.0 & 57.5 & 1.48  &$\pm$0.02 & 8.51  &$\pm$0.09 & 0.646  &$\pm$0.012 & 1.15  &$\pm$0.02 & 1.31  &$\pm$0.03 & 24.64  &$\pm$0.57 \\
\phantom{}17a & 64.0 & 71.8 & \multicolumn{4}{c}{(contaminated)} & 0.032  &$\pm$0.018 & 0.06  &$\pm$0.03 & 0.19  &$\pm$0.03 & 12.13  &$\pm$1.74 \\
\phantom{}18a & 77.0 & 81.0 & 0.88  &$\pm$0.03 & 5.07  &$\pm$0.14 & 0.320  &$\pm$0.014 & 0.57  &$\pm$0.02 & 0.12  &$\pm$0.02 & 7.37  &$\pm$1.23 \\
\phantom{}19a & 83.0 & 85.0 & 0.26  &$\pm$0.02 & 1.47  &$\pm$0.10 & 0.148  &$\pm$0.010 & 0.27  &$\pm$0.02 & 0.07  &$\pm$0.02 & 4.68  &$\pm$1.62 \\
\phantom{}19b & 86.7 & 87.7 & \multicolumn{4}{c}{(contaminated)} & 0.031  &$\pm$0.007 & 0.06 &$\pm$0.01 & 0.05  &$\pm$0.02 & 3.36  &$\pm$1.62 \\
\phantom{}20a & 91.5 & 96.2 & 0.78  &$\pm$0.03 & 2.70  &$\pm$0.09 & 0.166  &$\pm$0.014 & 0.30  &$\pm$0.03 & 0.00  &$\pm$0.05 & 0.01 &$\pm$3.55 \\
\phantom{}21a & 100.7 & 101.2 & 0.00  &$\pm$0.01 & 0.02  &$\pm$0.05 & 0.007  &$\pm$0.005 & 0.01  &$\pm$0.01 & \multicolumn{4}{c}{(contaminated)} \\
\hline
\end{tabular}
\end{table}

\begin{figure}[htb!]
\epsscale{0.5}
\plotone{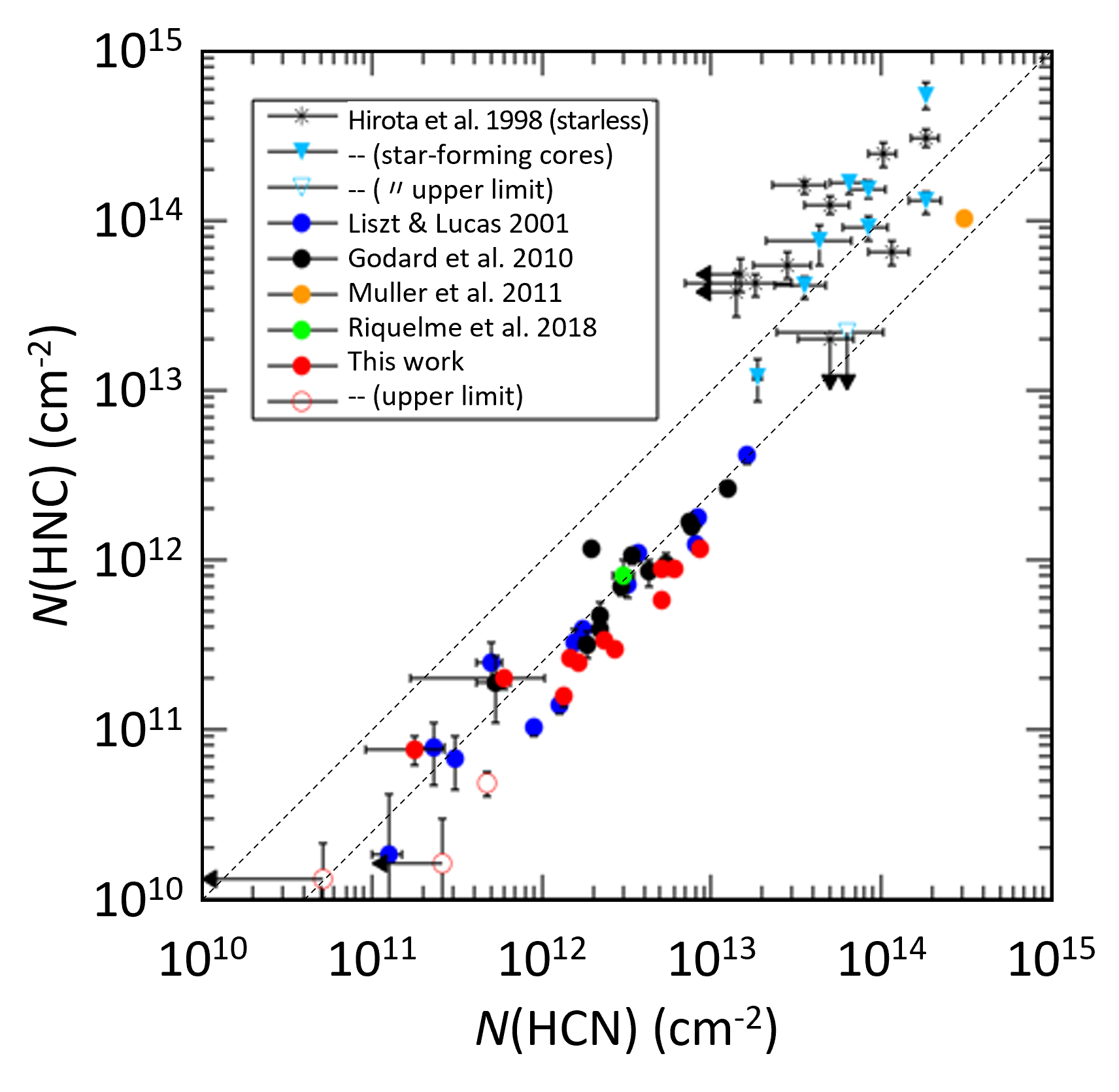}
\caption{Comparison of the column densities of HCN and HNC. 
The data for Galactic absorption systems \citep{2001A&A...370..576L,2010A&A...520A..20G,2018A&A...610A..43R} and $z$=0.89 QSO absorption system \citep{2011A&A...535A.103M} are also plotted. 
Those obtained toward dark cloud cores \citep{1998ApJ...503..717H} have significantly smaller $N$(HCN)/$N$(HNC) ratio close to unity. 
The diagonal lines indicate the constant $N$(HCN)/$N$(HNC) ratios of 1 and 4. }
\label{fig:hcn_hnc}
\end{figure}

A correlation plot of column densities of HCN and HNC based on Table \ref{table:col_cn} is shown in Figure \ref{fig:hcn_hnc}, together with the results of previous studies for the Galactic absorption system \citep{2001A&A...370..576L,2010A&A...520A..20G,2018A&A...610A..43R} and for $z$=0.89 QSO absorption system \citep{2011A&A...535A.103M}. 
The results obtained from emission line observations for dark cloud cores \citep{1998ApJ...503..717H} are also overlaid. 
The diagonal lines represent ratios of 1 and 4. 

In the present study and the other absorption studies toward Galactic diffuse clouds, the HCN/HNC ratio is nearly constant ($\gtrsim$4). 
On the other hand, the HCN/HNC ratio for dark cloud cores is about 1, indicating that the ratios are very different. 
The systematic difference between star-forming and starless cores claimed by \cite{1998ApJ...503..717H} is not clearly seen. 

This difference in the HCN/HNC ratio may be understood as a difference in kinetic temperature, as has been suggested by previous studies \citep{1998ApJ...503..717H, 2020A&A...635A...4H}. 
The empirical relation by \cite{2020A&A...635A...4H} indicates that the clouds explored in this study have a kinetic temperature of $\gtrsim$40 K, and the dark cloud cores with a ratio of about 1 have a kinetic temperature of 10 K. 

\begin{figure}[htb!]
\epsscale{0.5}
\plotone{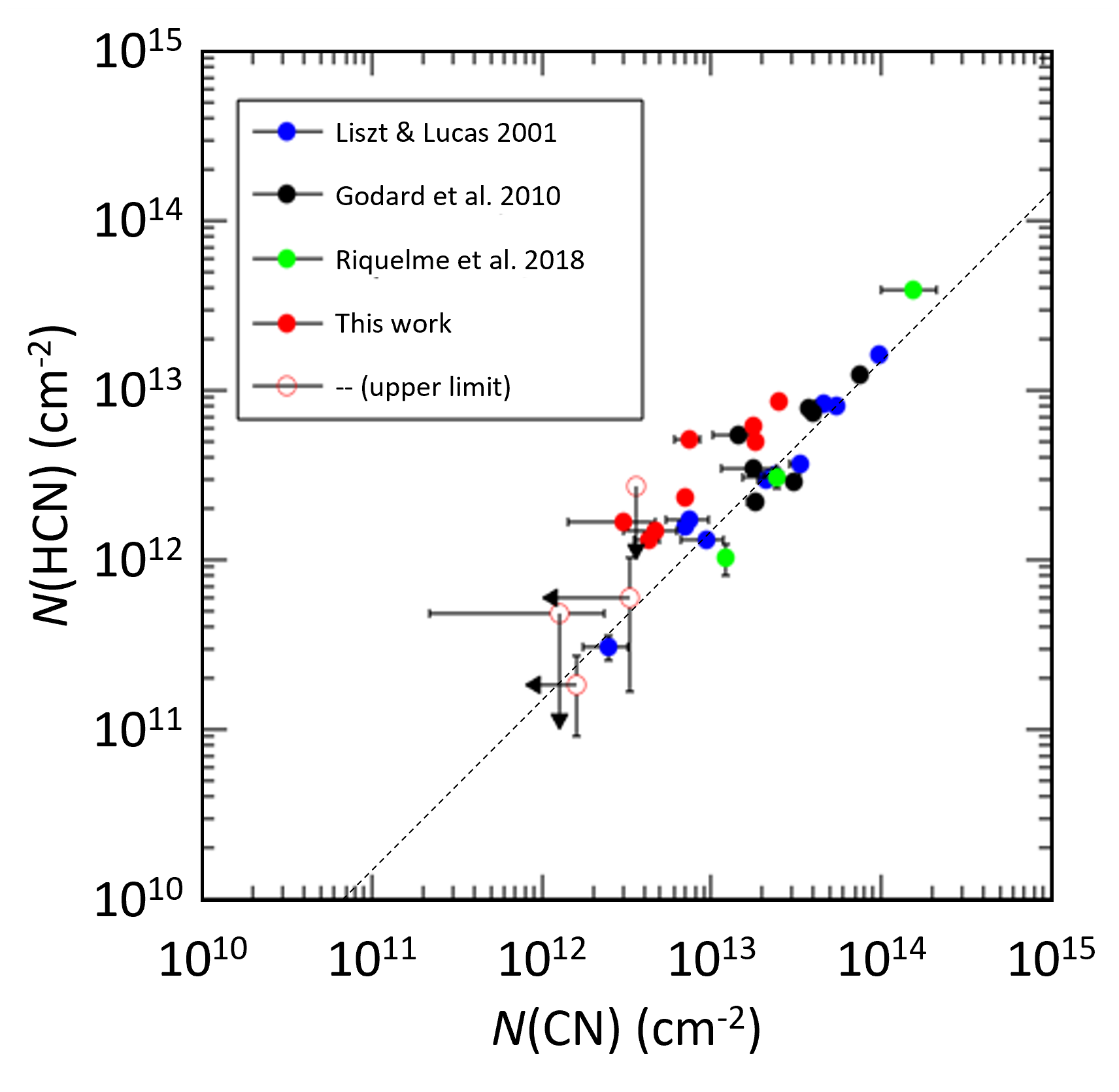}
\caption{Comparison of the column densities of CN and HCN. The data for Galactic absorption systems \citep{2001A&A...370..576L,2010A&A...520A..20G,2018A&A...610A..43R}  are also plotted. 
The diagonal line indicates the constant $N$(CN)/$N$(HCN) ratio of 6.8, which is the best fit obtained by \cite{2001A&A...370..576L}. 
\label{fig:cn_hcn}}
\end{figure}

A correlation plot of the column densities of CN and HCN obtained in this study is shown in Figure \ref{fig:cn_hcn}, together with the results of previous studies \citep{2001A&A...370..576L,2010A&A...520A..20G,2018A&A...610A..43R}. 
The correlation between CN and HCN is good and in agreement with the previous studies, though our data tend to have a lower CN/HCN ratio than the best fit (6.8) obtained by \cite{2001A&A...370..576L}. 
Here we reiterate that the velocity range for this analysis of the CN group in Table \ref{table:col_cn} was set narrower than that in Table \ref{table:col_hco+_hnc} to reduce the effect of blending by hyperfine components. 
The column densities shown here are thus smaller than those in Table \ref{table:col_hco+_hnc} by about 10\%. 
This slight difference does not affect the overall conclusion of this work. 

\begin{figure}[htb!]
\epsscale{0.5}
\plotone{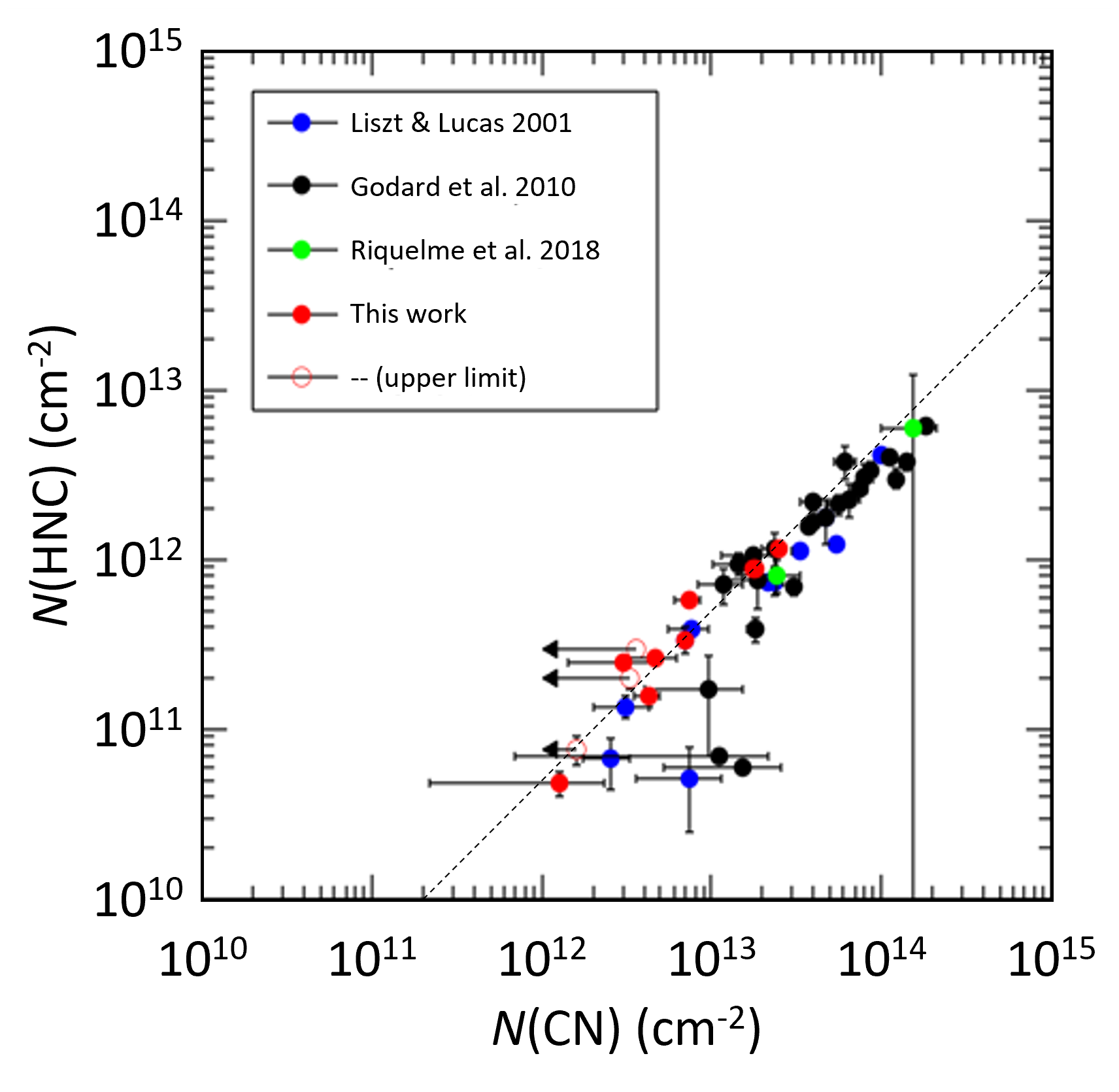}
\caption{Comparison of the column densities of CN and HNC. 
The data for Galactic absorption systems \citep{2001A&A...370..576L,2010A&A...520A..20G,2018A&A...610A..43R} are also plotted. 
The diagonal line indicates the constant $N$(CN)/$N$(HNC) ratio of 20. 
\label{fig:cn_hnc}}
\end{figure}

Similarly, a correlation plot of the column densities of CN and HNC obtained in this study is shown in Figure \ref{fig:cn_hnc}. 
At large $N$(CN) $> 3 \times 10^{12}$ cm$^{-2}$, CN and HNC show a tight and linear correlation with the HNC/CN abundance ratio $\simeq$1/20. 
The ratio starts to show a large and significant scatter below this threshold. 
%Those with a significantly small HNC/CN abundance ratio are XX. 

\subsubsection{Other N-bearing molecules}
\label{sec:nitrides}

We obtain 1$\sigma$ upper limits of $N$(N$_2$H$^+$)/$N$(HCO$^+$) $< 0.14$ and $< 0.09$ toward Components 7 and 16, which are presumably dark clouds. 
These values are much smaller than the previous results --- $N$(N$_2$H$^+$)/$N$(HCO$^+$) = 0.25 in TMC-1 \citep{1992IAUS..150..171O}. 
Since N$_2$H$^+$ is considered to form only very late in the chemical evolution of dark clouds \citep{1999A&A...350..204N}, this low upper limit supports that these clouds are still in the early phase of evolution. 

Regarding ND, there is not much work on the measurement of its abundance in interstellar space. 
\cite{2016A&A...587A..26B} detected the ND emission from the prestellar core 16293E in L1689N and estimated $X$(ND) = $1.6 \times 10^{-9}$ in its central $15''$ region. 
In this work, due to the high noise level at ALMA Band 8, the 1$\sigma$ upper limit ($X$(ND) $< 3 \times 10^{-8}$) was not sufficient for direct comparison with the above value. 
It may still be useful as a reference value for the Galactic disk clouds. 

CN$^-$ is one of the smallest anions observable in interstellar space and will provide a key to understanding the formation mechanism of anions in interstellar space. 
The past observations demonstrate that the anion-to-neutral abundance ratio of C$_{2n-1}$N$^-$ (with $n$ = 1, 2, 3, and 4) increases with increasing molecular size \citep{doi:10.1021/acs.chemrev.6b00480}, as was expected if the radiative electron attachment to the neutral counterpart of the anion is the formation mechanism \citep{1981Natur.289..656H}. 
The only claimed detection of interstellar CN$^-$ so far is toward the carbon star IRC+10216, presumably in the outer regions of the envelope with very large anion/neutral ratio \citep[$N$(CN$^-$)/$N$(CN) = $2.5 \times 10^{-3}$;][]{2010A&A...517L...2A} but with severe contamination in the $J$=1--0 and $J$=2--1 lines. 
We set very robust 1$\sigma$ upper limits of $N$(CN$^-$)/$N$(CN) $<$ 0.19 and $<$ 0.29 (or $\log X$(CN$^-$) at $-8.53$ and $-8.50$) for Components 7 and 16, respectively. 

\subsubsection{Hydrocarbons}
\label{sec:hydrocarbons}

\paragraph{c-{\rm C$_3$H$_2$} and {\rm HCO$^+$}}

\begin{figure}[htb!]
\epsscale{0.5}
\plotone{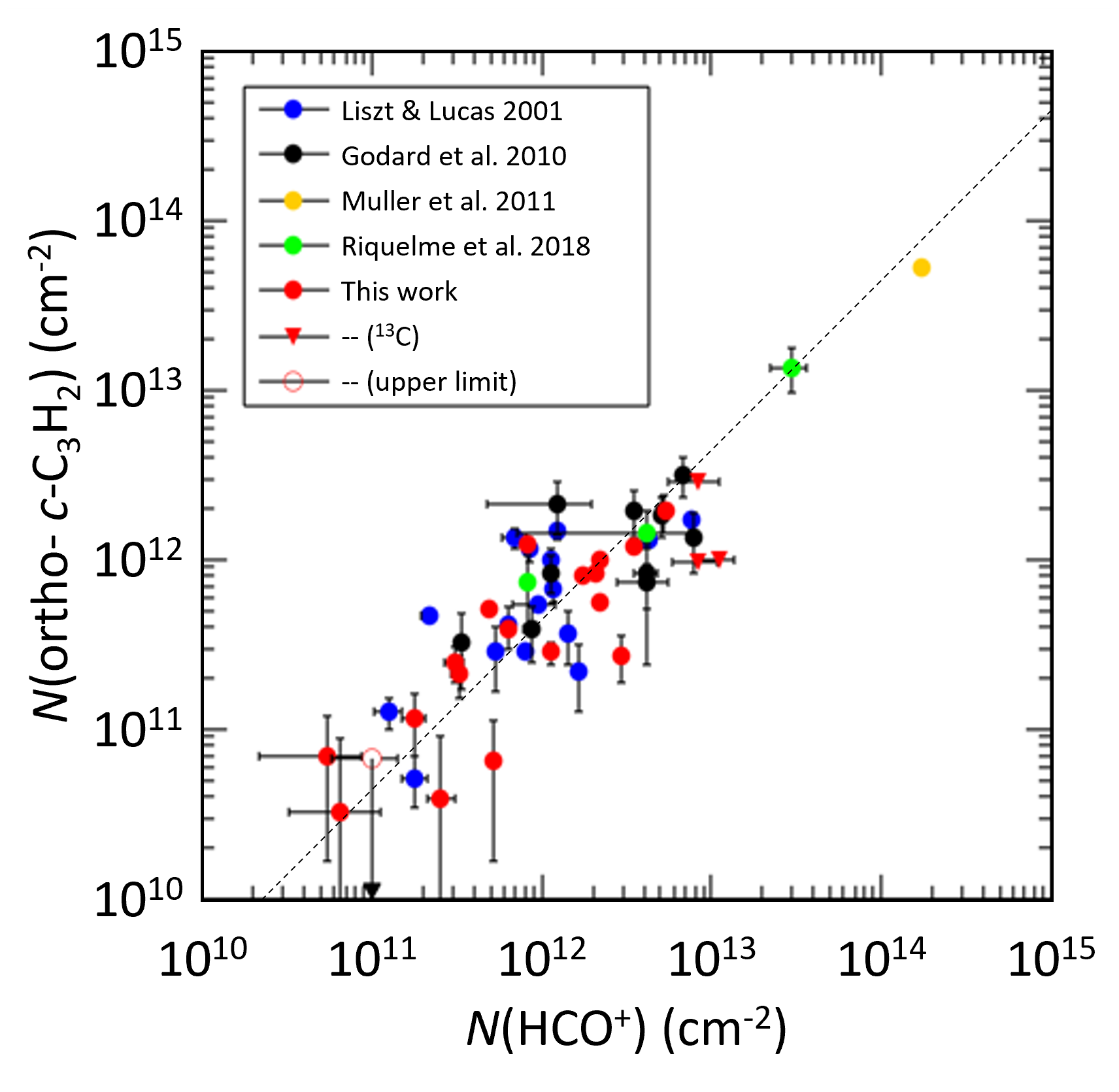}
\caption{Comparison of the column densities of HCO$^+$ and ortho- $c$-C$_3$H$_2$. 
The column density of HCO$^+$ marked with red triangles was calculated from that of H$^{13}$CO$^+$ by assuming the HCO$^+$/H$^{13}$CO$^+$ isotopologue ratio of 50. 
The data from previous studies \citep{2001A&A...370..576L,2010A&A...520A..20G,2011A&A...525A.116G,2011A&A...535A.103M,2018A&A...610A..43R} are also plotted. 
The diagonal line indicates the constant $N$($c$-C$_3$H$_2$)/$N$(HCO$^+$) ratio of 0.5}
\label{fig:hco+_c2h}
\end{figure}

A correlation plot of $c$-C$_3$H$_2$ and HCO$^+$ column densities obtained in this study is shown in Figure \ref{fig:hco+_c2h}, together with the results of the previous studies \citep{2001A&A...370..576L,2010A&A...520A..20G,2011A&A...525A.116G,2011A&A...535A.103M,2018A&A...610A..43R}. 
A relatively tight linear correlation was observed over the wide range $N$(HCO$^+$) = (0.05--200) $\times 10^{12}$ cm$^{-2}$. 
This suggests that $c$-C$_3$H$_2$ is quite ubiquitous in diffuse molecular gas and may be a good tracer of H$_2$ in diffuse gas as has been pointed out in previous studies \citep{2000A&A...358.1069L}. 

\paragraph{{\rm C$_2$H}, c-{\rm C$_3$H}, and c-{\rm C$_3$H$_2$}}

\begin{figure}[htb!]
\epsscale{0.5}
\plotone{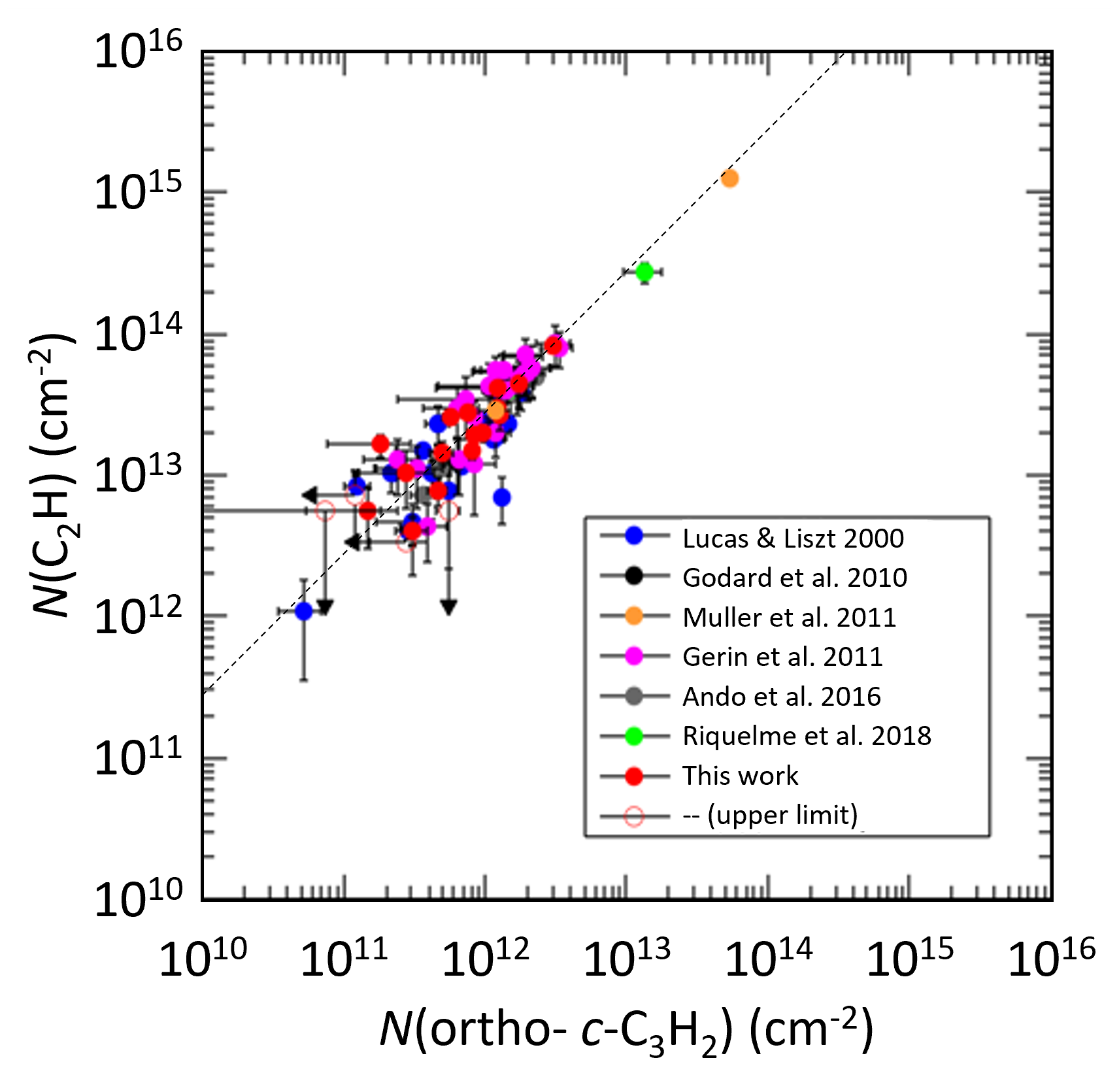}
\caption{Comparison of the column densities of C$_2$H and ortho- $c$-C$_3$H$_2$. 
The data from previous studies \citep{2000A&A...358.1069L,2010A&A...520A..20G,2011A&A...525A.116G,2011A&A...535A.103M,2016PASJ...68....6A,2018A&A...610A..43R} are also plotted. 
The diagonal line indicates the constant $N$(C$_2$H)/$N$($c$-C$_3$H$_2$) ratio of 28, which is the best fit obtained by \cite{2000A&A...358.1069L}.}
\label{fig:c2h_c3h2}
\end{figure}

A correlation plot of C$_2$H and $c$-C$_3$H$_2$ column densities obtained in this study is shown in Figure \ref{fig:c2h_c3h2} together with the results of previous studies \citep{2000A&A...358.1069L,2010A&A...520A..20G,2011A&A...525A.116G,2011A&A...535A.103M,2016PASJ...68....6A,2018A&A...610A..43R}. 
A tight linear relationship was observed in this study, at least for higher column density components (Figure \ref{fig:c2h_c3h2}). 
This supports the tight correlation between the abundances of C$_2$H and $c$-C$_3$H$_2$ in diffuse clouds \citep{2000A&A...358.1069L,2011A&A...525A.116G}. 

\begin{figure}[htb!]
\epsscale{0.5}
\plotone{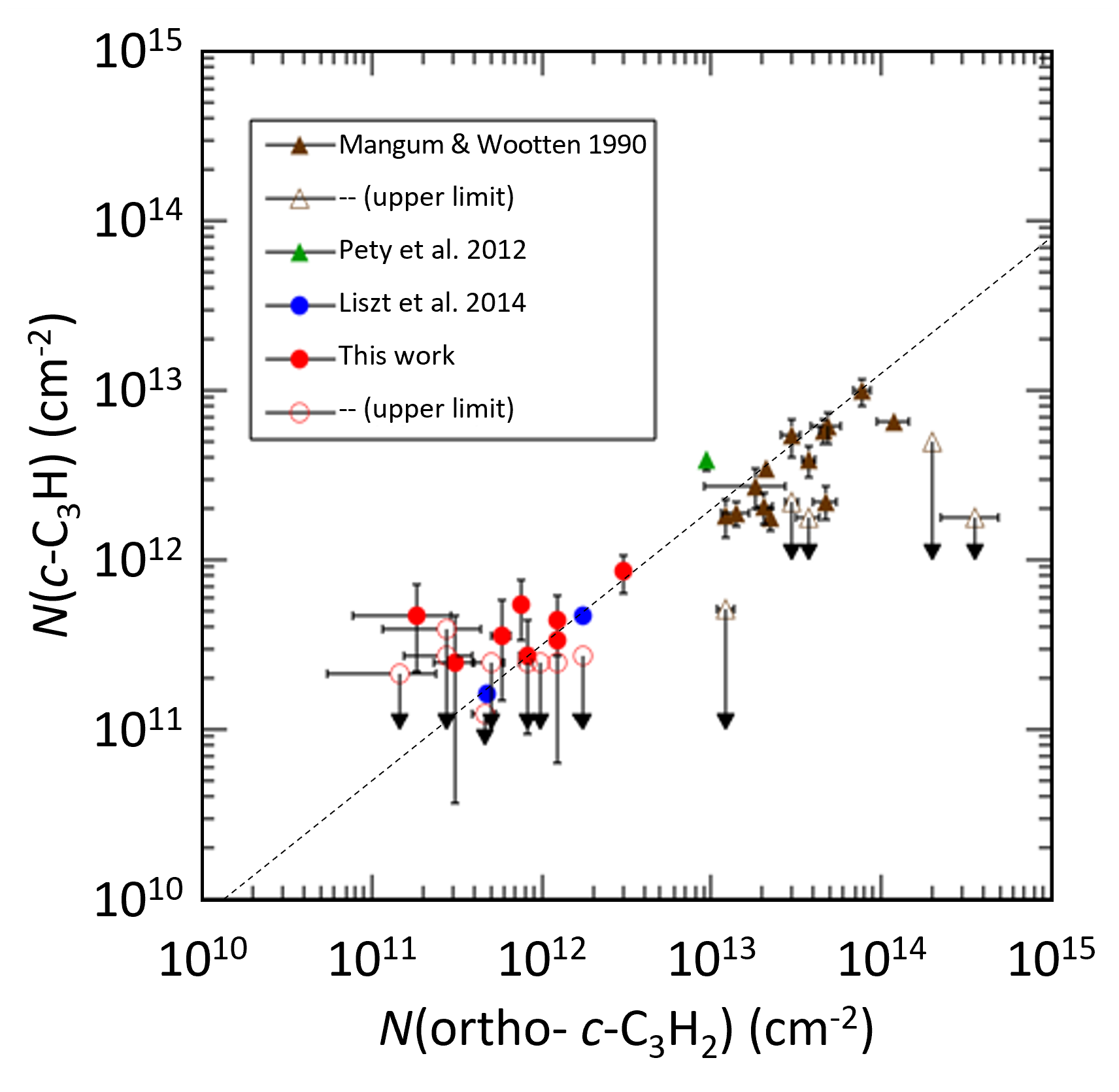}
\caption{Comparison of the column densities of ortho- $c$-C$_3$H$_2$ and $c$-C$_3$H. 
The data from previous studies \citep{1990A&A...239..319M,2012A&A...548A..68P,2014A&A...564A..64L} are also plotted. 
The diagonal lines indicate the power-law slope of 1/1.25, which is the best fit obtained by \cite{2014A&A...564A..64L}. }
\label{fig:c3h_c3h2}
\end{figure}

On the contrary, the correlation between ortho- $c$-C$_3$H$_2$ and $c$-C$_3$H is much poorer (Figure \ref{fig:c3h_c3h2}) though \cite{2014A&A...564A..64L} claimed that the $c$-C$_3$H/$c$-C$_3$H$_2$ ratio is near constant at $\simeq$0.1 for diffuse gas. 
This difference comes from the increase of the data sampled at lower column density and the inclusion of the upper limit data from \cite{1990A&A...239..319M}. 
The fitting by \cite{2014A&A...564A..64L} may be affected by the small number of data points in low column density regions. 
Nevertheless, with only a few exceptions, ortho- $c$-C$_3$H$_2$ is more abundant than $c$-C$_3$H. 

The ubiquity of $c$-C$_3$H$_2$ in diffuse environments is surprising because the gas phase formation reaction of this molecule in diffuse environments seems difficult. 
The formation reaction of $c$-C$_3$H$_2$ has been speculated by \cite{1993ApJ...417..181M} to be the dissociative recombination of $c$-C$_3$H$_3^+$,  
\begin{equation}
\label{eq:c3h2}
{\rm C_3H_3^+} + e^- \rightarrow {\rm C_3H_2} + {\rm H}.
\end{equation}
On the other hand, $c$-C$_3$H$_2$ is generally destroyed by reactions between ions and neutral molecules, and any species of the type HX$^+$ can react to convert the $c$-C$_3$H$_2$ back to $c$-C$_3$H$_3^+$ \citep{1993ApJ...417..181M}. 
The most important reactants for the destruction of $c$-C$_3$H$_2$ are HCO$^+$, H$_3^+$, and H$_3$O$^+$ like: 
\begin{equation}
\label{eq:c3h3+}
{\rm C_3H_2} + {\rm HCO^+} \rightarrow {\rm C_3H_3^+} + {\rm CO}, 
\end{equation}
and thus $c$-C$_3$H$_2$ is mostly destroyed by converting it back to an even larger molecule, $c$-C$_3$H$_3^+$ \citep{1997A&AS..121..139M}. 

The ubiquity of C$_2$H and $c$-C$_3$H$_2$ supports a top-down formation mechanism by release from hydrogenated amorphous carbon, rather than a bottom-up synthesis by gas phase reaction \citep{2000ESASP.456...91F,2003A&A...406..899F,2004A&A...417..135T,2005A&A...435..885P,2015ApJ...800L..33G,2020ApJ...896..130C} even in dark and diffuse clouds.
Recent experimental \citep{2015MNRAS.447.1242D} and theoretical \citep{2022MNRAS.511.3832A} studies also support this scenario. 

%The basic hydrocarbon chemistry in small translucent molecular clouds was studied by \cite{2000ApJS..126..427T}. On the other hand, comparing the column densities of the hydrocarbon groups to HCO$^+$, we find that at $N$(HCO$^+$) $\lesssim$ 10$^{12}$ cm$^{-2}$, the hydrocarbon groups increase relative to HCO$^+$ with a dispersion (Figure \ref{fig:hco+_c2h}).

%The observed ubiquity of C$_2$H and $c$-C$_3$H$_2$ may be explained if the photo erosion of UV-irradiated PAHs efficiently feeds them in the interstellar space (\cite{2000ESASP.456...91F}; \cite{2003A&A...406..899F}; \cite{2004A&A...417..135T}; \cite{2005A&A...435..885P}; \cite{2015ApJ...800L..33G}; \cite{2020ApJ...896..130C}) even in dark and diffuse clouds. 

\subsubsection{{\rm CO}, {\rm HCO}$^+$, and ``{\rm CO}-poor'' molecular gas}
\label{sec:co_hco+}

\begin{figure}[htb!]
\epsscale{0.5}
\plotone{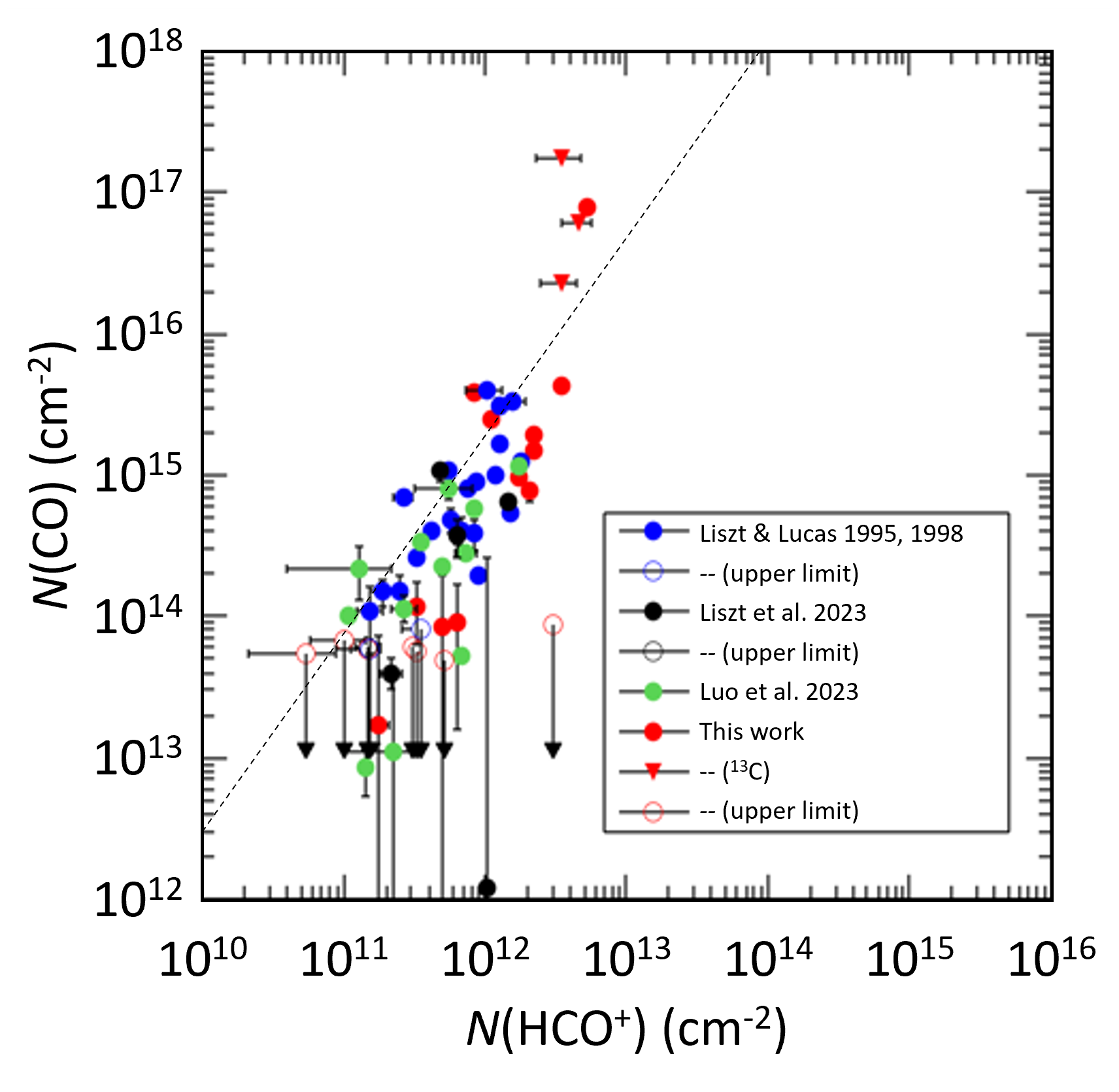}
\caption{Comparison of the column densities of HCO$^+$ and CO through absorption studies. 
The column densities of HCO$^+$ and CO marked with red triangles were calculated from those of H$^{13}$CO$^+$ and $^{13}$CO by assuming the H$^{13}$CO$^+$/HCO$^+$ ratio of 50 and CO/$^{13}$CO ratio of 25. 
The data for Galactic QSO absorption systems \citep{1998A&A...339..561L,2019A&A...627A..95L,2023ApJ...943..172L} are also plotted. 
The diagonal line is a fit by \cite{1998A&A...339..561L} and has a power-law slope of 1.4. 
\label{fig:hco+_co}}
\end{figure}

A correlation plot of the column densities of CO and HCO$^+$ obtained in this study is shown in Figure \ref{fig:hco+_co} together with the results of the previous studies \citep{1998A&A...339..561L,2019A&A...627A..95L,2023ApJ...943..172L}. 
In the previous studies, CO and HCO$^+$ were reported to be loosely correlated with a power-law slope of 1.4. 
Our data extended the plot toward the dense end ($N$(HCO$^+$) $\sim$ 10$^{13}$ cm$^{-2}$, or $A_{\rm V}$ $\sim$ 3 mag) and prefers an even steeper power-law slope close to 2. 

Given that all major species analyzed in this study except CO (i.e., HCO$^+$, HCN, HNC, CN, C$_2$H, $c$-C$_3$H, and $c$-C$_3$H$_2$) show a good positive correlation, only the behavior of CO is totally different. 
This is thus understood as a reduction of CO rather than an enhancement of HCO$^+$ and all the other major species. 

\begin{figure}[htb!]
%\epsscale{0.5}
\plotone{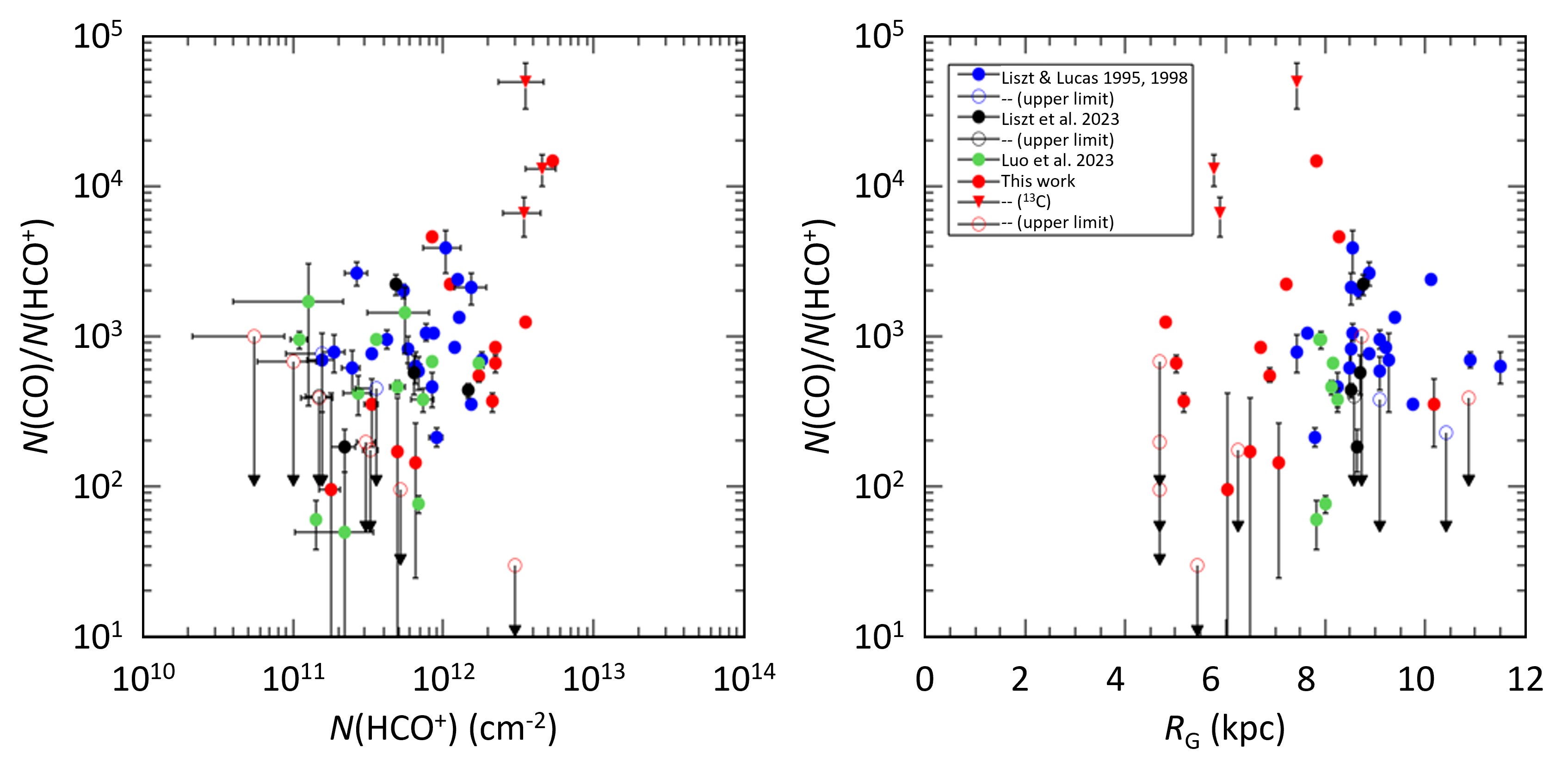}
\caption{The CO/HCO$^+$ fractional abundance as functions of the HCO$^+$ column density and the Galactocentric distance $R_{\rm G}$. 
The data marked with red triangles were calculated from those of H$^{13}$CO$^+$ and $^{13}$CO by assuming the H$^{13}$CO$^+$/HCO$^+$ ratio of 50 and CO/$^{13}$CO ratio of 25. 
The data for Galactic QSO absorption systems \citep{1998A&A...339..561L,2019A&A...627A..95L,2023ApJ...943..172L} are also plotted. }
\label{fig:co_hco+_ratio}
\end{figure}

To understand the controlling factor of the CO/HCO$^+$ fractional abundance, the CO/HCO$^+$ abundance ratio was plotted as a function of the HCO$^+$ column density and the Galactocentric distance $R_{\rm G}$ in Figure \ref{fig:co_hco+_ratio}. 
Its dependence on the HCO$^+$ column density, or $A_{\rm V}$, is quite visible with a large scatter, while its dependence on the Galactocentric distance is not apparent. 

The presence of ``CO-dark'' (or we call it here more precisely as ``CO-poor'') molecular gas has been pointed out in UV-irradiated environments \citep[e.g.,][]{1988ApJ...334..771V,2010ApJ...716.1191W} and in diffuse environments  \citep[e.g.,][]{2007ApJS..168...58S,2008ApJ...687.1075S,2018A&A...617A..54L}. 
This relative reduction of the CO abundance was also pointed out by \cite{2019A&A...627A..95L} and \cite{2020ApJ...889L...4L}, but the amplitude of the reduction was not much (up to a factor of 10) for their data. 
Thanks to the addition of the measurements toward dark clouds, the plot more clearly shows the possible dependence of the CO/HCO$^+$ abundance ratio on the HCO$^+$ column density. 

%In this study, in addition to components with correlated HCO$^+$ and CO abundance, a CO-deficient components were found. 
%A previous study by Luo et al. showed that HCO$^+$ abundance is strongly correlated with H$_2$ abundance and that HCO$^+$ is a better indicator of H$_2$ than CO in diffuse clouds (\cite{2020ApJ...889L...4L}).
%Luo et al. found that the abundance of CO increases with gas density and decreases with increasing Far-ultraviolet，(FUV) intensity, while the abundances of OH and HCO+ mostly have an opposite trend to that of CO. The CO-poor component of my observations is likely to be an environment of high FUV intensity.

\subsubsection{{\rm HCO}$^+$ and {\rm HOC}$^+$}
\label{sec:hco+_hoc+}

No HOC$^+$, the isomer of HCO$^+$, was detected in the present study, and the upper limit for the HCO$^+$/HOC$^+$ isomer ratio at the 1$\sigma$ level is approximately 200 for the dark clouds, Components 7 and 16. 
Note that the upper limit was obtained for the components with saturated HCO$^+$ absorption by multiplying the H$^{13}$CO$^+$ column density by a factor of 50 to obtain the HCO$^+$ column density. 

The present results are higher than those for diffuse clouds \citep[70$\pm$15;][]{2004A&A...428..117L,2019A&A...622A..26G} and lower than those for massive star-forming regions \citep[360--6000;][]{1995ApJ...455L..73Z,1997ApJ...481..800A}. 
This may be explained if these clouds are richer in molecular hydrogen than diffuse clouds but poorer than GMCs, and the conversion of HOC$^+$ to HCO$^+$ proceeds as in the following reaction equation \citep{1996ApJ...463L.113H}:
\begin{equation}
\label{eq:hoc+_h2}
{\rm HOC^+} + {\rm H_2} \rightarrow {\rm HCO^+} + {\rm H_2}.
\end{equation}

%HOC$^+$ is progressively converted to the isomer HCO$^+$ in reaction with H$_2$ as in the following equation.
%\begin{equation}
%HOC^+ + H_2 \rightarrow HCO^+ + H_2 
%\end{equation}
%For research on our Galaxy, HOC$^+$ has been detected in emission toward dense gas in Sgr B2 (OH), Orion-KL and several other massive star-forming regions (\cite{1995ApJ...455L..73Z}; \cite{1997ApJ...481..800A}) with $N$(HCO$^+$)/$N$(HOC$^+$) ratio of 360--6000. These results indicate that HOC$^+$ is much more abundant than predicted by previous ion-molecule models (\cite{1996ApJ...463L.113H}). 
%Regarding the diffuse gas, \cite{2004A&A...428..117L} and \cite{2019A&A...622A..26G} made sensitive absorption measurements of HCO$^+$ and HOC$^+$ toward bright continuum sources and found $N$(HCO$^+$)/$N$(HOC$^+$) ratio of $70\pm15$. 
%As for the study of external galaxies,  $X$(HOC$^+$) $\sim$ (0.15--6)$\times$10$^{-10}$ and $X$(HCO$^+$)/$X$(HOC$^+$) $\sim$ 10--150 are observed in the center of the starburst galaxy NGC 253 (\cite{2021ApJ...923...24H}), which indicates that the HOC$^+$ abundance in the center of NGC 253 is significantly higher than in quiescent Galactic spiral arm dark clouds and the Galactic Center clouds. 
%The lower limit values obtained for Components 7, 8, 17, and 18 in the current study is significantly larger than the above values for diffuse clouds and NGC 253, and is closer to the values for denser gas. 

\subsubsection{{\rm HCO}/{\rm H}$^{13}${\rm CO}$^+$ ratio and UV intensity}
\label{sec:hco/h13co+_ratio}

In emission line studies, the enhancement of the HCO/H$^{13}$CO$^+$ abundance ratio is often used as a probe of PDRs because both of them trace dense regions with very similar optical thicknesses and because their frequencies are extremely close ($\Delta$ = 23.172 MHz, or 80.07 km~s$^{-1}$ for HCO $N_K$=$1_{0,1}$--$0_{0,0}$, $J$=$\frac{1}{2}$--$\frac{1}{2}$, $F$=1--1), making it possible to measure very accurate ratios of column densities in simultaneous observations. 
Although the column densities of HCO and HCO$^+$ should be compared from the viewpoint of the chemical network and the isotopic fractionation, the above merits justify using the HCO/H$^{13}$CO$^+$ rather than HCO/HCO$^+$. 
We extend this analysis to diffuse gas whose density is much lower than the critical densities of HCO and H$^{13}$CO$^+$ \citep[$n$(H$_2$) $>$ $10^5$ cm$^{-3}$:][] {2002ApJ...575..950O,2009A&A...494..977G} through measurements of their absorption lines. 

\begin{figure}[htb!]
\epsscale{0.5}
\plotone{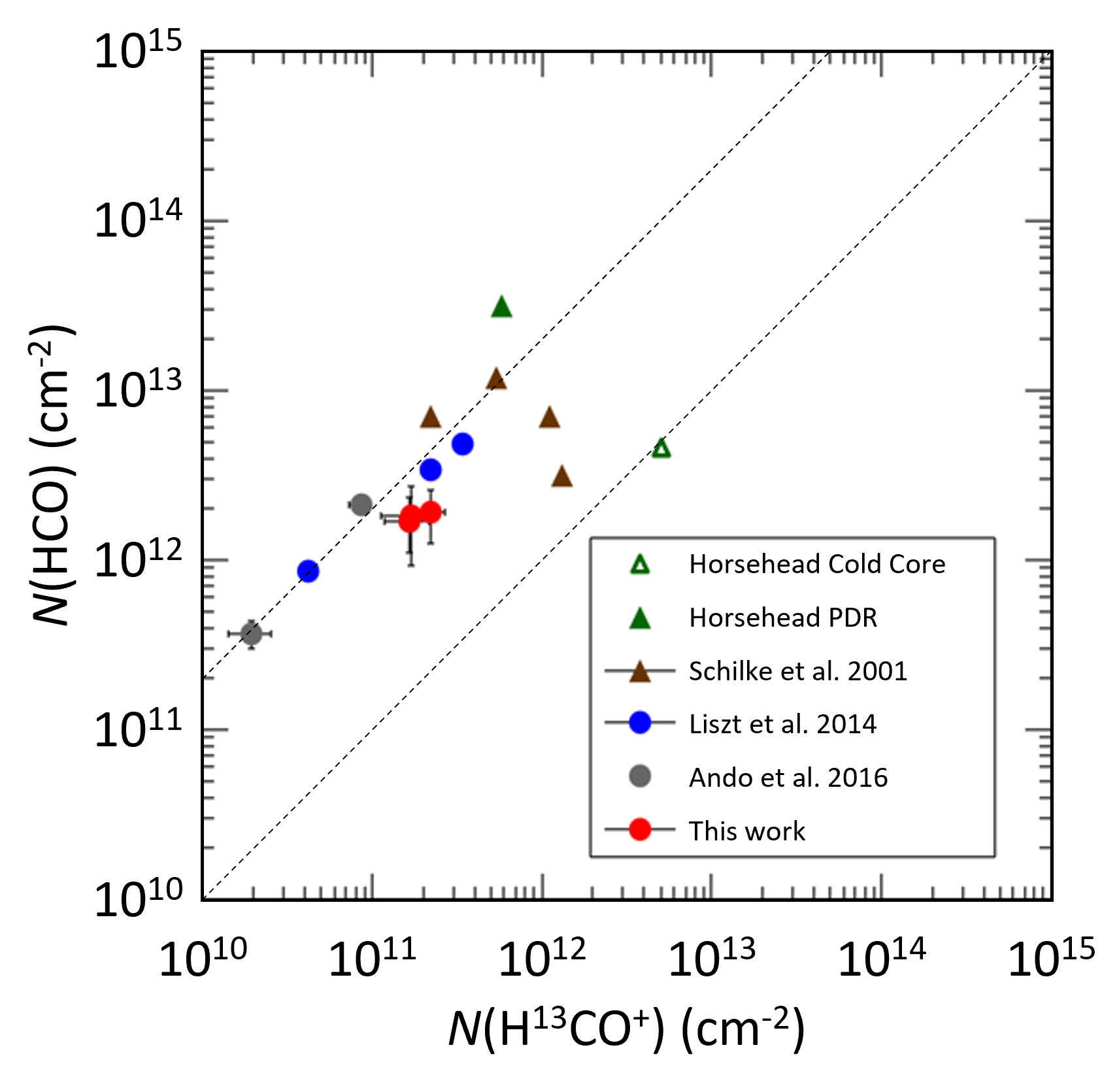}
\caption{Comparison of the column densities of H$^{13}$CO$^+$ and HCO, overlaid on the plots of those from the previous studies for the Galactic PDRs \citep{2001A&A...372..291S, 2009A&A...494..977G} and the Galactic molecular absorption systems \citep{2014A&A...564A..64L, 2016PASJ...68....6A}. 
The diagonal lines indicate the constant $N$(HCO)/$N$(H$^{13}$CO$^+$) ratios of 20 and 1. }
\label{fig:hco_hco+}
\end{figure}

A correlation plot of the column densities of HCO and H$^{13}$CO$^+$ obtained in this study is shown in Figure \ref{fig:hco_hco+}. 
In addition, the results of absorption measurements toward QSOs and Galactic radio continuum sources from previous studies \citep{2014A&A...564A..64L,2016PASJ...68....6A} and those of emission studies toward typical PDRs \citep{2001A&A...372..291S,2009A&A...494..977G} are plotted. 
The diagonal lines indicate the constant $N$(HCO)/$N$(H$^{13}$CO$^+$) ratios of 20 and 1. 
The PDRs tend to have larger ratios, while the associated cold, dense core has a small value close to unity. 

The possible processes that enhance the HCO/HCO$^+$ ratio in PDRs were examined by \cite{2009A&A...494..977G} and the references therein. 
In the FUV-shielded gas, the charge exchange reactions of HCO$^+$ with metals, like, 
\begin{equation}
\label{eq:hco+_ion_exchange}
{\rm Fe} + {\rm HCO^+} \rightarrow {\rm HCO} + {\rm Fe^+}, 
\end{equation}
are the main gas-phase formation routes of HCO. 
Since metals are depleted from the gas phase in dense cores, the HCO abundance remains low inside the cores. 

The formation of HCO in the PDR is dominated by the dissociative recombination of H$_2$CO$^+$, like, 
\begin{equation}
\label{eq:and C+_recomb}
{\rm H_2CO^+} + e^- \rightarrow {\rm HCO} + {\rm H}.
\end{equation}
In addition, the photodissociation of H$_2$CO contributes to the formation of HCO, like, 
\begin{equation}
\label{eq:h2co_photodissociation}
{\rm H_2CO} + h\nu \rightarrow {\rm HCO} + {\rm H}.
\end{equation}
Another effective pathway to produce HCO is to proceed with chemical reactions of atomic oxygen with carbon radicals that reach high abundances only in the PDR, 
\begin{equation}
\label{eq:hco_synthesis}
\ce{C+ ->C[H2] CH2+ ->C[H2] CH3+ ->C[e-]  CH2 ->C[O]  HCO}
\end{equation}
In both cases, the environment is of high UV intensity, which is common in PDRs. 

The above descriptions are made for dense PDRs but remain valid in a diffuse regime as long as the regions are C$^+$-rich.
The absorption line observations can measure the abundance ratio even when excitation is insufficient. 
Therefore, even at relatively low densities, the C$^+$-rich regions can be observed as a region of high $N$(HCO)/$N$(H$^{13}$CO$^+$) ratio. 

The previous studies showed that high $N$(HCO)/$N$(H$^{13}$CO$^+$) ratios are observed in the normal molecular gas in the Galactic plane \citep{2014A&A...564A..64L,2016PASJ...68....6A}. 
Our data for Components 6, 7, 15, and 16 showed that the ratio remains high even in dark clouds when observed in absorption. 
This suggests that from the viewpoint of interstellar chemical reactions involving C$^+$, most of the normal molecular gas in the Galactic plane is in an environment closer to the PDRs than to the cold, dense cores.
This may be understood as our absorption measurements toward dark clouds still unbiasedly sample diffuse surrounding material. In contrast, the ratio through emission measurements is strongly biased to the regions with the highest density. 

%HCO is well known to be a PDR tracer (\cite{1988ApJ...328..785S}; \cite{2009A&A...494..977G}; \cite{2009ApJ...706.1323M}; \cite{2016PASJ...68....6A}). In a diffuse medium, HCO can be produced the following chemical process.
%\begin{equation}
%\label{eq:xx}
%O + CH_2 \rightarrow HCO + H
%\end{equation}
%$CH_2$ is made from C$^+$ and H$_2$. The carbon can be ionized by FUV radiation. 
%The high $N$(HCO)/$N$(H$^{13}$CO$^+$) ratio suggests that the Galactic diffuse gas is a PDR-like environment and thus the UV photochemistry reaction can proceed. 
%In the diffuse cloud, the low molecular gas density ($n({\rm H_2})$ $\sim$ 10$^{1}$--10$^{2}$ cm$^{-3}$) (Draine 2011) makes it easy for FUV radiation from distant OB stars to reach (\cite{1999RvMP...71..173H}), ionizing carbon more easily to form HCO.

\subsubsection{{\rm H$_2$CO} and {\rm C}}
\label{sec:h2co+_carbon}
\paragraph{\rm H$_2$CO}
Although the sensitivity is limited, we see marginal detection of H$_2$CO in Components 7 and 16. 
Components 6, 15, 18, and 20 are barely visible in this $J_{K_a,K_c}$=$3_{0,3}$--$2_{0,2}$ line. 
Observations in $J_{K_a,K_c}$=$2_{0,2}$--$1_{0,1}$ at 145.602949 GHz are needed for a better constraint of the H$_2$CO column density. 
It is known that H$_2$CO has a rapid increase in abundance around $N$(HCO$^+$) $\sim 1 \times 10^{12}$ cm$^{-2}$ \citep{1995A&A...299..847L}. 
This study obtained the H$_{2}$CO/HCO$^{+}$ ratio for the four saturated CO absorbers ($N$(HCO$^+$) $\sim 1 \times 10^{12}$ cm$^{-2}$ ), which are dark clouds. 
The H$_{2}$CO/HCO$^{+}$ ratio is $<$ 0.36, 1.36$\pm$0.65, $<$ 0.41, and 0.54$\pm$0.36 for Components 6, 7, 15, and 16, respectively. 
On the other hand, H$_{2}$CO was not detected in diffuse clouds, implying the ratio was less than approximately 10. 
Previous studies have obtained values of about 2 for dark clouds, TMC1 \citep{1992IAUS..150..171O}, and 2.3 for diffuse and translucent clouds \citep{2006A&A...448..253L}, indicating that the composition is constant regardless of cloud type. 
In these dark cloud results, it is noteworthy that the ratios are small. 
The high $^{13}$CO/C$^{18}$O and HCO/H$^{13}$CO$^{+}$ also suggest a PDR-like environment, implying the photodissociation of H$_{2}$CO.

\paragraph{\rm C}
A similar trend is also visible for C in the $^3P_1$--$^3P_0$ transition, though the poor signal-to-noise ratio of the archived data prevents further study. 
A comparison of $N$(C) and $N$(CO) in Table \ref{table:non-detection} shows an equal amount of neutral gas-phase carbon and CO for Components 7 and 16. 

\paragraph{\rm CO$^+$}
%CO$^+$ is known to be the main chemical precursor of HCO$^+$ and CO in thermal, cosmic-ray, and UV-driven chemistry. 
%Although 
CO$^+$ is routinely observed in dense PDRs, and the CO$^+$ abundance was $7.2 \times 10^{-11}$ toward the PDR3 position of Mon R2 and $2.0 \times 10^{-11}$ toward the IF position of Mon R2 \citep{2016A&A...593L..12T}. On the other hand, while HCO$^+$ is abundant, there are no reported detections of CO$^+$ in the diffuse and translucent interstellar medium or dark clouds. 
The best constraint in the diffuse and translucent interstellar medium so far was given by \cite{2021A&A...648A..38G}, and the 1$\sigma$ upper limit of the CO$^+$ abundance is $<4 \times 10^{-11}$. 
As discussed in \cite{2021A&A...648A..38G}, the non-detection of CO$^+$ at this level in diffuse clouds suggests that CO$^+$ is not the precursor of HCO$^+$ that recombines to form CO.
The present result constrained the 1$\sigma$ upper limits in dark clouds to be $<7.2 \times 10^{-11}$ and $<8.3 \times 10^{-11}$, respectively, for Components 7 and 16.

\subsubsection{Sulfur-bearing molecules and shock tracers}
\label{sec:s-bearing_molecules}

\paragraph{{\rm CS}, {\rm SO}, and {\rm SiO}}
Sulfur and silicon are elements with low volatility; thus, sulfur- and silicon-bearing molecules are rare in the gas phase. 
Molecules like SiO and SO are thus used as tracers of interstellar shocks that spatter these molecules out from dust grains. 

\begin{figure}[htb!]
\epsscale{0.5}
\plotone{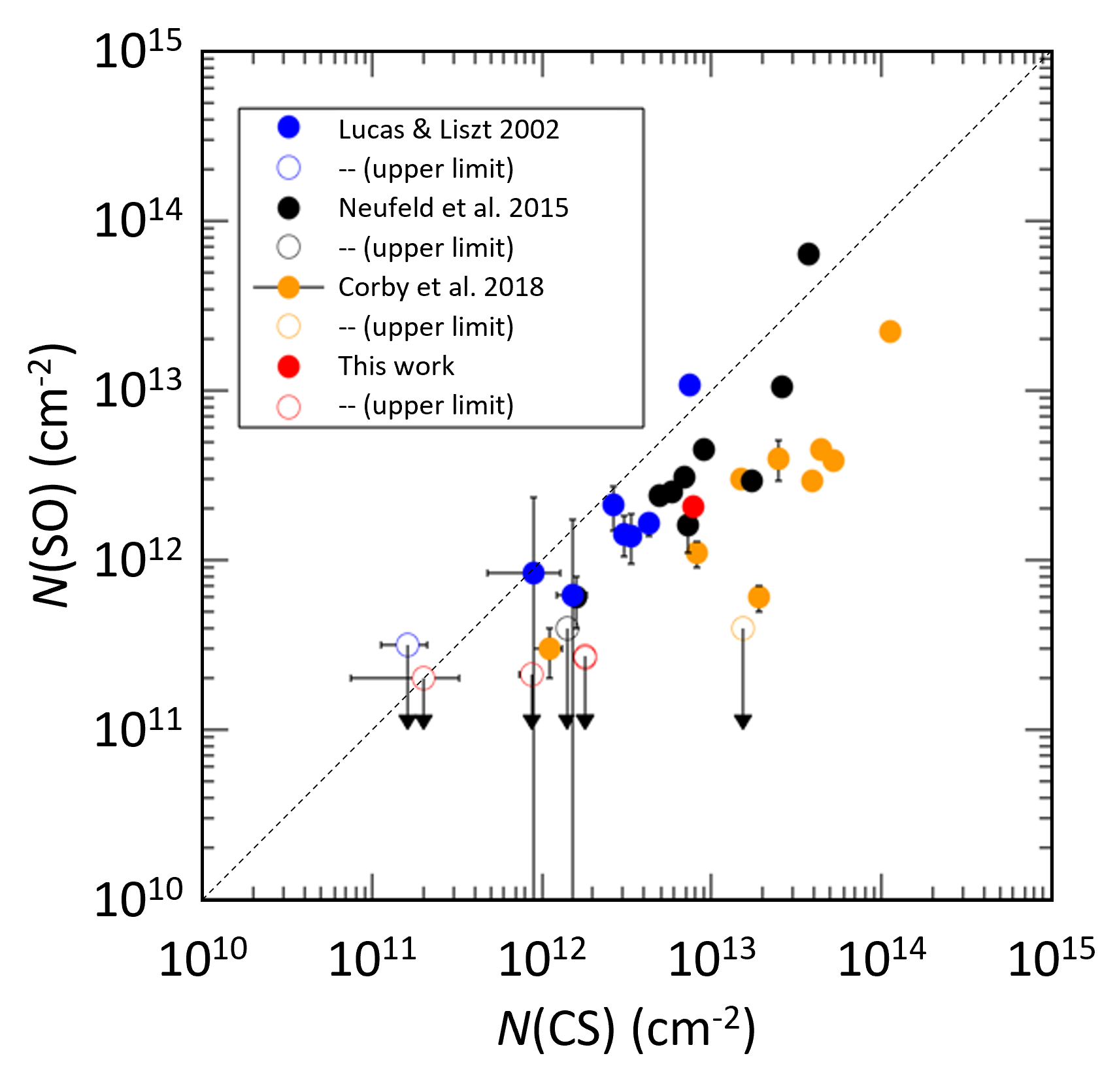}
\caption{Comparison of the column densities of CS and SO. 
The data from previous studies \citep{2002A&A...384.1054L,2015A&A...577A..49N,2018A&A...610A..10C} are also plotted. 
The diagonal line indicates the constant $N$(SO)/$N$(CS) ratio of 1.}
\label{fig:cs_so}
\end{figure}

\begin{figure}[htb!]
\epsscale{0.5}
\plotone{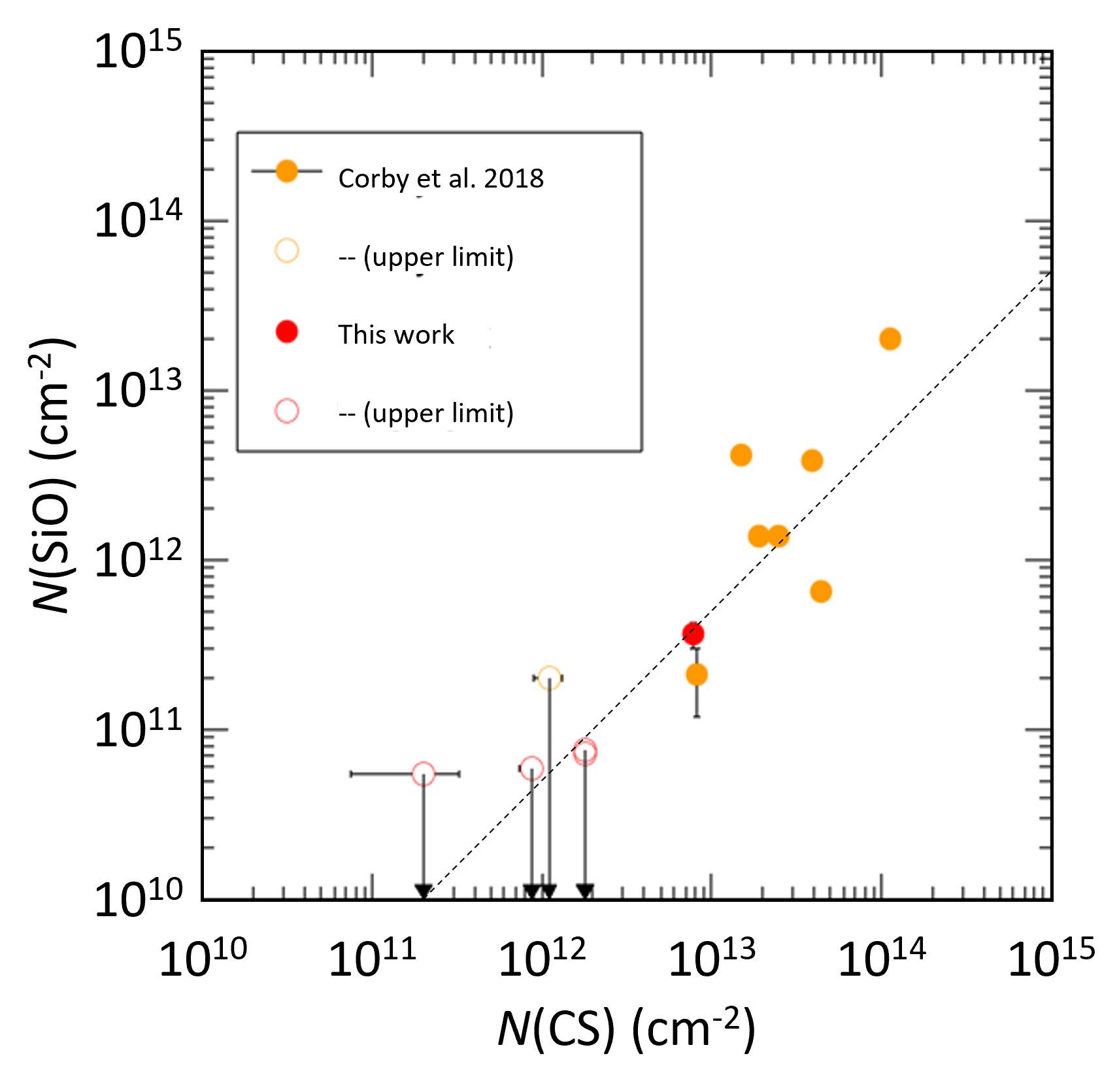}
\caption{Comparison of the column densities of CS and SiO. 
The data from the previous study \citep{2018A&A...610A..10C} are also plotted. 
The diagonal line indicates the constant $N$(SiO)/$N$(CS) ratio of 0.05.}
\label{fig:cs_sio}
\end{figure}

In this study, one of the velocity components showed absorption of SO, SiO, and CS. 
The column densities of SO and SiO are plotted against that of CS in Figures \ref{fig:cs_so} and \ref{fig:cs_sio}, respectively, overlaid on the data from previous studies \citep{2002A&A...384.1054L,2015A&A...577A..49N,2018A&A...610A..10C}. 

In Figure \ref{fig:cs_so}, we identified an upper envelope that corresponds to the $N$(SO)/$N$(CS) ratio close to unity. 
In some cases, $N$(CS) and $N$(SO) have significant errors, and only upper limits are obtained. 
However, there is generally a good correlation between $N$(CS) and $N$(SO). 

Considering that the $X$(CS) values obtained for the four dark clouds with saturated CO absorption lines vary by as much as two orders of magnitude, they both change closely following each other.

Regarding the correlation between CS and SiO, the present result shows the SiO/CS abundance ratio of 0.046$\pm$ 0.009, similar to the ones obtained in the previous study \citep{2018A&A...610A..10C}. 

In the component we obtained a SiO/CS ratio, the CO excitation temperature was only slightly higher than the other components, and no YSO-like excitation sources were found. 
This suggests that the shock wave's impact was not so large, or the gas cooled rapidly before re-agglomeration into dust.

\paragraph{{\rm HCS$^+$}, {\rm H$_2$CS}, and {\rm C$_2$S}}

We only obtained 1$\sigma$ upper limits of the abundances of HCS$^+$, H$_2$CS, and C$_2$S at $<5 \times 10^{-8}$, $<5 \times 10^{-9}$, and $<6 \times 10^{-9}$, respectively, for Component 7, which is presumably a dark cloud. 
These values do not provide meaningful limits since the previous study showed that $X$(HCS$^+$) $\sim 2 \times 10^{-10}$, $X$(H$_2$CS) $\sim 3 \times 10^{-10}$, and $X$(C$_2$S) $\sim 1 \times 10^{-10}$ in diffuse and translucent clouds in the line-of-sight to Sgr B2 \citep{2018A&A...610A..10C}. 

\paragraph{{\rm SH$^+$}}

Although we did not detect SH$^+$ this time, we could put a 1$\sigma$ upper limit at $3 \times 10^{-8}$. 
This value is more than an order of magnitude lower than the previous studies \citep{2000A&A...355..327L,2002A&A...384.1054L}.

In this study, the 1$\sigma$ upper limits of $\log X$(SH$^+$) obtained for Components 7 and 16 were $-8.60$ and $-8.74$.
On the other hand, \cite{2011A&A...525A..77M} and \cite{2012A&A...540A..87G} found SH$^+$ absorption in Galactic diffuse clouds.
\cite{2011A&A...525A..77M} sampled the gas on the Galactic plane toward Sgr B2(M) to detect 6 components with $\log$(SH$^+$)/(H+2H$_2$) ranging from $-9.48$ to $-7.96$. 
Similarly, \cite{2012A&A...540A..87G} sampled the gas toward Sgr B2(N) and Sgr A$^*$+50 and obtained $\log X$(SH$^+$) to be $-6.15$ and $-6.22$, respectively, which are considerably higher than the present results. 

This significant difference in $X$(SH$^+$) may reflect the effect of galactic shock waves in these clouds because SH$^+$ is known to be produced by the following endothermic chemical reaction \citep{1986ApJ...310..408D,1986MNRAS.221..673M},
\begin{equation}
\label{eq:sh+}
{\rm S}^+ + {\rm H_2} \rightarrow {\rm SH}^+ + {\rm H}. 
\end{equation}
The lower SH$^+$ abundance than the results of \cite{2012A&A...540A..87G} suggests that the contribution of shock waves is less than in the region of their samples.

\paragraph{{\rm CH$_3$OH}}

Complex organic molecules (COMs) are thought to be tracers of interstellar shocks because these species form on grain surfaces and are sputtered out to the gas phase. 
CH$_3$OH is one of the simplest among such species. 

CH$_3$OH was searched in absorption toward diffuse clouds by \cite{2008A&A...486..493L} without success. 
The 1$\sigma$ upper limit of $N$(CH$_3$OH)/$N$(HCO$^+$) toward the three targets ranged from 0.06 to 0.18, which is even smaller than the value in TMC-1 \citep{1992IAUS..150..171O}. 
On the other hand, CH$_3$OH detected in absorption toward the $z$=0.89 system \citep{2021A&A...652A...5M} had $\log X$(CH$_3$OH) = $-8.07$. 
Our upper limits of the CH$_3$OH abundance ($\log X$(CH$_3$OH) $< -8.72$ and $< -8.70$ for Components 7 and 16) are much smaller than the above values. 
These dark clouds may be still young in the chemical evolution that leads to COMs.  

\subsubsection{Isotopologue ratios involving {\rm C}, {\rm O}, and {\rm S}} \label{sec:co_isotopic_abundance_ratios}

The isotopologue ratio in the interstellar medium reflects the history of nucleosynthesis and is used to understand such history in the Galaxy. 
It is widely known that there are gradients of isotopic ratios as a function of Galactocentric distance. 
The isotopologue ratios of molecules often show deviation from the elemental isotopic ratio, and such an effect is called chemical fractionation. 
Chemical fractionation is more prominent in diffuse clouds than in cold, dense clouds \citep[e.g.,][]{2011ApJ...728...36R}. 
In this section, we try to extract the abundance ratios of some of the isotopologues. 

\paragraph{{\rm CO/$^{13}$CO}}

CO is the main form of gas-phase carbon in molecular clouds, and thus the CO/$^{13}$CO isotopologue ratio is essential because it controls the C/$^{13}$C ratio of the gas-phase carbon reservoir. 
Because of the CO absorption saturation in dark clouds and giant molecular clouds and the limited $^{13}$CO sensitivity toward diffuse clouds and cloud edges, however, precise measurement of the CO/$^{13}$CO isotopologue ratio over a wide column density range is difficult. 

In this study, a meaningful CO/$^{13}$CO isotopologue ratio was observed toward cloud edges and some of the dark and diffuse clouds, which have 30.1$\pm$4.1, 62.5$\pm$18.8, 85.5$\pm$48.8, $>$ 91.9 and 135.4$\pm$80.6 for Components 5, 8, 11, 19, and 20, respectively. 

\begin{figure}[htb!]
\epsscale{0.5}
\plotone{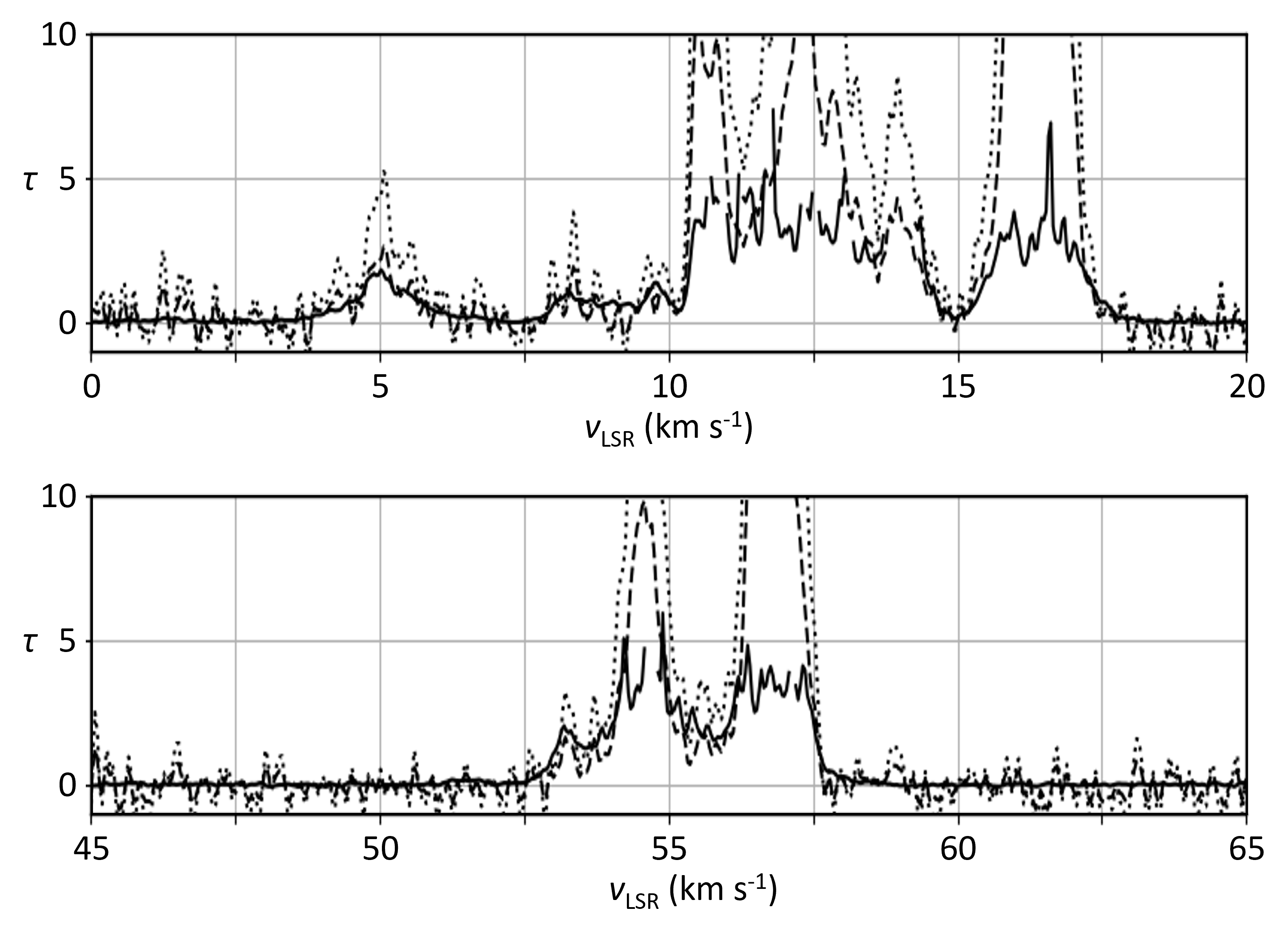}
\caption{Comparison of the CO opacity (solid lines) with the $^{13}$CO opacity multiplied by 25 (dashed lines) and 50 (dotted lines). 
Note that the CO opacity exceeding 3 ($e^{-\tau} < 0.05$) contains large errors and thus cannot be fully relied on.}
\label{fig:tau_dumped}
\end{figure}

For saturated CO absorbers, the slopes of the line wings are used to estimate the isotopologue ratio. 
Figure \ref{fig:tau_dumped} compares the CO opacity with the $^{13}$CO opacity multiplied by 25 and 50. 
Components 5, 6, 7, 15, and 16 have significantly small CO/$^{13}$CO ratios that favor 25 rather than 50. 

\paragraph{{\rm $^{13}$CO/C$^{18}$O} double isotope ratio}

In this study, the $^{13}$CO/C$^{18}$O double isotope ratio was obtained for the four saturated CO absorbers. 
The $^{13}$CO/C$^{18}$O isotopologue ratio is $>$ 78.3, 175.2$\pm$119.8, $>$ 31.8, and $>$ 76.8 for Components 6, 7, 15, and 16, respectively. 
On the other hand, the isotope ratios as elements are $^{12}$C/$^{13}$C = 89 \citep{1987ApJ...317..926H} and $^{16}$O/$^{18}$O = 498 \citep{1997A&A...326..139W} in the solar system. 
If this is directly reflected in the CO isotope ratio, the $^{13}$CO/C$^{18}$O isotope ratio is 5.5. 
In addition, $^{13}$CO and C$^{18}$O are 1/59 and 1/560 compared to CO in the local interstellar medium \citep{1994ARA&A..32..191W,1998A&A...337..246L}, expecting $^{13}$CO/C$^{18}$O isotope ratio of 9.5. 
The CO/$^{13}$CO ratios in Components 5, 6, 7, 15, and 16 are low at about 25, which would suggest that the $^{13}$CO/C$^{18}$O ratio is at most 20. 
However, the results obtained for the dark clouds are much larger than this. 

%These chemical fractionations are explained by selective photodissociation. 
%This is because $^{13}$CO and C$^{18}$O are less abundant than $^{12}$CO, resulting in less self-shielding effect and more photodissociation of C$^{18}$O than $^{13}$CO. 
These chemical fractionations may be explained by the selective far-UV photodissociation of the C$^{18}$O reported in previous studies \citep{2016ApJ...826..193L,2018A&A...617A..14P}.
Observations with the NRO 45 m telescope toward the Orion A giant molecular cloud \citep{2014A&A...564A..68S} show an average $N$($^{13}$CO)/$N$(C$^{18}$O) value of 16.47$\pm$0.10 in the nearly edge-on photon-dominated region. 
The average $X$($^{13}$CO)/$X$(C$^{18}$O) value in the other regions is 12.29$\pm$0.02. 
The values obtained in the dark clouds in this study are higher than those in such photodissociation regions (PDRs). 
Furthermore, the ratios of HCO to H$^{13}$CO$^{+}$ are comparable to those in typical PDR environments. 
 Interestingly, evidence for selective photodissociation, i.e., an influence of UV radiation, is seen even in normal Galactic disk clouds. 

\paragraph{{\rm CN/$^{13}$CN}}

CN is expected to be fractionated in the opposite sense compared to CO. CO is the most abundant C-bearing molecule and will regulate the availability of $^{13}$C atoms from the carbon reservoir if both molecules coexist.

Because of the weakness of $^{13}$CN absorption, only the lower limit of the CN/$^{13}$CN isotopologue ratio was obtained toward some of the saturated CO absorbers. 
The ratio is large, and $>$ 340 and $>$ 240 for Components 7 and 16, respectively. 
Note that these components' CO/$^{13}$CO isotopologue ratio is smaller than 50.

The observed large values toward dark clouds in the inner Galaxy have considerable deviations from previous studies. 
For instance, \cite{2005ApJ...634.1126M} measured the CN/$^{13}$CN isotopologue ratio in dense clouds to find the Galactic gradient CN/$^{13}$CN = (6.01$\pm$1.19)$R_{\rm G}$ + 12.28$\pm$9.33. 
\cite{2011ApJ...728...36R} obtained the CN/$^{13}$CN isotopologue ratio toward the 11 diffuse clouds ranges from 36 to 134, having a much larger deviation than the local interstellar C/$^{13}$C ratio \cite[70$\pm$7;][]{2007ApJ...667.1002S}. 
They compared these CN/$^{13}$CN isotopologue ratios with the CO/$^{13}$CO isotopologue ratio for the same lines of sight to find these ratios are generally fractionated in the opposite sense. 
The present results support the above scenario. 

\paragraph{{\rm HCN/H$^{13}$CN}}

Because of the weakness of H$^{13}$CN absorption, only the lower limit of the HCN/H$^{13}$CN isotopologue ratio was obtained toward some of the saturated CO absorbers. 
The ratio is quite large, $>$ 106 and $>$ 94 for Components 7 and 16, respectively. 
This trend, similar to CN, may be explained if HCN is fractionated in the opposite sense compared to CO.  

\paragraph{{\rm CS/C$^{34}$S}}

The CS abundance is estimated to be $\log X$(CS) = $-$8.41 and $-$9.36 for Components 7 and 16, respectively. 
This demonstrates that the CS abundance has a large scatter that reaches one order of magnitude in dark clouds. 

We observed a high CS/C$^{34}$S isotopologue ratio ($>$ 41) toward Component 7 ($R_{\rm{G}}$ = 7.44 kpc). 
This is significantly higher than that expected (19.06$\pm$2.43 at $R_{\rm{G}}$ = 7.44 kpc) from the Galactic gradient of sulfur isotope ratio $^{32}$S/$^{34}$S = (1.50$\pm$0.17) $R_{\rm{G}}$ + (7.90$\pm$1.17) obtained through observations of a large sample of massive star-forming regions in CS and its isotopologues \citep{2020ApJ...899..145Y}. 
The present result is also much larger than the CS/C$^{34}$S isotopologue ratio of diffuse and translucent clouds in the line-of-sight to Sgr B2 measured by \cite{2018A&A...610A..10C}, which ranges from 5 to 10 except for the $-92$ km~s$^{-1}$ cloud with 29$\pm$14. 

The upper limit for the CS/C$^{34}$S isotopologue ratio in this study is the largest known. 
In the mapping spectral line survey toward W3(OH), including CS and C$^{34}$S lines \citep{2017ApJ...848...17N}, we also see a hint of an increase of CS/C$^{34}$S toward the cloud edge. 
We find no reasonable explanation for this high CS/C$^{34}$S isotopologue ratio since the self-shielding of the main species will not be effective for such rare species. 

\section{Discussion} 
\label{sec:discussion}

\subsection{Substructure of molecular clouds deduced from {\rm CO} emission/absorption study}
\label{sec:substructure_of_molecular_clouds}

This study provides the following seven noteworthy results for general Galactic disk molecular clouds.
\begin{enumerate}
\item[(1)] Warm environment -- The nearly constant and high HCN/HNC isomer ratio ($\gtrsim$4) suggests a nearly uniform kinetic temperature at $\gtrsim$40 K.
\item[(2)] Narrow- and broad-line components -- Absorption lines consist of ``narrow-line components'' ($\sim$0.1 km~s$^{-1}$) and ``broad-line components'' ($\gtrsim$0.3 km~s$^{-1}$). 
The narrow line width is comparable with the thermal line width. 
\item[(3)] Compactness of narrow-line components -- By comparing the ALMA absorption profiles with the corresponding single-dish emission profiles, we found some of the narrow-line absorption components with high excitation temperatures are not prominent in the single-dish emission profile. 
This is understood if they occupy only a small fraction of the single-dish beam. 
\item[(4)] CO-poor gas -- The CO/HCO$^+$ abundance ratio has a large scatter of $\sim$3 orders, ranging from $\sim2 \times 10^4$ to $\sim30$. 
There is a trend that the low $N$(HCO$^+$) regions have low CO/HCO$^+$ abundance ratios. 
\item[(5)] Mismatch of emission and absorption profiles -- Some of the significant CO emission components show no counterparts in CO absorption.
This may be explained if the cloud has CO-emitting substructures unresolved by the single-dish beam and the line of sight toward the QSO misses CO-containing molecular gas. 
\item[(6)] Saturated CO absorption -- ``Saturated CO absorbers'' were detected. 
The QSO continuum toward these components was fully saturated over a wide ($\sim$4 km~s$^{-1}$) range and has $N$(H$_2$) $\gtrsim 1 \times 10^{21}$ cm$^{-2}$ (km~s$^{-1}$)$^{-1}$, which means that the line is optically thick throughout the velocity range.
\item[(7)] C$^+$-rich environments -- Even for velocity components with saturated CO absorption, large values of $N$(HCO)/$N$(H$^{13}$CO$^+$) $\sim$ 20 were observed. 
These values are comparable to typical PDRs in the Galaxy and suggest C$^+$-rich environments. 
The large $^{13}$CO/C$^{18}$O abundance ratio, possibly caused by selective photodissociation, also supports this argument. 
\end{enumerate}

The presence of compact narrow-line components, spatial heterogeneity in the single dish beam, and a C$^+$-rich environment require a clumpy structure in the cloud. 
The narrow-line components have near-thermal line widths and a small filling factor. 
Regarding the broad-line components, the close matching of the synthesized and emission line profiles that correspond to the broad-line components suggests a filling factor close to unity. 
%The presence of the broad-line CO-poor (2--3 orders of magnitude lower in $N$(CO)/$N$(HCO$^+$) than normal) components requires extended diffuse gas surrounds the clumpy substructures.
Similar narrow-line components are detected near the dense core, while broad-line components are seen outside the core in the nearby molecular clouds \citep{2010ApJ...712L.116P,2012A&A...546A.103S}. 

\subsection{Physical conditions and CO abundance of substructures} 
\label{sec:substructure-1}

Here we deduce the physical conditions and CO abundance of some of the substructures identified by absorption measurements with the aid of \texttt{ndradex} \citep{ndradex} developed based on \texttt{RADEX} \citep{2007A&A...468..627V}. 
Both \texttt{ndradex} and \texttt{RADEX} calculate $T_{\rm ex}$ and $\tau$ as functions of $T_{\rm k}$, $n({\rm H_2})$, and $X({\rm CO})/(dv/dr)$. 
Since $T_{\rm ex}$ and $\tau$ of the individual subcomponents are listed in Tables \ref{table:gaussfit12} and \ref{table:gaussfit13}, it is possible to estimate the physical conditions and the CO abundance of these subcomponents. 
Note that the saturated CO absorbers are not included in Table \ref{table:gaussfit12} since Gaussian fitting was not possible for these saturated absorption features. 
Thus only the subcomponents with $\tau_{1-0}$ $\lesssim$ 1 (Tables \ref{table:gaussfit12} and \ref{table:gaussfit13}) are analyzed. 
As illustrated in Figures \ref{fig:radex_norm} and \ref{fig:radex_poor}, the standard assumption of canonical $X$(CO) value and $dv/dr$ = 1 km s$^{-1}$ pc$^{-1}$ cannot reproduce such low $\tau_{1-0}$. 
This means that a significant reduction of $X({\rm CO})/(dv/dr)$ is needed to explain the observations, supporting our findings on the ubiquity of the CO-poor molecular gas. 
We thus do not assume the canonical $X$(CO) value in the present analysis. 

Instead, we assume $T_{\rm k}$ = 50 K suggested from the large HCN/HNC ratio toward these velocity components throughout this subsection. 
Very weak dependences of $T_{\rm ex}$ and $\tau$ on $T_{\rm k}$ in low-density regime justify this assumption. 
Then, the other two free parameters that determine $T_{\rm ex}$ and $\tau$ are $n({\rm H_2})$ and $X({\rm CO})/(dv/dr)$. 

\begin{figure}[htb!]
\epsscale{0.5}
\plotone{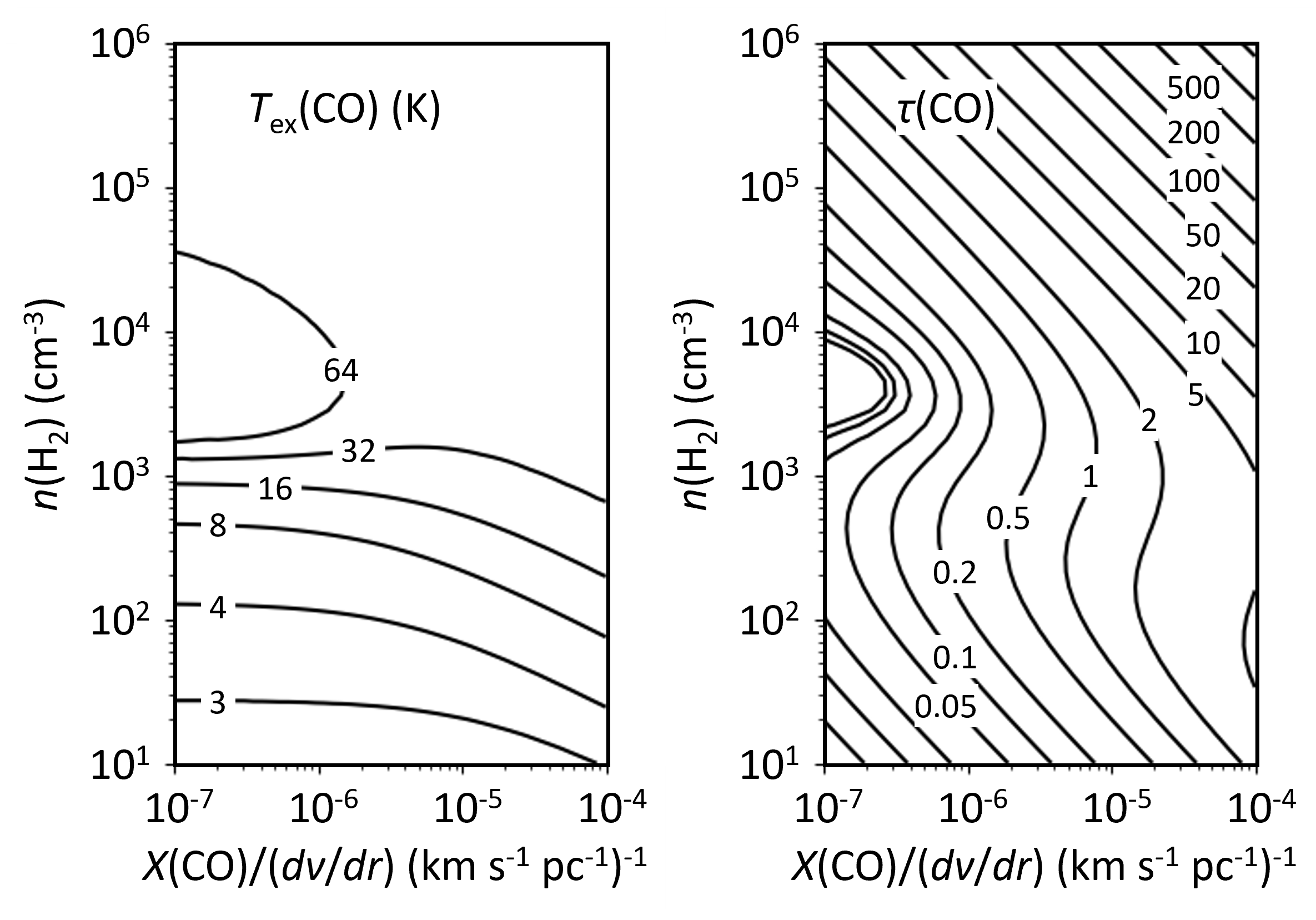}
\caption{The CO excitation temperature $T_{\rm ex}$(CO), and opacity $\tau$(CO) calculated as functions of $X({\rm CO})/(dv/dr)$ and $n({\rm H_2})$ using \texttt{ndradex} \citep{ndradex} developed based on \texttt{RADEX} \citep{2007A&A...468..627V} assuming the kinetic temperature of 50 K. 
The coutours for $\tau$(CO) are at $(1, 2, 5) \times 10^{n}$ starting from 0.005.}
\label{fig:radex_Xco_abundance}
\end{figure}

The dependence of $T_{\rm ex}$ and $\tau$ on $n({\rm H_2})$ and $X({\rm CO})/(dv/dr)$ are plotted in Figure \ref{fig:radex_Xco_abundance}. 
For most of the components identified in CO in Table \ref{table:gaussfit12}, $\tau_{1-0} \gtrsim 1$ and $\tau_{2-1}$/$\tau_{1-0}$ $\simeq$ 1 ($T_{\rm ex}$ $\simeq$ 5.5 K from Equation \ref{eq:tex_deduction4}). 
For this optically thin (low-$\tau$) subthermal (low-$T_{\rm ex}$) regime, $T_{\rm ex}$ is determined by $n$(H$_2$)$T_{\rm k}^{1/2}$ as shown in \texttt{RADEX} results (Figure \ref{fig:radex_poor}). The range of variation of $T_{\rm k}$ is also narrower than $n$(H$_2$). Therefore $T_{\rm k}$ does not contribute strongly to $T_{\rm ex}$ and both $n$(H$_2$) and $X({\rm CO})/(dv/dr)$ can be estimated with sufficient reliability as a cross point of the lines with constant $T_{\rm ex}$ and constant $\tau$.

We start the analysis from Subcomponent 5-1 as an example of the narrow-line components and deduce its physical conditions and CO abundance. 
For this subcomponent, $T_{\rm ex}$ $\simeq$ 8.2 K and $\tau_{1-0}$ $\simeq$ 0.65 requires $n$(H$_2$) $\simeq 3 \times 10^2$ cm$^{-3}$ and $X({\rm CO})/(dv/dr)$ $\simeq 4 \times 10^{-6}$ $(\rm km~s^{-1}~pc^{-1})^{-1}$. 
The values for the other subcomponents were deduced similarly and were listed in Tables  \ref{table:gaussfit12} and \ref{table:gaussfit13}. 
Since $dv/dr$ is an order of 1 $\rm km~s^{-1}~pc^{-1}$, $X({\rm CO})$ may be estimated from $X({\rm CO})/(dv/dr)$. 
The $X({\rm CO})$ values support the finding in Subsection \ref{sec:co_hco+} that the CO abundance significantly decreases in lower column density components, while it is comparable with the canonical value in saturated CO absorbers. 

The excitation temperature of the subcomponents traced by CO tends to be higher than that traced by $^{13}$CO as shown in Tables \ref{table:gaussfit12} and \ref{table:gaussfit13}. 
This is understood as a result of the photon-trapping effect that reduces the effectiveness of radiative cooling for optically thicker lines and keeps the excitation temperature higher. 
This is also visible in the left panel of Figure \ref{fig:radex_Xco_abundance}. 

\subsubsection{Properties of molecular gas at the diffuse end}
\label{sec:diffuse_end}

Through the sensitive absorption measurements with ALMA, we detected very faint CO absorption at the level of $\tau_{\rm max}$ $\simeq$ 0.1, or $N$(CO)/$dv$ $\simeq 2 \times 10^{14}$ cm$^{-2}$ (km~s$^{-1}$)$^{-1}$, as shown in Figure \ref{fig:gaussfit12b} and Table \ref{table:gaussfit12}. 
Most of them have very low $\tau_{2-1}$/$\tau_{1-0}$ $\simeq$ 0.3, suggesting very low $T_{\rm ex}$ $\simeq$ $T_{\rm CMB}$. 
If these components have $T_{\rm ex}$ comparable to the others ($T_{\rm ex} - T_{\rm CMB}$ $\sim$ 3 K) and $f$ $\simeq$ 1, $\tau_{1-0}$ $\simeq$ 0.1 predicts the measurable main beam antenna temperature at the level of $T_{\rm mb}$ $\gtrsim$ 0.3 K. 
The lack of corresponding emission at that level supports the very low $T_{\rm ex}$. 
The exception is the faint Subcomponent 18-1, which has high $\tau_{2-1}$/$\tau_{1-0}$, indicating $T_{\rm ex}$ = 6.52$\pm$2.72 K. 
The corresponding $T_{\rm mb}$ of the CO $J$=1--0 emission at the level of 1.5 K with $\tau_{1-0}$ $\simeq$ 0.1 suggests even higher $T_{\rm ex}$ exceeding 10 K. 
Interestingly, this subcomponent with very faint CO absorption has comparable column densities of HCO$^+$, CN-bearing molecules, and hydrocarbons with Components 8, 11, 19, and 20, as shown in Tables \ref{table:col_hco+_hnc} and \ref{table:col_c2h_c3h_c3h2}. 

We classify these faint CO absorption features into three based on the nature of the parent cloud and the location of the line of sight in the cloud. 

The faint components in 20--50 km~s$^{-1}$ (e.g., Subcomponents 10-1 and 10-2) are seen as surrounding material around scattered globules or small clouds in the longitude-velocity diagram (Figure \ref{fig:arm}) and the CO channel maps (Figure \ref{fig:fugin_12co}). 
This may be the molecular gas left unstructured or dispersing from structured molecular clouds. 

The faint components near 80, 84, and 101 km~s$^{-1}$ (i.e., Subcomponents 18-1, 19-1, and 21-1) are found near the edge of prominent clouds and thus are considered cloud edges. 

The faintest components near $-19$ and $0$ km~s$^{-1}$ (i.e., Subcomponents 2-1, 4-1, and 4-2) cannot be seen in either figure. 
The $\tau_{2-1}$/$\tau_{1-0}$ ratio suggests very low $T_{\rm ex}$ close to $T_{\rm CMB}$, accounting for the extremely low brightness of these subcomponents.
For the $-19$ km~s$^{-1}$ components, a very small structure is seen close to the line of sight, but it is not as bright as globules. 
Notably, these components already have narrow-line components, and these subcomponents are considered to be in the earliest stage of molecular cloud formation. 

\subsection{Sizes of substructures} 
\label{sec:substructure-2}

Although we did not image the subcomponents, the sizes of these subcomponents can be estimated as the filling factor of the individual subcomponent within the single-dish beam through a comparison of the actual main beam antenna temperature from single-dish emission-line observations with the ``synthesized main beam antenna temperature'' calculated from $f (1 - \exp(-\tau)) (T_{\rm ex} - T_{\rm CMB})$ through absorption measurements. 
Because $T_{\rm ex}$ and $\tau$ are known for the absorption components, the difference in the shape of the emission and absorption profiles is understood as the heterogeneity (difference in the filling factor) in the single-dish beam. 
As shown in Figure \ref{fig:gaussfit12}, the main beam antenna temperature corresponding to Subcomponents 5-2 and 10-2 are consistent with the one calculated from $f (1 - e^{-\tau}) (T_{\rm ex} - T_{\rm CMB})$ by assuming $f$ = 1. 
On the other hand, there are many subcomponents with high $T_{\rm ex}$ in absorption and low $T_{\rm mb}$ for their opacity in single-dish emission spectra, and they shall be compact. 
Examples of such compact components include Subcomponents 5-1, 11-1, 11-2, and 19-1. 
Since $f = (2R/\theta D)^2$ with the clump radius $R$, the single-dish FWHM beam size $\theta$, and the distance to the cloud $D$, these components which are weaker in the observed main beam antenna temperature than expected from $T_{\rm ex}$ and $\tau$ by more than one order of magnitude shall be smaller than one-third of the $15''$ beam of the Nobeyama 45 m telescope, or $<$0.1 pc at 5 kpc. 

\begin{figure}[htb!]
%\epsscale{0.5}
\plotone{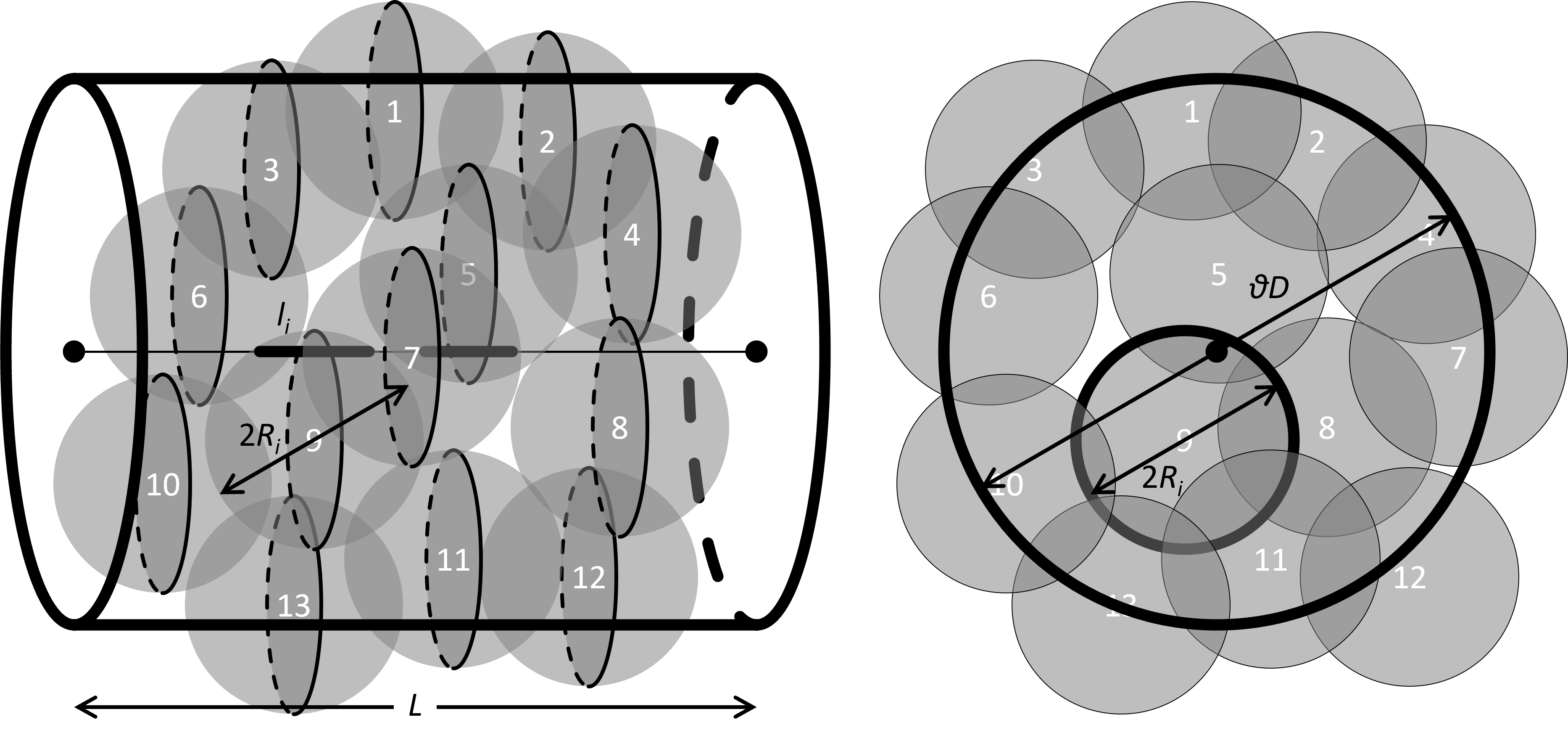}
\caption{A schematic view of a molecular cloud consisting of spherical clumps.
Left: A side view of a section of the molecular cloud in a single-dish telescope beam. 
Each clump $i$ has a radius $R_i$, and the line of sight toward the QSO penetrates each clump for the length $l_i$. 
The QSO's line of sight penetrates only a few clumps in the column. 
The length $L$ represents the depth of the molecular cloud. 
Inter-clump space may be filled with other media, which are not shown in this figure. 
Right: The same as the left figure, but projected on the sky. 
The linear scale of the diameter of the single-dish telescope beam is calculated as a product of the telescope beam width $\theta$ and the distance to the cloud $D$.}
\label{fig:pic_abs_em}
\end{figure}

\subsection{Other examples of small low-density substructures}
\label{sec:thermal_instability}

The current results indirectly show that there are small ($\lesssim$0.3 pc) substructures even in diffuse clouds and cloud envelopes. 
Such structures have been known through sensitive and high-resolution single-dish observations to resolve small-scale diffuse structures in nearby molecular clouds. 
For instance, \cite{1997ApJ...481..302S} obtained extensive CO $J$=1--0 images along the minor axis of the L1641 cloud in Orion with a $15''$ (0.03 pc) beam spaced by $34''$. 
Besides bright ridge-like emission in the middle of the cloud, CO emission was detected almost everywhere within the outermost boundary of the cloud. 
At least two distinct populations of molecular gas were found: well-defined ``clumps'' with small size ($\sim$ 0.3 pc), high main beam antenna temperature ($\sim$ 25 K), and relatively small line width ($\simeq$ 1.5 km~s$^{-1}$) and ``extended components'' with low main beam antenna temperature ($\sim$ 2.5 K) and broader line width ($\simeq$ 2.5 km~s$^{-1}$). 
In addition, \cite{2002ApJ...565.1050S} resolved clumps with sizes from $\sim$0.2 pc down to $\sim$0.02 pc through extensive strip-scan observations of high-latitude clouds MBM 32, MBM 54, and MBM 55 in the CO $J$=1--0 emission. 
\cite{2003ApJ...594..340S} further discovered several small ($\lesssim$ 0.1 pc) low-density structures in the envelope of Heiles Cloud 2 in Taurus. 
Toward a filamentary cloud complex facing the Sh 2--27 HII region, \cite{2012ApJ...754...95T} found that many small clumps are moving at supersonic velocities. 

Considering that the Jeans length $\lambda_{\rm J} = c_{\rm s} (G\rho)^{-1/2}$ corresponds to 0.6 ($T_{\rm k}$/10 K)$^{1/2}$ [$n$(H$_2$)/$10^3$ cm$^{-3}$]$^{-1/2}$ pc, most of these small substructures with lower density and higher kinetic temperature are unlikely to be formed by gravitational contraction. 
We will examine their formation mechanisms in our forthcoming papers. 
%Theoretically, two-dimensional fluid simulations by \cite{2002Ap&SS.281...67I} show that external perturbations and shock-wave compression form a thermally unstable, clumpy, cold, dense gas cloud and surrounding shock-heated warm atomic gas in a multiple-phase turbulent flow state. 
%Since the turbulence of the warm gas at about the speed of sound is supersonic for the low-temperature cloud, the low-temperature gas cloud moves through the warm gas at a supersonic velocity. 
%This gas motion is supersonic turbulence (\cite{1981MNRAS.194..809L}) in gas cloud tracers such as CO (\cite{2002Ap&SS.281...67I}). 
%These small low-temperature clouds grow by accumulation and coalescence of atomic gases.

\subsection{Molecular cloud model with small substructures and CO-poor gas}

\begin{figure}[htb!]
\epsscale{0.5}
\plotone{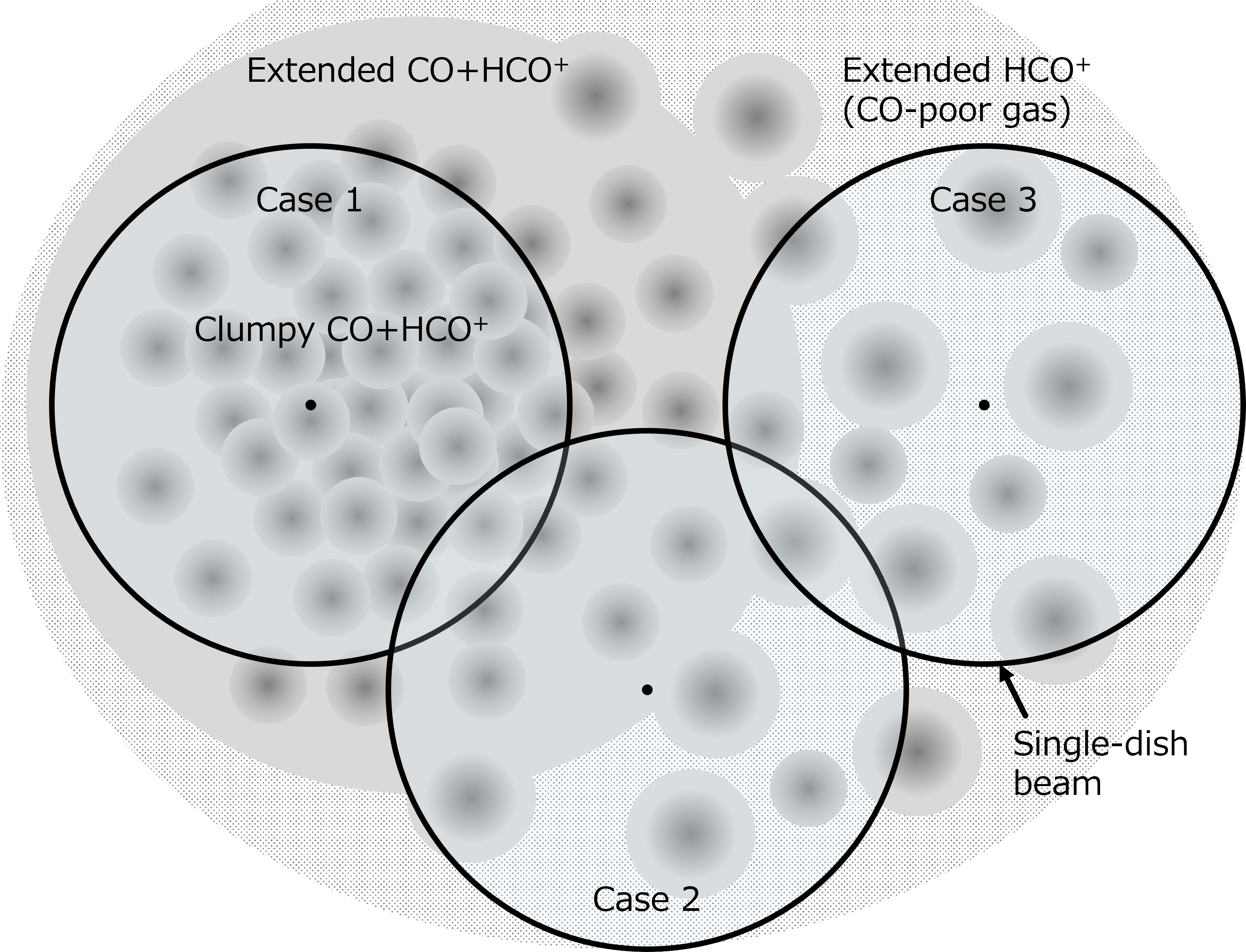}
\caption{Proposed clumpy molecular gas with CO-poor components. 
\label{fig:clump}}
\end{figure}

Based on the above, we propose a model of molecular clouds with the structure shown in Figure \ref{fig:clump}.
Since molecular clouds are a non-uniform medium due to their internal structures, the detected velocity components and chemical composition may differ depending on how the line of sight penetrates the cloud. 
The following three cases would exist.

Case 1: Both CO and HCO$^+$ absorption lines have narrow- and broad-line components, and the CO emission lines have bright broad-line components. 
The narrow-line components observed in the absorption lines often have higher densities than average. 
They may be compact because their beam-filling factor is small. 
The narrow-line components are illustrated as clumps in Figure \ref{fig:clump}. 
The detection of multiple narrow-line components in the saturated CO absorbers suggests that many clumps are concentrated in the center of the cloud. 
On the other hand, some of the broad-line components observed in the absorption lines successfully reproduce the profile of the emission lines assuming a filling factor of unity. 
It means that some of the broad-line components are extended. \\
Case 2: Broad CO and HCO$^+$ absorption lines and weak and broad CO emission are detected. 
The absorption and emission lines trace the CO-emitting components within the 45 m telescope beam. 
The weak intensity of the emission lines suggests that the beam-filling factor is small.\\
Case 3: A broad-line components are observed in the HCO$^+$ absorption lines without prominent CO absorption lines. 
The line of sight of the absorption may have missed the clumps. 
This implies that the clumps are sparsely distributed. 
The CO abundance in the extended HCO$^+$-absorbing gas might be extremely low.

This model could explain sparse cloud structure or chemical composition heterogeneities in the beam of the NRO 45 m telescope.
If the clumps overlap on the line of sight in the QSO direction, then narrow-line components were detected in absorption separately; the component in Figure \ref{fig:co_abs}, which extends from 20.0--34.5 km~s$^{-1}$, is considered such a region.

To examine where Cases 1, 2, or 3 are situated and which type of molecular clouds they correspond to, we used the FUGIN galactic plane survey shown in Figures \ref{fig:fugin_12co} and \ref{fig:fugin_13co}. 
Table \ref{table:velcomp} summarizes the cloud types observed in absorption and emission lines and the position of the QSO line of sight in the clouds. 
Next, we examined which types of individual molecular clouds were classified.
Case 1 corresponds to Components 5, 6, 7, 15, 16, and 20, and the line of sight is considered to penetrate centers of dark clouds; 
Case 2 corresponds to Components 2, 4, 8, 10, 11, 12, 18, and 19; 
Case 3 corresponds to Components 1, 9, 13, 14, 17, 21, 22, and 23. 
These classifications are also reflected in Table \ref{table:velcomp}. 

\subsection{Search for redshifted absorption lines}
\label{sec:redshifted_lines}

Because QSO J1851+0035 is heavily obscured by the Galactic plane, its optical redshift is unknown. 
We thus searched for the redshifted absorption lines associated with the QSO host galaxy and intergalactic components as a blind search by using the consolidated spectra in Figure \ref{fig:band35678} and the line identification in Tables \ref{table:mol_data-1}--\ref{table:interference}. 
Only features near 213.57 GHz, 231.61 GHz, 242.28 GHz, 331.06 GHz, and 478.83 GHz remain unidentified. 
We calculated their frequency ratios to identify the possible pair of redshifted lines since the ratio remains unchanged after the redshift. 
After comparing them with those of strong absorption lines of CO, HCO$^+$, CN, HCN, and C, we found that none of the possible pairs matched them. 

The lack of notable absorption features at $\int \tau dv$ $\sim$ 1 km~s$^{-1}$ level in the continuous 84--116 GHz frequency range demonstrates that there is no significant CO absorber in front of this QSO at $0.001 < z < 0.37$ for CO $J$=1--0 and at $0.99 < z < 1.74$ for CO $J$=2--1 with $N$(CO) $\gtrsim 1 \times 10^{14}$ cm$^{-2}$. 

\section{Conclusions} 
\label{sec:conclusions}

Using ALMA calibration source data and archival data from the NRO 45 m telescope emission line observations, we have reported the results of physical state and chemical composition analyses for about 20 velocity components in the QSO J1851+0035 ($l$=$33.498^{\circ}$, $b$=$+0.194^{\circ}$) direction behind the Galactic plane. Our results are summarized as follows.
\begin{enumerate}
\item A total of 17 species (CO, $^{13}$CO, C$^{18}$O, HCO$^+$, H$^{13}$CO$^+$, HCO, H$_2$CO, HCN, HNC, CN, C$_2$H, $c$-C$_3$H, $c$-C$_3$H$_2$, CS, SO, SiO, and C) were detected in absorption by Galactic molecular gas toward QSO J1851+0035. 
Upper limits were set for the column densities of CO$^+$, HC$^{18}$O$^+$, HOC$^+$, CH$_3$OH, $^{13}$CN, CN$^-$, H$^{13}$CN, CH$_3$CN, ND, N$_2$H$^+$, $l$-C$_3$H$^+$, SH$^+$, C$^{34}$S, HCS$^+$, H$_2$CS, C$_2$S, SO$_2$, and CF$^+$. 
\item More than 20 velocity components were identified with absorption and emission profiles of CO, $^{13}$CO, and HCO$^+$. 
Some were further resolved into subcomponents by the multiple-Gaussian fitting of the $J$=1--0 and $J$=2--1 line profiles of CO (and those of $^{13}$CO for components with saturated CO absorption) with fixed centers and line widths. 
As a result, narrow-line components ($\simeq$0.1 km~s$^{-1}$) and broad-line components ($\gtrsim$0.3 km~s$^{-1}$) were identified. The excitation temperature of each identified component is approximately 5 K. Components with higher excitation temperatures are those with narrower line widths, which implies that the environment is dense enough for collisional excitation to predominate.
\item Based on the chemical compositions and the results of previous research, we found a very tight and linear correlation between the column densities within the CN-bearing species (HCN, HNC, CN) and the hydrocarbons (C$_2$H, $c$-C$_3$H, $c$-C$_3$H$_2$). 
Comparing the column densities of these groups with that of HCO$^+$, we found that the CN-bearing species tend to have reduced abundances at $N$(HCO$^+$) $\lesssim$ 10$^{12}$ cm$^{-2}$. 
\item A large dispersion (up to three orders of magnitude) was observed in the CO and HCO$^+$ column density ratios. 
This dispersion is attributed to the CO-poor gas, which was predominantly found in low-column-density regions. 
Such component is often observed as the velocity components with significant HCO$^+$ absorption lines but without significant CO absorption lines in ALMA data, while significant CO emission lines are visible in the Nobeyama 45 m data. 
The \texttt{RADEX} analysis of the absorption data also supports the ubiquity of the CO-poor molecular gas. 
\item The results of the excitation analysis for CO and $^{13}$CO indicate that the excitation temperature of each cloud is below 7 K, meaning that the observed gases are low density. 
The emission and absorption studies with the Nobeyama 45 m telescope indicate that the excitation temperature of dense gas tracers (HCO$^+$, HCN) toward this line of sight is very low ($\sim$2.9 K). 
\item The $N$(HCO)/$N$(H$^{13}$CO$^+$) ratio, which is thought to increase with UV light, exceeds 10 in the region where $N$(HCO$^+$) is relatively high ($\sim$$10^{12}$ cm$^{-2}$) as in the photodissociation region, suggesting that these molecular gases are in an environment where H$_2$ and C$^+$ are mixed.
\item One of the velocity components showed absorption of shock wave tracers (SiO and SO) while no YSO-like excitation source was found. 
The CO excitation temperature was just 6.75$\pm$0.11 K, slightly higher than the other components. 
This suggests that the shock wave's impact was not so significant or that the gas	cooled rapidly before re-agglomeration into dust. 
\item Discrete narrow-line component, spatial heterogeneity in the beam, and a C$^+$-rich environment require a clumpy structure in the cloud. 
The presence of broader-line components implies extended gas. 
\end{enumerate}

\begin{acknowledgments}
This paper uses the ALMA data tabulated in Appendix \ref{sec:archive}. 
ALMA is a partnership of ESO (representing its member states), NSF (USA), and NINS (Japan), together with NRC (Canada), MOST and ASIAA (Taiwan), and KASI (Republic of Korea), in cooperation with the Republic of Chile. The Joint ALMA Observatory is operated by ESO, AUI/NRAO and NAOJ. The National Radio Astronomy Observatory is a facility of the National Science
 Foundation operated under cooperative agreement by Associated Universities, Inc.
The data are partially obtained from Nobeyama 45m and ASTE Science Data Archive (https://nobeyama-archive.nao.ac.jp/). 
The Nobeyama 45-m radio telescope is operated by the Nobeyama Radio Observatory, a branch of the National Astronomical Observatory of Japan. 
Data analysis was carried out on the Multi-wavelength Data Analysis System operated by the Astronomy Data Center (ADC), National Astronomical Observatory of Japan.
The ATLASGAL project is a collaboration between the Max-Planck-Gesellschaft, the European Southern Observatory (ESO), and the Universidad de Chile. It includes projects E-181.C-0885, E-078.F-9040(A), M-079.C-9501(A), M-081.C-9501(A) plus Chilean data.
JK acknowledges support from NSF through grant AST-2006600.
KK acknowledges the support by JSPS KAKENHI Grant Numbers JP17H06130 and JP23K20035.
K.N. and Y.Y. were supported by the International Graduate Program for Excellence in Earth-Space Science (IGPEES), The University of Tokyo.

We thank Yuri Aikawa, Shu-ichiro Inutsuka, Yuto Komichi, Daisuke Taniguchi, and Satoshi Yamamoto for fruitful discussions, and Fumiya Maeda for the support of the NRO 45 m telescope observations. 
\end{acknowledgments}


\vspace{5mm}
\facilities{ALMA, NRO 45 m telescope}


\software{CASA}



\appendix

\section{ALMA archive data used for this study} 
\label{sec:archive}

The list of ALMA archive data used for this study is shown in Tables \ref{table:archive-1a}--\ref{table:archive-3}. 
In these tables, $f_{\rm min}$, $f_{\rm max}$, $\Delta f$, $\theta$, and $\Delta S$ mean the minimum and maximum frequency, the frequency resolution, the angular resolution, and the line sensitivity of the data, respectively. 

\begin{table}[htb!]
 \caption{List of ALMA Band 3 data used for this study}
 \label{table:archive-1a}
 \centering
  \begin{tabular}{ccrrrccc}
   \hline \hline
  Project Code & ALMA & \multicolumn{1}{c}{$f_{\rm min}$} & \multicolumn{1}{c}{$f_{\rm max}$} & \multicolumn{1}{c}{$\Delta f$} & $\theta$ & $\Delta S$ @10 km~s$^{-1}$ & Obs. date\\
               & Band & \multicolumn{1}{c}{(GHz)} & \multicolumn{1}{c}{(GHz)} & \multicolumn{1}{c}{(kHz)} & ($''$)          & (mJy beam$^{-1}$) & (yyyy-mm-dd)\\
   \hline 
2016.1.00010.S & 3 & 109.73 & 109.79 & 30.52 & 0.315 & 0.70 & 2016-10-11 \\
 &  & 110.15 & 110.21 & 30.52 & 0.315 & 0.70 & 2016-10-11 \\
 &  & 115.22 & 115.28 & 30.52 & 0.351 & 1.30 & 2016-10-12 \\
 &  & 115.19 & 115.31 & 61.04 & 0.351 & 1.30 & 2016-10-12 \\
2016.1.01363.S & 3 & 86.30 & 86.36 & 61.04 & 2.561 & 1.72 & 2017-01-21 \\
2016.1.01363.S & 3 & 86.30 & 86.36 & 61.04 & 2.561 & 1.72 & 2017-01-21 \\
 &  & 86.71 & 86.77 & 61.04 & 2.561 & 1.72 & 2017-01-21 \\
 &  & 86.81 & 86.86 & 61.04 & 2.561 & 1.72 & 2017-01-21 \\
 &  & 96.70 & 96.76 & 61.04 & 2.561 & 1.60 & 2017-01-21 \\
2017.1.00501.S & 3 & 84.05 & 84.99 & 564.45 & 0.780 & 1.83 & 2018-01-18 \\
 &  & 84.91 & 85.85 & 564.45 & 0.780 & 1.82 & 2018-01-18 \\
 &  & 85.82 & 86.75 & 564.45 & 0.780 & 1.87 & 2018-01-18 \\
 &  & 86.72 & 87.66 & 564.45 & 0.780 & 1.86 & 2018-01-18 \\
 &  & 87.63 & 88.57 & 564.45 & 0.699 & 2.01 & 2018-01-27 \\
 &  & 88.53 & 89.47 & 564.45 & 0.699 & 2.02 & 2018-01-27 \\
 &  & 89.44 & 90.38 & 564.45 & 0.699 & 2.01 & 2018-01-27 \\
 &  & 90.35 & 91.28 & 564.45 & 0.699 & 2.00 & 2018-01-27 \\
 &  & 91.25 & 92.19 & 564.45 & 1.149 & 1.80 & 2018-03-17 \\
 &  & 92.16 & 93.10 & 564.45 & 1.149 & 1.82 & 2018-03-17 \\
 &  & 93.06 & 94.00 & 564.45 & 1.149 & 1.81 & 2018-03-17 \\
 &  & 93.97 & 94.91 & 564.45 & 1.149 & 1.81 & 2018-03-17 \\
 &  & 94.88 & 95.81 & 564.45 & 1.074 & 1.62 & 2018-03-12 \\
 &  & 95.78 & 96.72 & 564.45 & 1.074 & 1.63 & 2018-03-12 \\
 &  & 96.69 & 97.62 & 564.45 & 1.074 & 1.63 & 2018-03-12 \\
 &  & 97.59 & 98.53 & 564.45 & 1.074 & 1.63 & 2018-03-12 \\
 &  & 98.50 & 99.44 & 564.45 & 1.085 & 2.02 & 2018-03-14 \\
 &  & 99.41 & 100.34 & 564.45 & 1.085 & 1.97 & 2018-03-14 \\
 &  & 100.31 & 101.25 & 564.45 & 1.085 & 1.96 & 2018-03-14 \\
 &  & 101.22 & 102.15 & 564.45 & 1.085 & 1.95 & 2018-03-14 \\
 &  & 102.12 & 103.06 & 564.45 & 1.016 & 1.98 & 2018-03-14 \\
 &  & 103.03 & 103.97 & 564.45 & 1.016 & 1.93 & 2018-03-14 \\
 &  & 103.94 & 104.87 & 564.45 & 1.016 & 1.92 & 2018-03-14 \\
 &  & 104.84 & 105.78 & 564.45 & 1.016 & 1.92 & 2018-03-14 \\
 &  & 105.75 & 106.68 & 564.45 & 1.040 & 2.22 & 2018-03-14 \\
 &  & 106.65 & 107.59 & 564.45 & 1.040 & 2.21 & 2018-03-14 \\
 &  & 107.56 & 108.50 & 564.45 & 1.040 & 2.22 & 2018-03-14 \\
 &  & 108.47 & 109.40 & 564.45 & 1.040 & 2.21 & 2018-03-14 \\
 &  & 109.37 & 110.31 & 564.45 & 0.966 & 2.28 & 2018-03-14 \\
 &  & 110.28 & 111.21 & 564.45 & 0.966 & 2.17 & 2018-03-14 \\
 &  & 111.18 & 112.12 & 564.45 & 0.966 & 2.16 & 2018-03-14 \\
 &  & 112.09 & 113.03 & 564.45 & 0.966 & 2.16 & 2018-03-14 \\
 &  & 112.90 & 113.83 & 564.45 & 0.909 & 2.85 & 2018-03-21 \\
 &  & 113.70 & 114.64 & 564.45 & 0.909 & 2.84 & 2018-03-21 \\
 &  & 114.47 & 115.41 & 564.45 & 0.909 & 3.04 & 2018-03-21 \\
 &  & 114.97 & 115.91 & 564.45 & 0.909 & 3.03 & 2018-03-21 \\
  \hline
  \end{tabular}
\end{table}

\begin{table}[htb!]
 \caption{List of ALMA Band 3 and 5 data used for this study (continued)}
 \label{table:archive-1b}
 \centering
  \begin{tabular}{ccrrrccc}
   \hline \hline
  Project Code & ALMA & \multicolumn{1}{c}{$f_{\rm min}$} & \multicolumn{1}{c}{$f_{\rm max}$} & \multicolumn{1}{c}{$\Delta f$} & $\theta$ & $\Delta S$ @10 km~s$^{-1}$ & Obs. date\\
               & Band & \multicolumn{1}{c}{(GHz)} & \multicolumn{1}{c}{(GHz)} & \multicolumn{1}{c}{(kHz)} & ($''$)          & (mJy beam$^{-1}$) & (yyyy-mm-dd)\\
   \hline 
2018.1.00010.S & 3 & 112.89 & 113.36 & 282.23 & 0.759 & 2.18 & 2018-11-27 \\
2018.1.00101.S & 3 & 88.59 & 88.65 & 61.04 & 3.099 & 2.36 & 2019-03-10 \\
2018.1.00424.S & 3 & 86.77 & 86.89 & 282.23 & 0.669 & 2.82 & 2019-11-09 \\
 &  & 87.24 & 87.36 & 282.23 & 0.669 & 2.87 & 2018-10-28 \\
 &  & 93.10 & 93.21 & 282.23 & 0.658 & 2.83 & 2018-10-27 \\
 &  & 97.90 & 98.02 & 282.23 & 0.669 & 2.58 & 2018-10-28 \\
 &  & 102.96 & 103.08 & 282.23 & 0.658 & 2.55 & 2018-10-27 \\
 &  & 103.95 & 104.07 & 282.23 & 0.658 & 2.54 & 2018-10-27 \\
2018.1.00443.S & 3 & 89.10 & 89.22 & 70.56 & 1.982 & 2.03 & 2018-12-20 \\
2018.1.00850.S & 3 & 90.67 & 90.61 & 61.04 & 2.771 & 2.21 & 2019-01-16 \\
2018.1.01242.S & 3 & 113.47 & 113.53 & 70.56 & 0.588 & 2.42 & 2018-10-30 \\
2019.1.00685.S & 3 & 86.72 & 86.78 & 122.07 & 1.487 & 3.15 & 2019-10-31 \\
2017.1.01248.S & 5 & 191.28 & 191.52 & 564.45 & 0.923 & 1.44 & 2018-05-13 \\
 &  & 191.55 & 191.79 & 564.45 & 0.923 & 1.44 & 2018-05-13 \\
 &  & 191.78 & 192.01 & 564.45 & 0.923 & 1.44 & 2018-05-13 \\
 &  & 192.03 & 192.27 & 564.45 & 0.923 & 1.44 & 2018-05-13 \\
 &  & 192.29 & 192.52 & 564.45 & 0.923 & 1.44 & 2018-05-13 \\
 &  & 192.44 & 192.68 & 564.45 & 0.923 & 1.44 & 2018-05-13 \\
 &  & 193.23 & 193.47 & 564.45 & 0.923 & 1.44 & 2018-05-13 \\
  \hline
  \end{tabular}
\end{table}

\begin{table}[htb!]
 \caption{List of ALMA Band 6 data used for this study}
 \label{table:archive-2}
 \centering
  \begin{tabular}{ccrrrccc}
   \hline \hline
  Project Code & ALMA & \multicolumn{1}{c}{$f_{\rm min}$} & \multicolumn{1}{c}{$f_{\rm max}$} & \multicolumn{1}{c}{$\Delta f$} & $\theta$ & $\Delta S$ @10 km~s$^{-1}$ & Obs. date\\
               & Band & \multicolumn{1}{c}{(GHz)} & \multicolumn{1}{c}{(GHz)} & \multicolumn{1}{c}{(kHz)} & ($''$)          & (mJy beam$^{-1}$) & (yyyy-mm-dd)\\
   \hline 
2015.1.00072.S & 6 & 237.75 & 238.69 & 564.45 & 0.205 & 1.97 & 2016-08-04 \\
 & & 256.11 & 256.34 & 484.62 & 0.205 & 1.95 & 2016-08-04 \\
 & & 256.83 & 257.30 & 484.62 & 0.205 & 1.95 & 2016-08-04 \\
2015.1.00615.S & 6 & 219.47 & 219.71 & 564.45 & 0.157 & 1.95 & 2016-09-05 \\
 & & 231.74 & 231.97 & 564.45 & 0.157 & 1.91 & 2016-09-05 \\
2015.1.01454.S & 6 & 220.22 & 220.28 & 141.11 & 0.175 & 2.06 & 2016-09-11 \\
 & & 231.59 & 232.06 & 564.45 & 0.175 & 2.02 & 2016-09-11 \\
2015.1.01571.S & 6 & 212.87 & 213.81 & 969.24 & 0.254 & 1.26 & 2016-08-15 \\
2016.1.00005.S & 6 & 226.09 & 227.96 & 1938.48 & 1.526 & 2.76 & 2017-03-28 \\
 & & 229.09 & 230.96 & 1938.48 & 1.526 & 2.59 & 2017-03-28 \\
 & & 242.07 & 243.94 & 1938.48 & 1.526 & 2.54 & 2017-03-28 \\
 & & 244.08 & 245.95 & 1938.48 & 1.526 & 2.53 & 2017-03-28 \\
2016.1.00010.S & 6 & 219.46 & 219.58 & 61.04 & 0.138 & 0.71 & 2016-09-30 \\
 &  & 220.30 & 220.42 & 61.04 & 0.138 & 0.70 & 2016-09-30 \\
 &  & 230.44 & 230.55 & 61.04 & 0.138 & 0.72 & 2016-09-30 \\
2016.1.01284.S & 6 & 232.56 & 234.43 & 976.56 & 0.536 & 2.47 & 2016-11-22 \\
 & & 234.56 & 236.43 & 976.56 & 0.536 & 2.46 & 2016-11-22 \\
2016.1.01346.S & 6 & 215.93 & 217.80 & 488.28 & 1.194 & 2.91 & 2018-05-14 \\
 & & 217.80 & 219.67 & 488.28 & 1.194 & 2.90 & 2018-05-14 \\
2017.1.00101.S & 6 & 242.51 & 244.38 & 1953.13 & 0.254 & 1.54 & 2018-09-24 \\
 & & 244.51 & 246.38 & 1953.13 & 0.254 & 1.52 & 2018-09-24 \\
 & & 256.51 & 258.38 & 1953.13 & 0.254 & 1.52 & 2018-09-24 \\
2017.1.01248.S & 6 & 261.60 & 261.84 & 564.45 & 1.040 & 2.37 & 2018-05-15 \\
 & & 263.78 & 264.02 & 564.45 & 1.040 & 2.37 & 2018-05-15 \\
 & & 263.97 & 264.20 & 564.45 & 1.040 & 2.37 & 2018-05-15 \\
 & & 264.25 & 264.49 & 564.45 & 1.040 & 2.37 & 2018-05-15 \\
 & & 264.55 & 264.78 & 564.45 & 1.040 & 2.37 & 2018-05-15 \\
2018.1.00632.S & 6 & 225.05 & 226.93 & 2257.81 & 0.043 & 0.89 & 2019-07-18 \\
 & & 241.05 & 242.93 & 2257.81 & 0.043 & 0.88 & 2019-07-18 \\
2018.1.00659.L & 6 & 213.83 & 215.70 & 1128.91 & 0.293 & 3.51 & 2018-11-18 \\
 & & 220.23 & 222.10 & 1128.91 & 0.287 & 3.55 & 2018-11-18 \\
 & & 223.62 & 225.50 & 1128.91 & 0.287 & 3.52 & 2018-11-18 \\
 & & 227.22 & 229.10 & 1128.91 & 0.293 & 3.55 & 2018-11-18 \\
 & & 229.58 & 231.45 & 1128.91 & 0.293 & 3.50 & 2018-11-18 \\
 & & 235.43 & 237.30 & 1128.91 & 0.287 & 3.41 & 2018-11-18 \\
 & & 245.34 & 247.21 & 1128.91 & 0.239 & 3.21 & 2018-11-18 \\
 & & 251.57 & 253.45 & 1128.91 & 0.268 & 3.75 & 2018-11-18 \\
 & & 253.94 & 255.81 & 1128.91 & 0.268 & 3.74 & 2018-11-18 \\
 & & 262.09 & 263.03 & 564.45 & 0.239 & 3.00 & 2018-11-18 \\
 & & 265.52 & 267.39 & 1128.91 & 0.268 & 3.61 & 2018-11-18 \\
 & & 267.77 & 269.64 & 1128.91 & 0.268 & 3.68 & 2018-11-18 \\
2019.1.00246.S & 6 & 237.95 & 238.89 & 488.28 & 0.274 & 0.84 & 2021-03-28 \\
 & & 238.88 & 239.82 & 488.28 & 0.274 & 0.83 & 2021-03-28 \\
 & & 239.81 & 240.75 & 488.28 & 0.274 & 0.82 & 2021-03-28 \\
 & & 240.74 & 241.68 & 488.28 & 0.274 & 0.82 & 2021-03-28 \\
 & & 257.95 & 258.88 & 488.28 & 0.246 & 0.96 & 2021-03-31 \\
 & & 258.88 & 259.81 & 488.28 & 0.246 & 0.96 & 2021-03-31 \\
 & & 259.81 & 260.74 & 488.28 & 0.246 & 0.96 & 2021-03-31 \\
 & & 260.74 & 261.67 & 488.28 & 0.246 & 0.96 & 2021-03-31 \\
  \hline
  \end{tabular}
\end{table}

\begin{table}[htb!]
 \caption{List of ALMA Band 7 and 8 data used for this study}
 \label{table:archive-3}
 \centering
  \begin{tabular}{ccrrrccc}
   \hline \hline
  Project Code & ALMA & \multicolumn{1}{c}{$f_{\rm min}$} & \multicolumn{1}{c}{$f_{\rm max}$} & \multicolumn{1}{c}{$\Delta f$} & $\theta$ & $\Delta S$ @10 km~s$^{-1}$ & Obs. date\\
               & Band & \multicolumn{1}{c}{(GHz)} & \multicolumn{1}{c}{(GHz)} & \multicolumn{1}{c}{(kHz)} & ($''$)          & (mJy beam$^{-1}$) & (yyyy-mm-dd)\\
   \hline 
2013.1.01035.S & 7 & 278.79 & 278.90 & 122.07 & 0.816 & 2.81 & 2015-04-28 \\
 & & 279.41 & 279.53 & 122.07 & 0.816 & 2.81 & 2015-04-28 \\
 & & 288.07 & 288.13 & 141.11 & 0.816 & 2.61 & 2015-04-28 \\
 & & 289.14 & 289.20 & 141.11 & 0.816 & 2.61 & 2015-04-28 \\
 & & 289.54 & 289.66 & 244.14 & 0.816 & 2.61 & 2015-04-28 \\
 & & 290.00 & 290.24 & 282.23 & 0.816 & 2.60 & 2015-04-28 \\
 & & 290.46 & 290.70 & 282.23 & 0.816 & 2.60 & 2015-04-28 \\
2015.1.00054.S & 7 & 328.99 & 330.86 & 1128.91 & 0.122 & 5.02 & 2016-08-30 \\
 &  & 330.80 & 331.73 & 3906.25 & 0.122 & 5.02 & 2016-08-30 \\
 &  & 339.92 & 341.79 & 1128.91 & 0.122 & 4.98 & 2016-08-30 \\
 &  & 341.75 & 343.62 & 1128.91 & 0.122 & 5.01 & 2016-08-30 \\
2015.1.00283.S & 7 & 334.04 & 335.91 & 1938.48 & 0.126 & 2.08 & 2016-08-17 \\
 & & 336.42 & 336.66 & 282.23 & 0.126 & 2.08 & 2016-08-17 \\
 & & 336.93 & 337.17 & 282.23 & 0.126 & 2.08 & 2016-08-17 \\
 & & 345.55 & 346.02 & 282.23 & 0.126 & 2.04 & 2016-08-17 \\
 & & 346.87 & 347.10 & 282.23 & 0.126 & 2.06 & 2016-08-17 \\
 & & 347.20 & 347.44 & 282.23 & 0.126 & 2.06 & 2016-08-17 \\
2015.1.00600.S & 7 & 334.98 & 336.85 & 976.56 & 0.180 & 1.62 & 2016-07-26 \\
 & & 336.86 & 338.74 & 976.56 & 0.180 & 1.65 & 2016-07-26 \\
 & & 346.90 & 348.78 & 976.56 & 0.180 & 1.57 & 2016-07-26 \\
 & & 348.86 & 350.74 & 976.56 & 0.180 & 1.57 & 2016-07-26 \\
2015.1.01535.S & 7 & 352.63 & 354.50 & 1938.48 & 0.178 & 9.05 & 2016-07-27 \\
 & & 352.63 & 354.50 & 1938.48 & 0.178 & 9.05 & 2016-07-27 \\
 & & 365.65 & 367.52 & 1938.48 & 0.178 & 5.73 & 2016-07-27 \\
2016.1.00071.S & 7 & 333.51 & 335.38 & 1953.13 & 0.811 & 1.79 & 2017-04-04 \\
 & & 335.47 & 337.34 & 1953.13 & 0.811 & 1.84 & 2017-04-04 \\
 & & 345.51 & 347.38 & 1953.13 & 0.811 & 1.83 & 2017-04-04 \\
 & & 347.51 & 349.38 & 1953.13 & 0.811 & 1.90 & 2017-04-04 \\
2016.1.00988.S & 7 & 358.87 & 360.74 & 1938.48 & 0.989 & 3.28 & 2017-04-04 \\
 & & 360.01 & 360.13 & 70.56 & 0.989 & 3.28 & 2017-04-04 \\
 & & 370.86 & 372.73 & 1938.48 & 0.989 & 2.36 & 2017-04-04 \\
 & & 372.29 & 372.35 & 70.56 & 0.989 & 2.36 & 2017-04-04 \\
 & & 372.54 & 372.60 & 70.56 & 0.989 & 2.36 & 2017-04-04 \\
2016.1.01548.S & 7 & 344.86 & 346.73 & 976.56 & 1.067 & 2.31 & 2017-03-24 \\
 & & 346.40 & 348.27 & 976.56 & 1.067 & 2.44 & 2017-03-24 \\
2017.1.00377.S & 7 & 336.31 & 337.25 & 564.45 & 0.470 & 2.37 & 2018-04-04 \\
 & & 337.24 & 338.18 & 564.45 & 0.470 & 2.30 & 2018-04-04 \\
 & & 338.17 & 339.11 & 564.45 & 0.470 & 2.29 & 2018-04-04 \\
 & & 339.16 & 340.10 & 564.45 & 0.470 & 2.29 & 2018-04-04 \\
2018.1.00024.S & 7 & 320.90 & 321.37 & 282.23 & 0.074 & 2.18 & 2019-08-20 \\
 & & 321.50 & 323.38 & 1128.91 & 0.074 & 2.55 & 2019-08-20 \\
 & & 333.00 & 334.88 & 1128.91 & 0.074 & 2.53 & 2019-08-20 \\
 & & 334.90 & 336.77 & 1128.91 & 0.074 & 2.52 & 2019-08-20 \\
 2018.1.00190.S & 8 & 476.98 & 477.92 & 1952.15 & 0.221 & 8.90 & 2019-04-27 \\
 & & 477.84 & 479.71 & 1938.48 & 0.221 & 8.75 & 2019-04-27 \\
 & & 489.84 & 491.71 & 1938.48 & 0.221 & 7.80 & 2019-04-27 \\
 & & 491.64 & 492.57 & 1952.15 & 0.221 & 7.80 & 2019-04-27 \\
  \hline
  \end{tabular}
\end{table}

\section{Absorption profiles of individual species}
\label{sec:absorption_profile}

\begin{figure}[htb!]
\epsscale{0.5}
\plotone{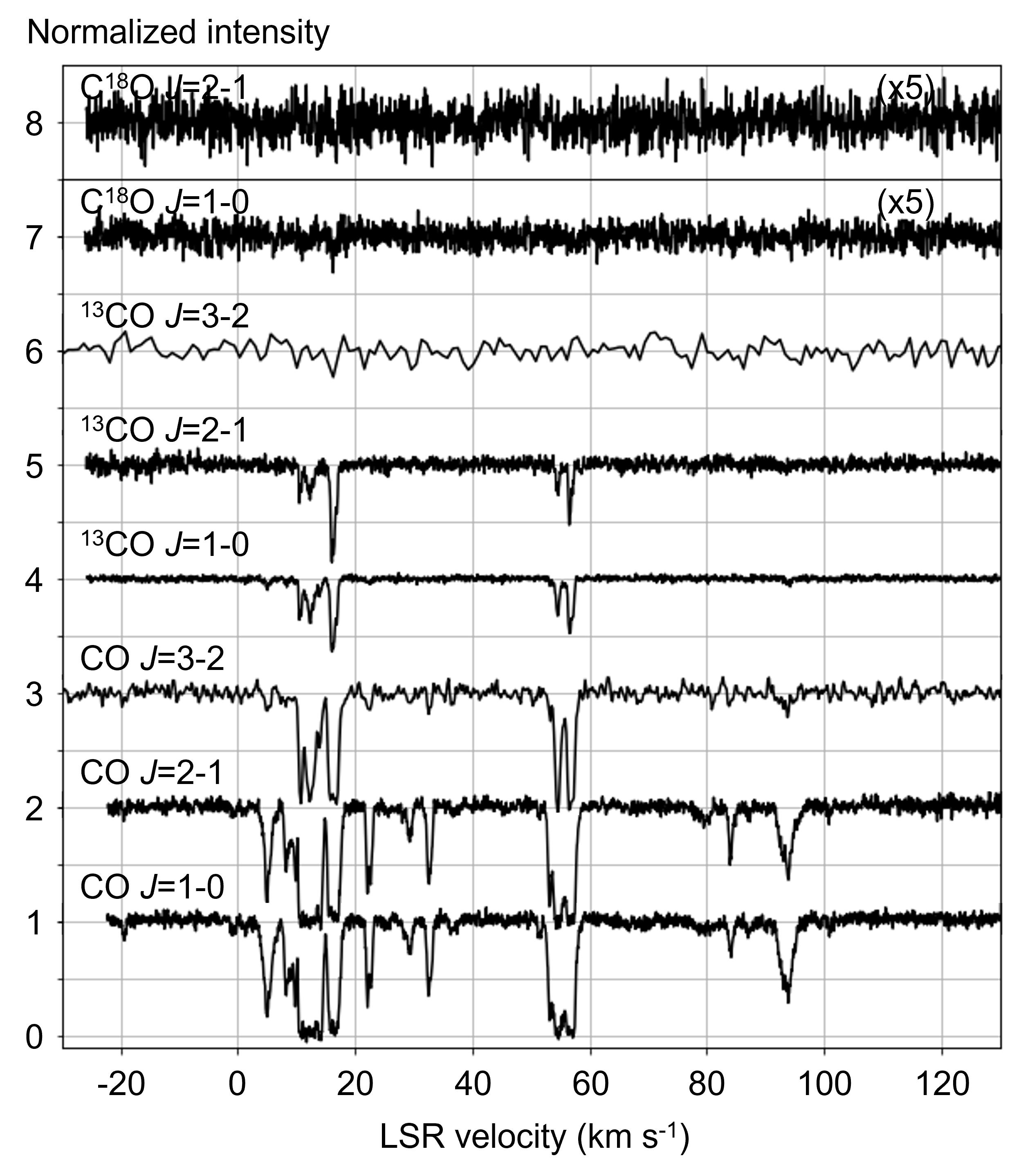}
\caption{Absorption profiles of CO and $^{13}$CO in $J$=1--0, $J$=2--1 and $J$=3--2 transitions, and C$^{18}$O in $J$=1--0 and $J$=2--1 transitions toward QSO J1851+0035. 
The intensity was normalized to the continuum level. 
The profiles with ``($\times5$)'' were multiplied by 5. 
The spectra were offset for clarity. 
\label{fig:co_abs}}
\end{figure}

\begin{figure}[htb!]
\epsscale{0.5}
\plotone{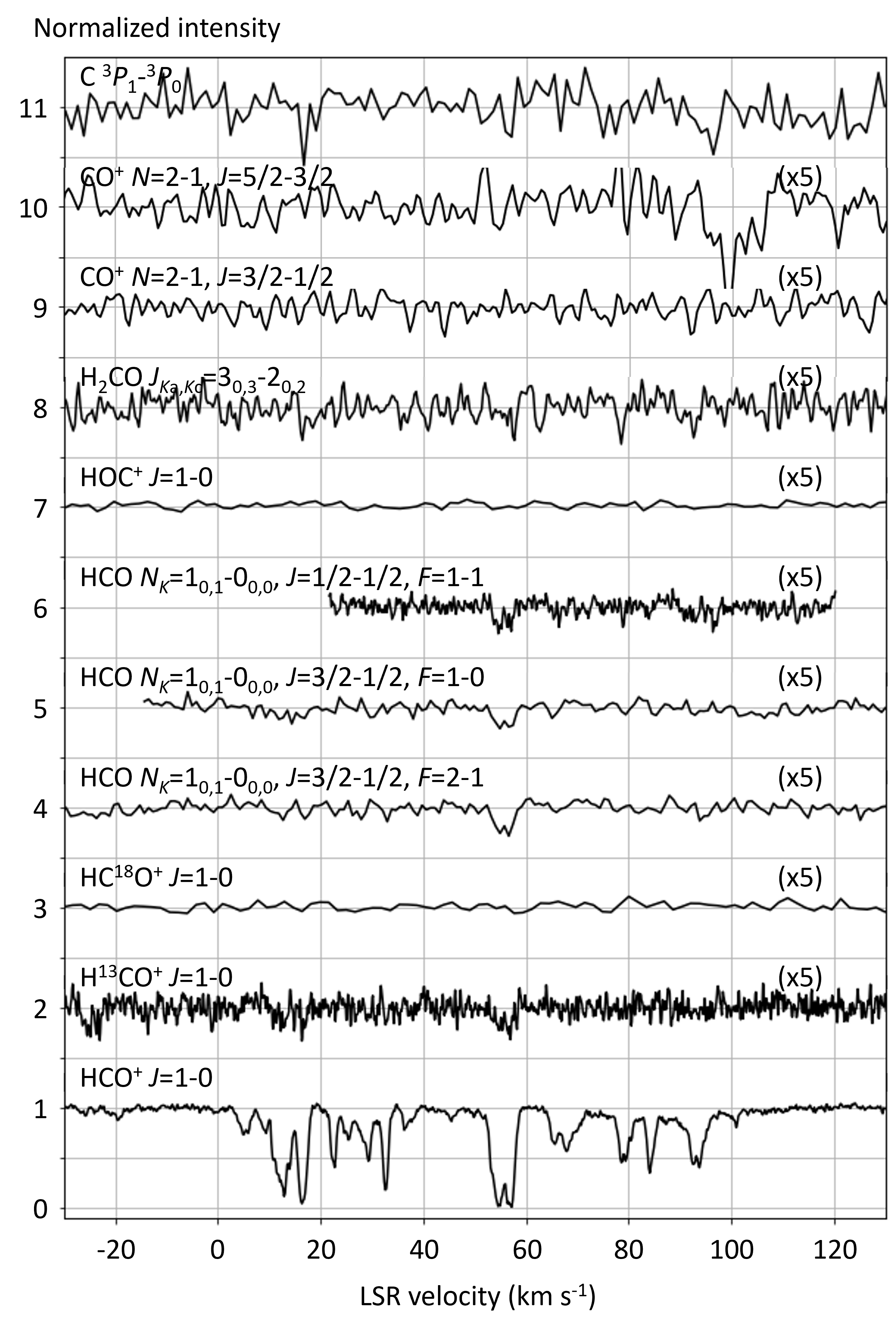}
\caption{The same as Figure \ref{fig:co_abs} but for HCO$^+$, H$^{13}$CO$^+$, HC$^{18}$O$^+$ and HOC$^+$ in $J$=1--0, HCO in $N$=1--0, H$_2$CO in $J_{K_a,K_c}$=$3_{0,3}$--$2_{0,2}$, CO$^+$ in $N$=1--0, and C in $^3P_1$--$^3P_0$ transitions. 
Note that the HCO and CO$^+$ lines have hyperfine components. 
The feature seen near 95 km~s$^{-1}$ of the HCO $N_K$=$1_{0,1}$--$0_{0,0}$, $J$=$\frac{1}{2}$--$\frac{1}{2}$, $F$=1--1 line profile is the Components 6 and 7 of H$^{13}$CO$^+$ $J$=1--0. 
Similarly, the feature seen near $-25$ km~s$^{-1}$ of the H$^{13}$CO$^+$ $J$=1--0 line profile is the Components 15 and 16 of HCO $N_K$=$1_{0,1}$--$0_{0,0}$, $J$=$\frac{1}{2}$--$\frac{1}{2}$, $F$=1--1. 
The feature near 100 km~s$^{-1}$ of the CO$^+$ $N$=2--1, $J$=$\frac{5}{2}$--$\frac{3}{2}$ line is a possible interference.
\label{fig:hco+_abs}}
\end{figure}

\begin{figure}[htb!]
\epsscale{0.5}
\plotone{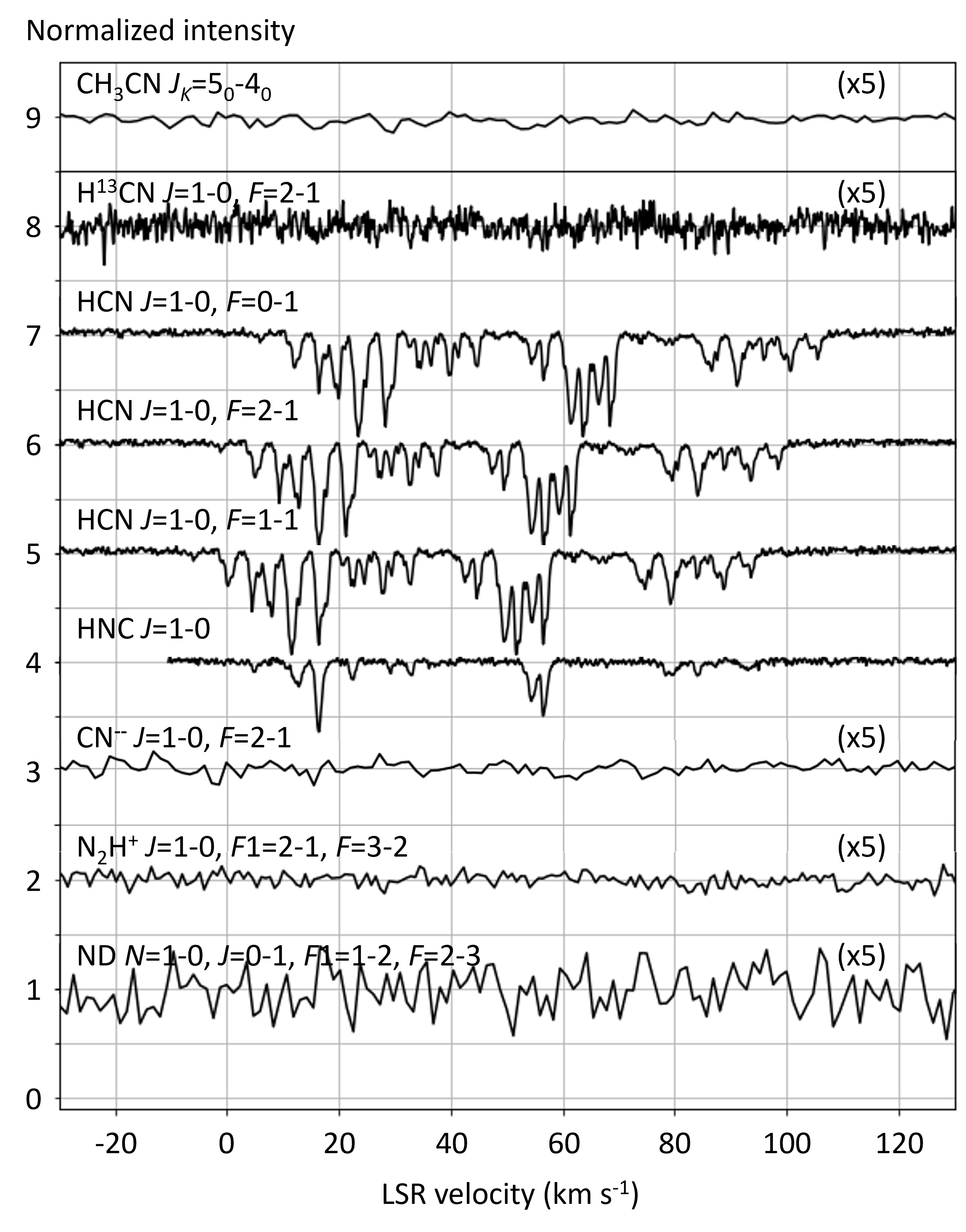}
\caption{The same as Figure \ref{fig:co_abs} but for HCN, H$^{13}$CN, HNC, and CN$^-$ in $J$=1--0, CH$_3$CN $J_K$=$5_0$--$4_0$, N$_2$H$^+$ $J$=1--0, and ND $N$=1--0 transition. Note that the HCN and CN$^-$ lines have hyperfine components. 
\label{fig:hcn_abs}}
\end{figure}

\begin{figure}[htb!]
\epsscale{0.5}
\plotone{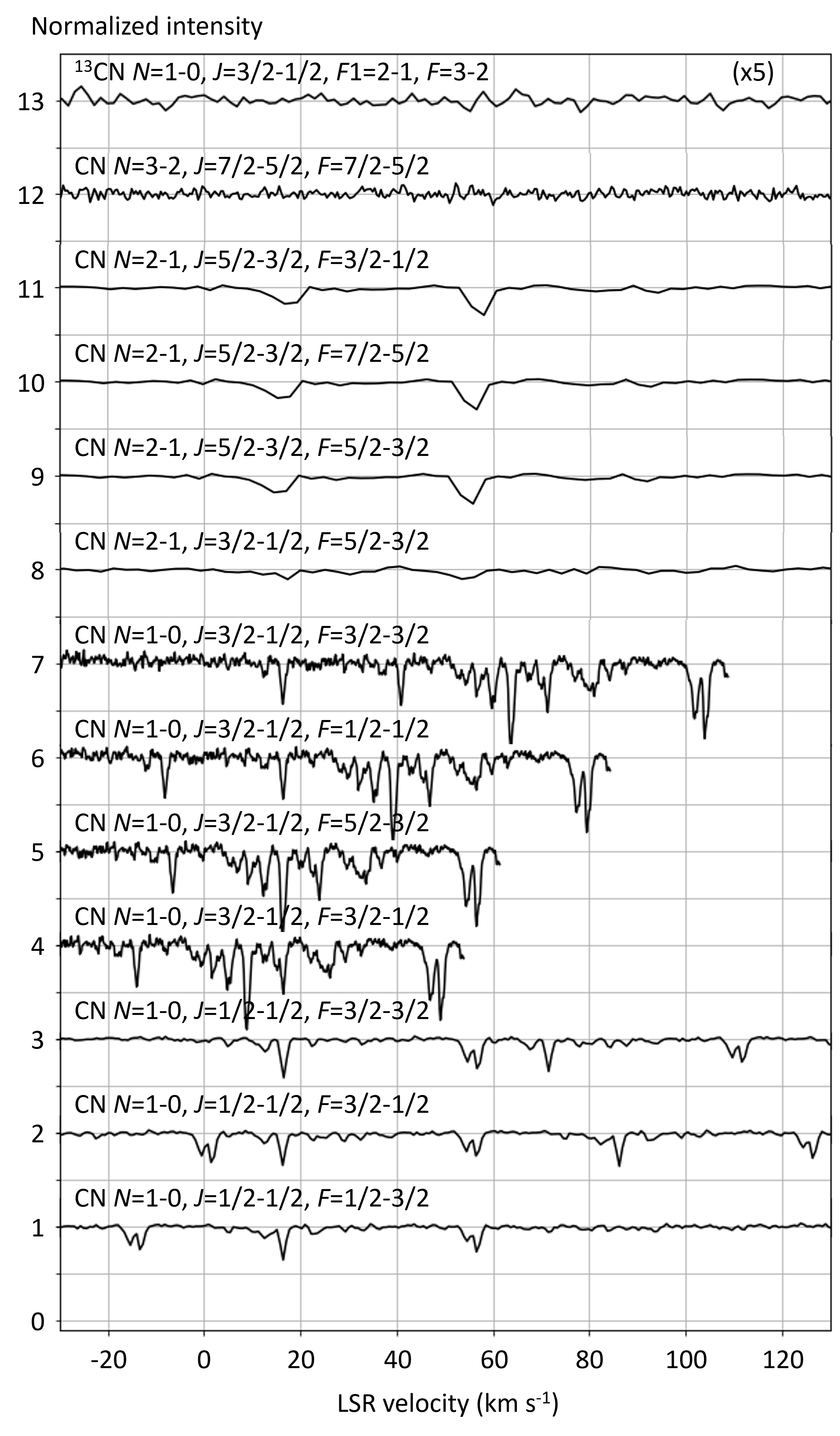}
\caption{The same as Figure \ref{fig:co_abs} but for CN in $N$=1--0, $N$=2--1, $N$=3--2 transitions, and $^{13}$CN in $N$=1--0 transition. 
Note that the CN and $^{13}$CN lines have hyperfine components. 
\label{fig:cn_abs}}
\end{figure}

\begin{figure}[htb!]
\epsscale{0.5}
\plotone{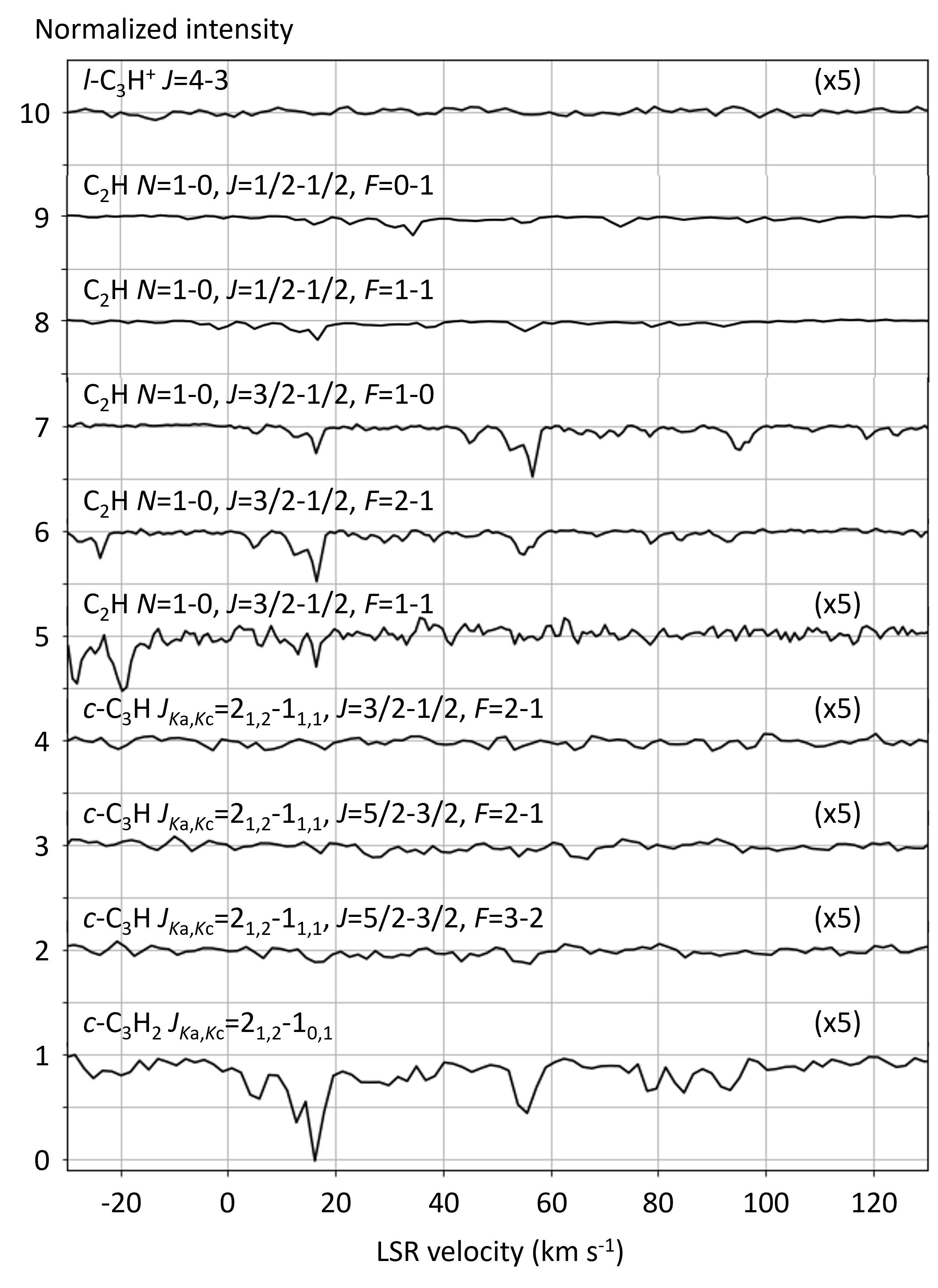}
\caption{The same as Figure \ref{fig:co_abs} but for $l$-C$_3$H$^+$ $J$=4--3, C$_2$H $N$=1--0, $c$-C$_3$H $J_{K_a,K_c}$=$2_{1,2}$--$1_{1,1}$, and $c$-C$_3$H$_2$ $J_{K_a,K_c}$=$2_{1,2}$--$1_{0,1}$ transitions. 
Note that the C$_2$H and $c$-C$_3$H lines have hyperfine components. 
\label{fig:c2h_abs}}
\end{figure}

\begin{figure}[htb!]
\epsscale{0.5}
\plotone{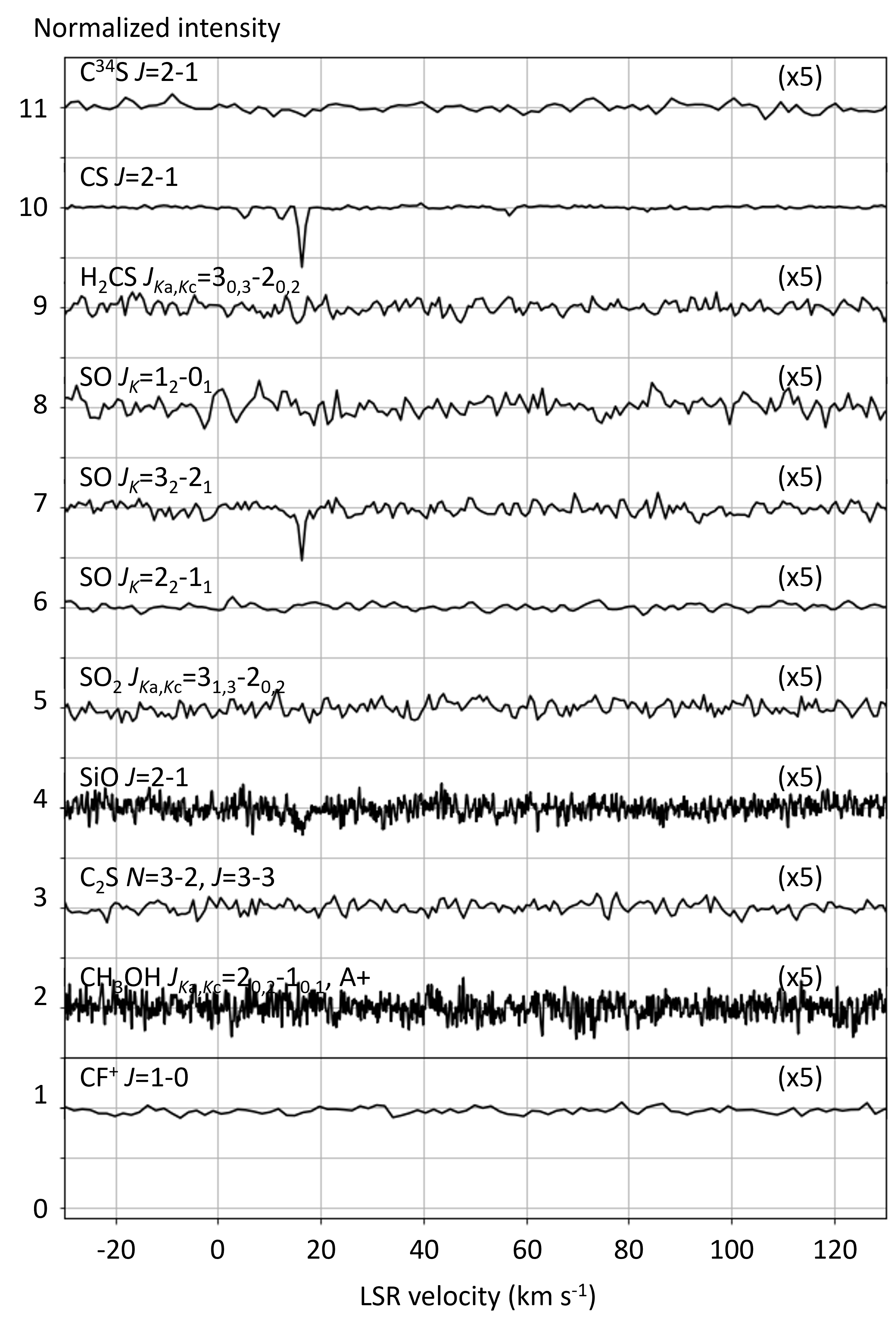}
\caption{The same as Figure \ref{fig:co_abs} but for CS $J$=2--1, SO $J_K$=$1_2$--$0_1$, $J_K$=$2_2$--$1_1$, $J_K$=$3_2$--$2_1$, SiO $J$=2--1, C$_2$S $N$=3--2, CH$_3$OH $J_{K_a,K_c}$=$2_{0,2}$--$1_{0,1}$, and CF$+$ $J$=1--0 transitions. 
\label{fig:cs_abs}}
\end{figure}

Here, we summarize the absorption profiles of individual species, including their hyperfine components in Figures \ref{fig:co_abs}--\ref{fig:cs_abs} for several species groups. 

\section{Nature of absorption host clouds}
\label{sec:host_cloud}

\subsection{Spatial distribution in {\rm CO} and $^{13}${\rm CO} emission lines}
\label{sec:channel_map}

\begin{figure}[htb!]
%\epsscale{0.5}
\plotone{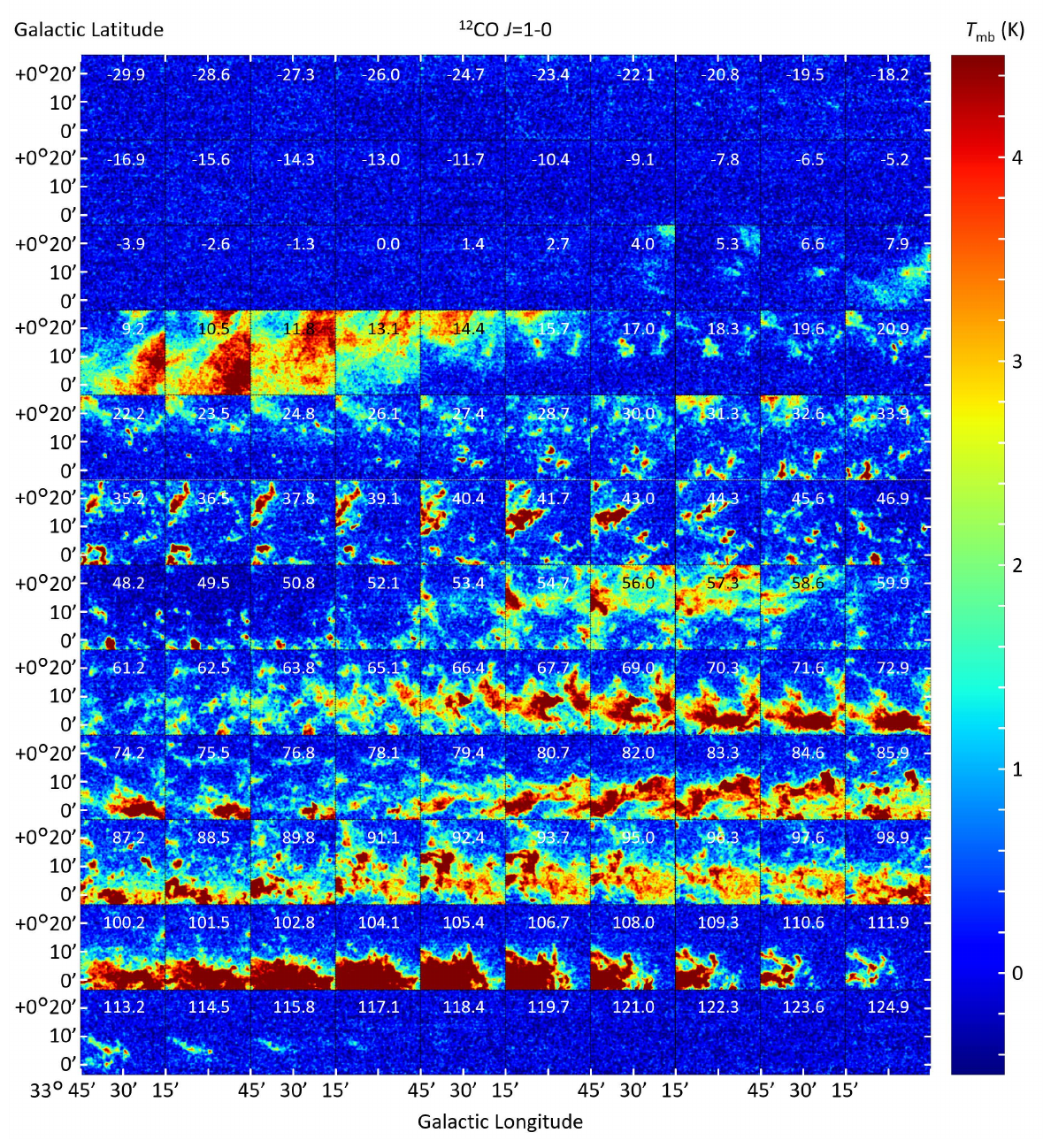}
\caption{Velocity channel maps in CO $J$=1--0 emission of the $0.5^{\circ}$$\times$$0.5^{\circ}$ area centered at the line of sight toward QSO J1851+0035. 
The data was obtained from the FUGIN Galactic Plane Survey with the NRO 45 m telescope \citep{2017PASJ...69...78U} and was smoothed with $40''$ Gaussian for better signal-to-noise ratio. 
The area corresponds to 43.6 pc$\times$43.6 pc area in a linear scale at the distance of 5 kpc. 
The number in each channel map denotes the central velocity of the channel, whose width is 1.3 km~s$^{-1}$. 
\label{fig:fugin_12co}}
\end{figure}

\begin{figure}[htb!]
%\epsscale{0.5}
\plotone{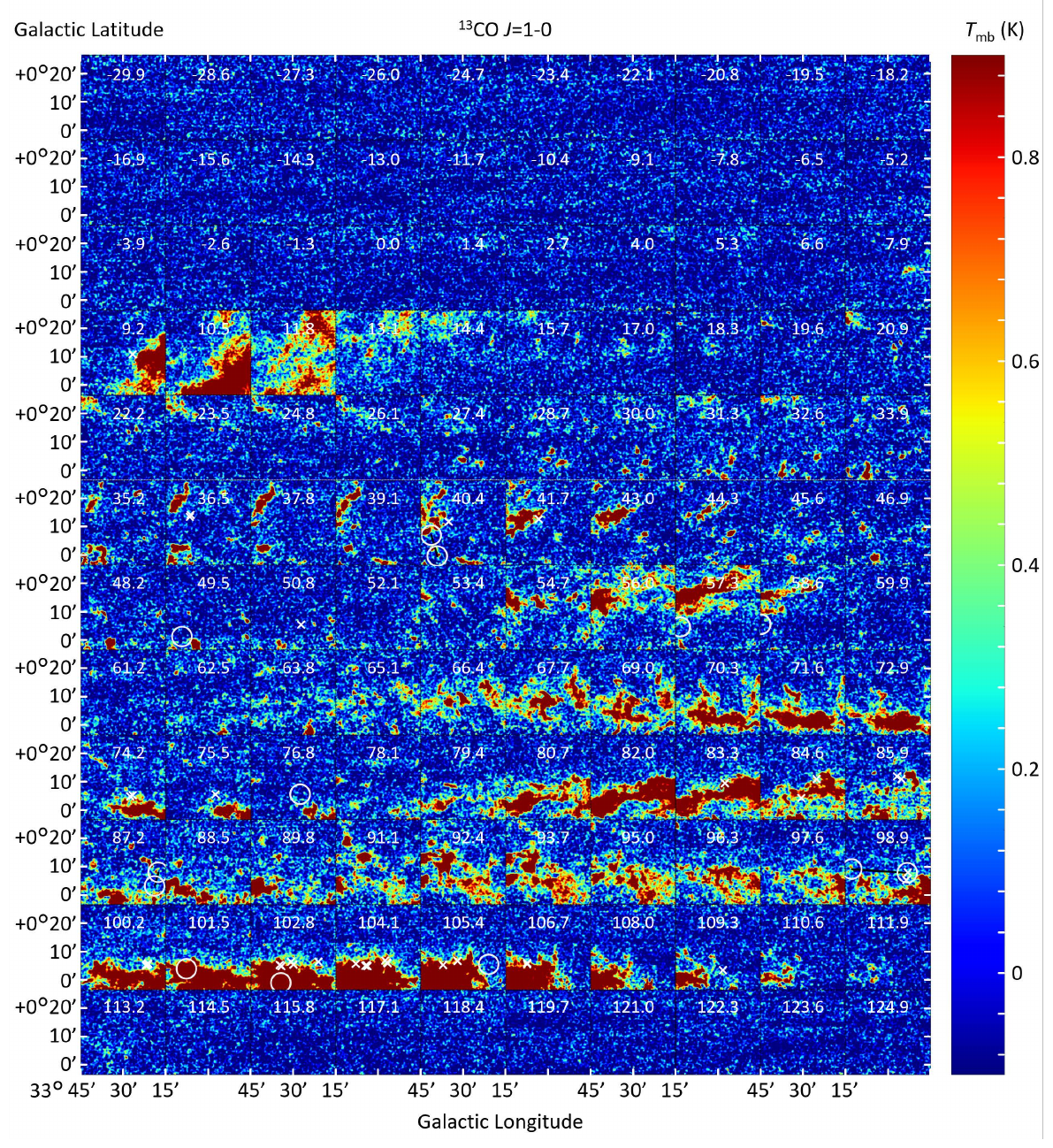}
\caption{The same as Figure \ref{fig:fugin_12co} but for $^{13}$CO $J$=1--0. 
The location of the HII regions listed in Table \ref{table:hrds} are indicated as white circles, whose diameter corresponds to 10 pc in a linear scale at the distance of 5 kpc. 
The location of the ATLASGAL sources listed in Table \ref{table:atlasgal} are indicated as white diagonal crosses. 
\label{fig:fugin_13co}}
\end{figure}

The spatial distribution of molecular gas in CO and $^{13}$CO emission lines is examined to understand the nature of the parent molecular clouds.
In Figures \ref{fig:fugin_12co}--\ref{fig:fugin_13co}, velocity channel maps of the $0.5^{\circ}$$\times$$0.5^{\circ}$ area centered at the line of sight toward QSO J1851+0035 are shown for CO and $^{13}$CO $J$=1--0 emission. 
The area corresponds to 43.6 pc$\times$43.6 pc in a linear scale at the distance of 5 kpc. 
Because the dense part of molecular clouds tends to occupy a small spacial fraction, the line of sight toward the QSO penetrates cloud envelopes in most cases. 
The brightness and the spatial distribution, including the degree of structurization, are used to classify the parent molecular clouds. 

\subsection{Physical conditions of molecular gas deduced from emission studies}
\label{sec:emission_line_ratio}
The observations with the FOREST receiver of the NRO 45 m telescope provided the line profiles of HCO$^+$, HCN, HNC, and C$_2$H toward three more lines of sight with offsets of ($\Delta$RA, $\Delta$DEC) = ($0''$, $+50''$), ($+50''$, $+50''$), ($+50''$, $0''$) from the pointing center (RA=18:51:46.723, DEC=+00:35:32.364). 
Since the line of sight toward the QSO is affected by continuum and is not ideal for emission line studies, we also analyzed the adjacent beams and compared the line profiles of CO, $^{13}$CO, HCO$^+$, HCN, HNC, and C$_2$H data (Figure \ref{fig:dense_core}). 
For the saturated CO absorbers, $T_{\rm b}$(CO)/$T_{\rm b}$(HCO$^+$) $>$ 100, $T_{\rm b}$(CO)/$T_{\rm b}$($^{13}$CO) $\sim$ 8, and $T_{\rm b}$(CO) $\sim$ 5 K. 
Comparison with the non-LTE calculation assuming an expanding sphere in \texttt{RADEX} suggests a number density of $\sim$300 cm$^{-3}$ at most. 

%In the same Beam 3, strong HCO$^+$, HCN, and HNC emission lines were observed in the 60--76 km~s$^{-1}$ range. 
%At this velocity component, CO/HCO$^+$ was $\sim$5, CO/$^{13}$CO was $\sim$50, and $T_{\rm mb}$(CO) was $\sim$2.5 K. 
%The large CO/$^{13}$CO intensity ratio suggests even CO is optically thin. 
%The moderate $T_{\rm mb}$(CO) under an optically thin regime suggests that $T_{\rm ex}$ is not very low. 
%The RADEX study suggests low $X$(CO)/$X$(HCO$^+$) to have the observed low CO/HCO$^+$ intensity ratio. 
%XX
%
\subsection{Association with HII regions}
\label{sec:hrds}

The position, velocity, and the near-far ambiguity solution of known HII regions taken from the HII Region Discovery Survey \citep[HRDS:][]{2011ApJS..194...32A,2012ApJ...754...62A} and references therein are summarized in Table \ref{table:hrds}. 

\begin{table}[htb!]
 \caption{HII regions near the line of sight toward QSO J1851+0035}
 \label{table:hrds}
 \centering
  \begin{tabular}{cccccl}
   \hline \hline
Name & $l$ & $b$ & $v_{\rm LSR}$ & N/F/T & References \\
& ($^{\circ}$) & ($^{\circ}$) & (km~s$^{-1}$) & & \\
\hline 
G033.810$-$00.190\phantom{} & 33.810 & $-$0.190 & \phantom{0}40.3 & & \cite{2003ApJ...587..714W} \\
G033.882$+$0.057\phantom{0} & 33.882 & $+$0.057 & \phantom{0}40.8 & F & HRDS \\
G033.810$-$0.154\phantom{0} & 33.810 & $-$0.154 & \phantom{0}50.0 & F & HRDS \\
G033.943$-$00.036\phantom{} & 33.943 & $-$0.036 & \phantom{0}57.2 & & \cite{1996ApJ...472..173L} \\
G033.994$-$0.006\phantom{0} & 33.994 & $-$0.006 & \phantom{0}59.0 & & HRDS \\
G033.418$-$00.004\phantom{} & 33.418 & $-$0.004 & \phantom{0}76.5 & & \cite{1989ApJS...71..469L} \\
G033.091$+$0.073\phantom{0} & 33.091 & $+$0.073 & \phantom{0}86.9 & F & HRDS \\
G033.130$-$00.090\phantom{} & 33.130 & $-$0.090 & \phantom{0}87.4 & & \cite{2002ApJS..138...63A} \\
G033.265$+$0.068\phantom{0} & 33.265 & $+$0.068 & \phantom{0}98.3 & T & HRDS \\
G033.920$+$00.110\phantom{} & 33.920 & $+$0.110 & \phantom{0}99.0 & & \cite{2002ApJS..138...63A} \\
G033.753$-$0.063\phantom{0} & 33.753 & $-$0.063 & \phantom{}101.7 & N & HRDS \\
G033.645$-$0.227\phantom{0} & 33.645 & $-$0.227 & \phantom{}102.9 & T & HRDS \\
G033.200$-$00.010\phantom{} & 33.200 & $-$0.010 & \phantom{}105.8 & & \cite{2003ApJ...587..714W} \\
\hline
\end{tabular}
\end{table}

\subsection{Association with ATLASGAL sources}
\label{sec:atlasgal}

The position and LSR velocity of ATLASGAL GaussClump Source Catalogue (GCSC) sources \citep{2014A&A...565A..75C} 
%Note that the ATLASGAL GCSC sources are candidate compact submillimeter sources that can potentially host precursors to high-mass stars. 
\begin{table}[htb!]
 \caption{ATLASGAL sources near the line of sight toward QSO J1851+0035}
 \label{table:atlasgal}
 \centering
  \begin{tabular}{cc}
   \hline \hline
  ATLASGAL & $v_{\rm LSR}$\\
  GCSC Name & (km~s$^{-1}$)\\
   \hline 
G033.3883$+$0.1670 & \phantom{00}9.43 \\
G033.7142$+$0.2557 & \phantom{0}36.27 \\
G033.7038$+$0.2832 & \phantom{0}37.23 \\
G033.6744$+$0.2028 & \phantom{0}39.90 \\
G033.6092$+$0.2305 & \phantom{0}41.90 \\
G033.4306$-$0.0143 & \phantom{0}73.76 \\
G033.4025$-$0.0032 & \phantom{0}73.78 \\
G033.4160$-$0.0030 & \phantom{0}75.10 \\
G033.4196$+$0.1249 & \phantom{0}83.10 \\
G033.5185$-$0.0575 & \phantom{0}84.68 \\
G033.3376$+$0.1655 & \phantom{0}84.80 \\
G033.3883$+$0.1984 & \phantom{0}85.29 \\
G033.2645$+$0.0670 & \phantom{0}98.79 \\
G033.2868$-$0.0190 & \phantom{0}99.20 \\
G033.2058$-$0.0088 & \phantom{0}99.96 \\
G033.2093$+$0.0023 & \phantom{0}99.96 \\
G033.2278$-$0.0179 & \phantom{}100.44 \\
G033.2371$-$0.0214 & \phantom{}100.44 \\
G033.2011$+$0.0194 & \phantom{}101.20 \\
G033.4941$-$0.0138 & \phantom{}102.20 \\
G033.6546$-$0.0204 & \phantom{}102.67 \\
G033.5236$+$0.0195 & \phantom{}103.30 \\
G033.6377$-$0.0345 & \phantom{}103.47 \\
G033.6218$-$0.0335 & \phantom{}103.60 \\
G033.6488$-$0.0273 & \phantom{}103.60 \\
G033.4173$+$0.0319 & \phantom{}103.70 \\
G033.3860$+$0.0020 & \phantom{}103.86 \\
G033.3917$+$0.0099 & \phantom{}103.86 \\
G033.6326$-$0.0227 & \phantom{}104.45 \\
G033.7578$-$0.0029 & \phantom{}104.54 \\
G033.7394$-$0.0197 & \phantom{}104.70 \\
G033.5685$+$0.0272 & \phantom{}105.65 \\
G033.7425$+$0.0029 & \phantom{}106.65 \\
G033.7440$-$0.0074 & \phantom{}106.65 \\
G033.4526$-$0.0810 & \phantom{}109.20 \\
   \hline
  \end{tabular}
\end{table}
are summarized in Table \ref{table:atlasgal}. 
The position and LSR velocity of ATLASGAL GaussClump Source Catalogue (GCSC) sources \citep{2014A&A...565A..75C} are summarized in Table \ref{table:atlasgal}. 
Note that the ATLASGAL GCSC sources are candidate compact submillimeter sources that can potentially host precursors to high-mass stars. 
\subsection{Association with dense gas}
\label{sec:dense_cores}

\begin{figure}[htb!]
\epsscale{0.5}
\plotone{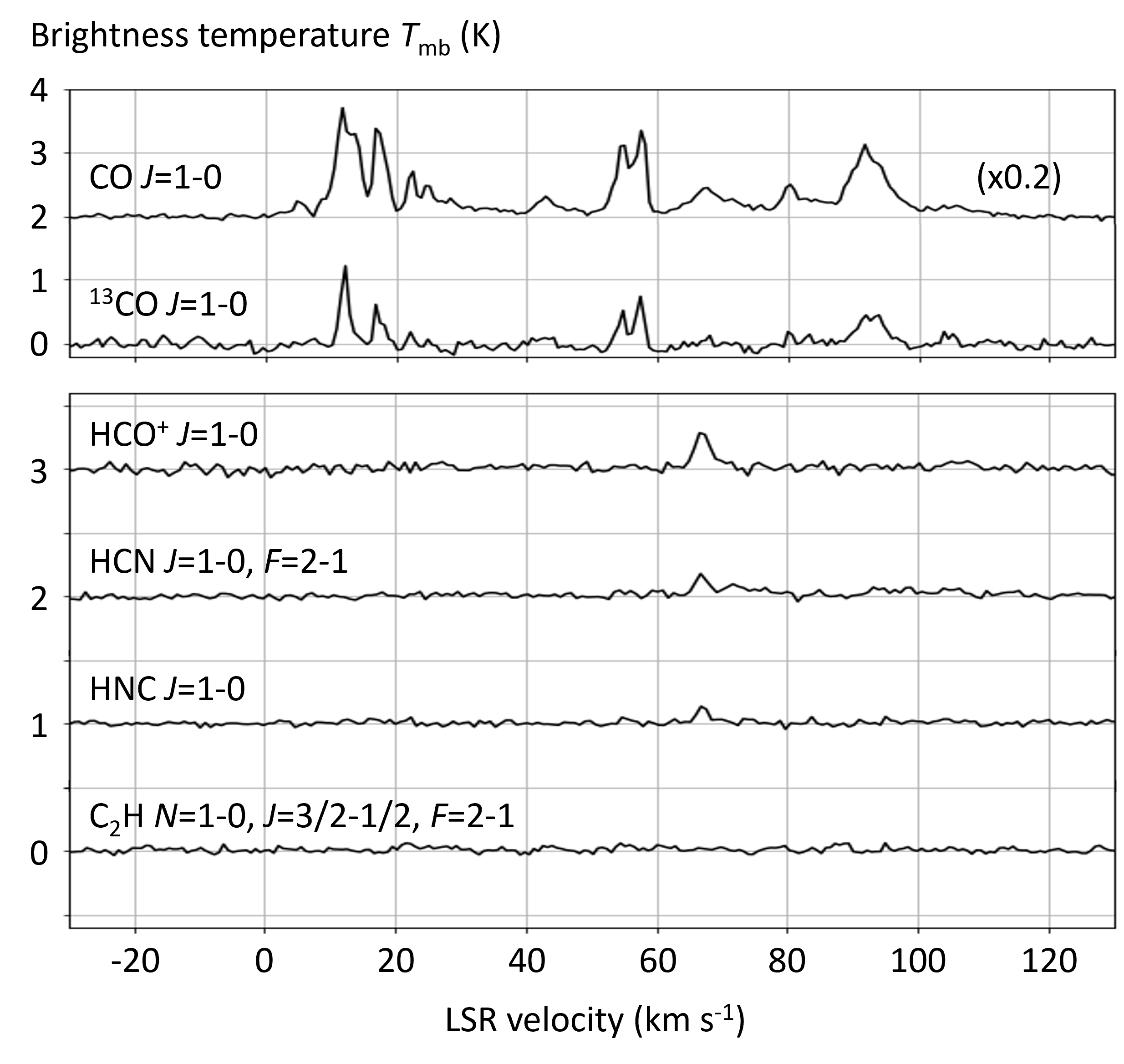}
\caption{Emission line profiles of CO, $^{13}$CO, HCO$^+$, HCN, HNC, and C$_2$H toward ($\Delta$RA, $\Delta$DEC) = ($+50''$, $+50''$) obtained with the NRO 45 m telescope. 
The spectra were offset for clarity. 
Note that the HCN and C$_2$H lines have hyperfine components. }
\label{fig:dense_core}
\end{figure}

Emission line profiles of CO, $^{13}$CO, HCO$^+$, HCN, HNC, and C$_2$H toward ($\Delta$RA, $\Delta$DEC) = ($+50''$, $+50''$) are shown in Figure \ref{fig:dense_core} as an example.
No emission was detected in these lines except in the velocity range corresponding to Component 17. 

\begin{figure}[htb!]
%\epsscale{0.5}
\plotone{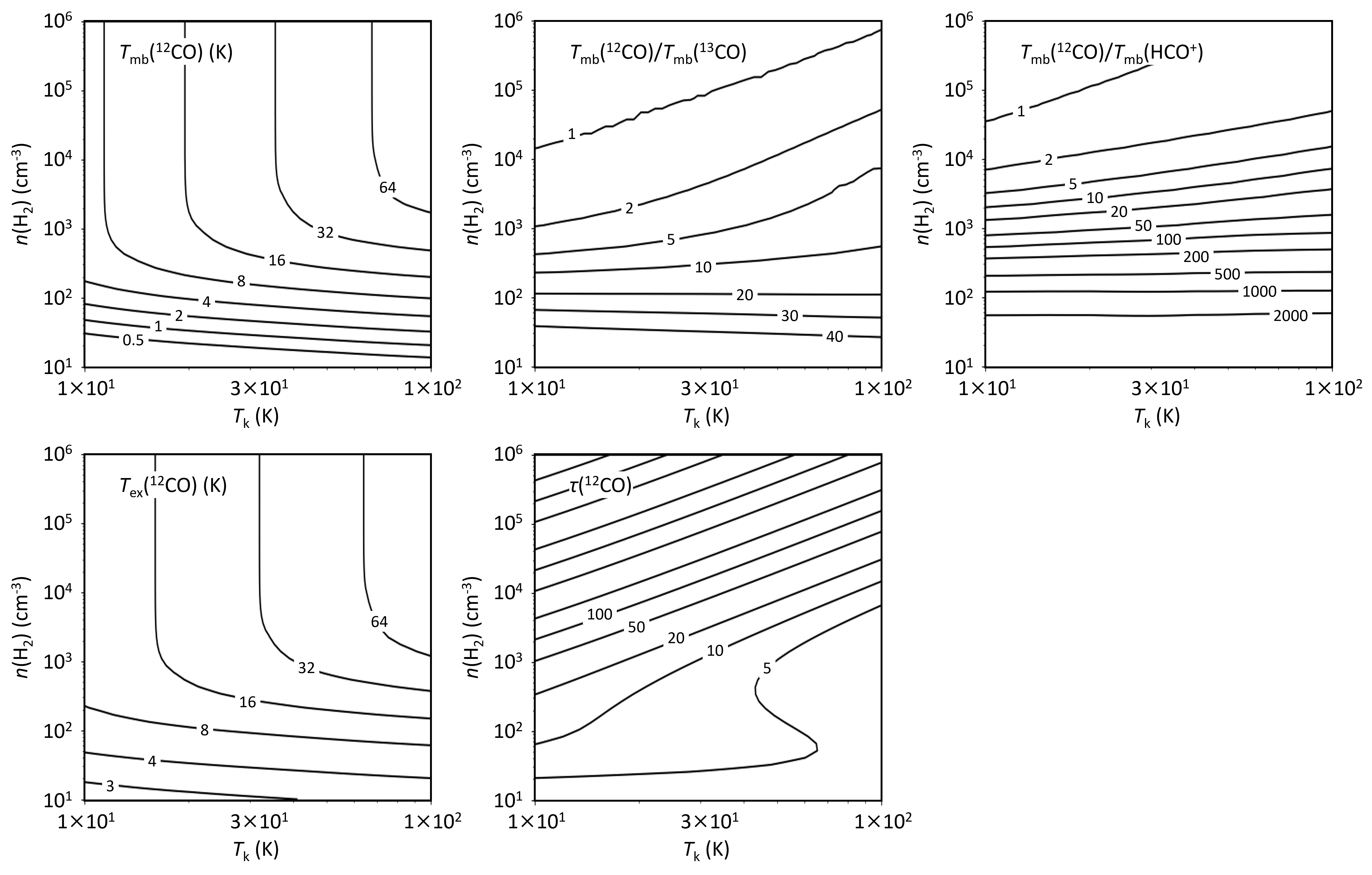}
\caption{The CO main beam antenna temperature ($T_{\rm mb}$(CO)), the line intensity ratios ($T_{\rm mb}$(CO)/$T_{\rm mb}$($^{13}$CO) and $T_{\rm mb}$(CO)/$T_{\rm mb}$(HCO$^+$)), CO excitation temperature ($T_{\rm ex}$(CO)), and opacity ($\tau$(CO)) calculated as functions of kinetic temperature and molecular hydrogen density using \texttt{ndradex} \citep{ndradex} developed based on \texttt{RADEX} \citep{2007A&A...468..627V}. 
The beam-filling factor was assumed to be unity. 
The abundances of CO and $^{13}$CO were set at $X$(CO) = $1 \times 10^{-4}$ and $X$($^{13}$CO) = $2 \times 10^{-6}$ as canonical values for Galactic disk clouds \citep[][and references therein]{2008ApJ...679..481P}. 
The abundance of HCO$^+$ was set at $X$(HCO$^+$) = $3 \times 10^{-9}$, which is the recommended value for diffuse gas in the Galactic disk \citep{2023ApJ...943..172L}. 
This value is consistent with the results for low-density ($<$$10^3$ cm$^{-3}$) regions in star-forming clouds \citep{2021A&A...648A.120R}. 
The velocity gradient $dv/dr$ was fixed at 1.0 $\rm km~s^{-1}~pc^{-1}$. 
The coutours for $\tau$(CO) are at $(1, 2, 5) \times 10^{n}$.}
\label{fig:radex_norm}
\end{figure}

\begin{figure}[htb!]
%\epsscale{0.5}
\plotone{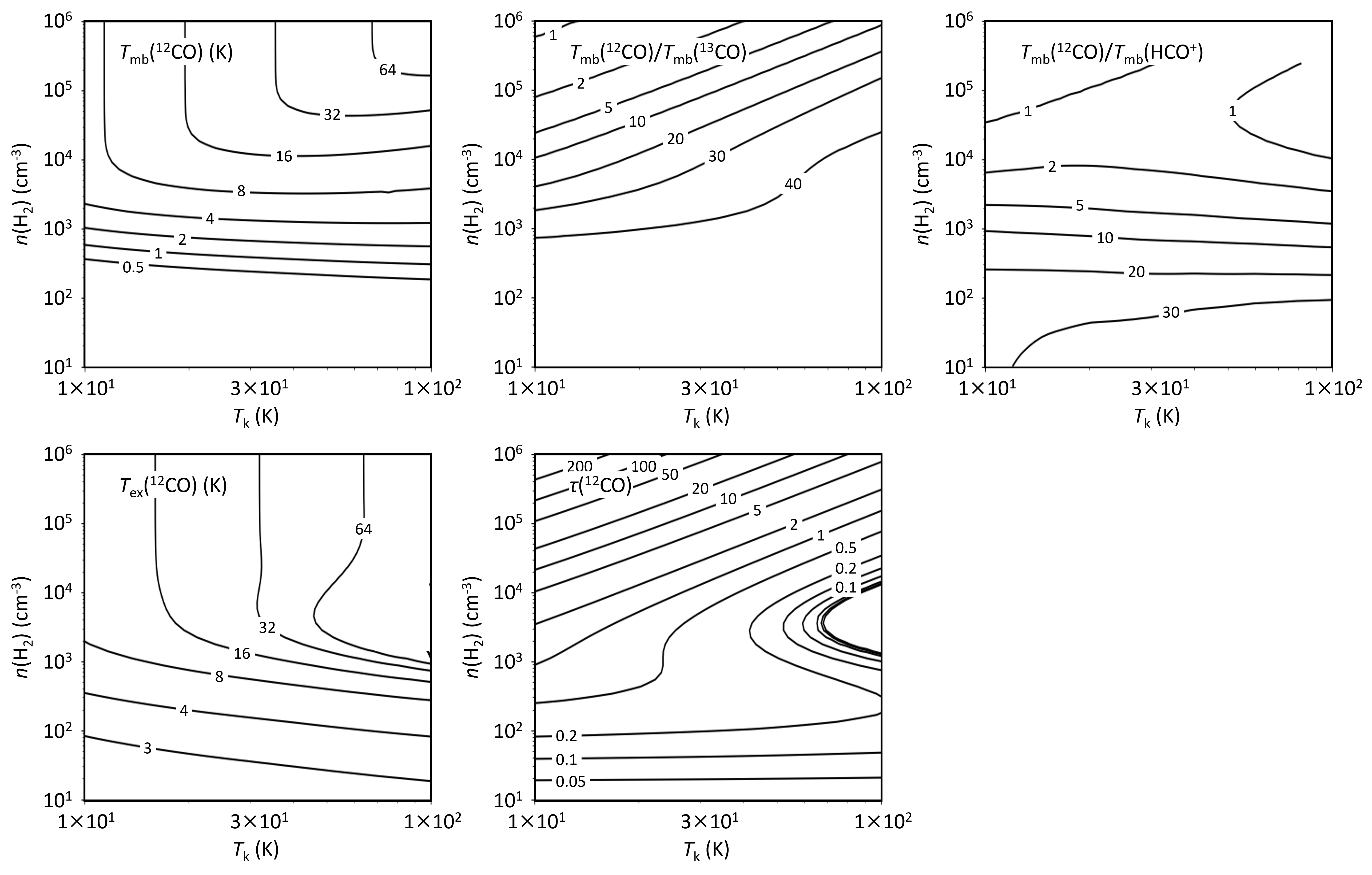}
\caption{Same as Figure \ref{fig:radex_norm} but for one of the CO-poor cases. 
The abundances of CO and $^{13}$CO were set at $X$(CO) = $1 \times 10^{-6}$ and $X$($^{13}$CO) = $2 \times 10^{-8}$, which are 100 times smaller than the canonical values for Galactic disk clouds \citep[][and references therein]{2008ApJ...679..481P}.}
\label{fig:radex_poor}
\end{figure}

Model calculation results of multi-line analysis of CO, $^{13}$CO, and HCO$^+$ are shown in Figures \ref{fig:radex_norm} and \ref{fig:radex_poor}. 
The CO main beam antenna temperature, $T_{\rm mb}$(CO), and the line intensity ratios, $T_{\rm mb}$(CO)/$T_{\rm mb}$($^{13}$CO) and $T_{\rm mb}$(CO)/$T_{\rm mb}$(HCO$^+$), were calculated as functions of $T_{\rm k}$ and $n$(H$_2$) for different values of $X$(CO)/($dv/dr$) using \texttt{ndradex} \citep{ndradex} developed based on \texttt{RADEX} \citep{2007A&A...468..627V}. 
The beam-filling factor was assumed to be unity. 

In Figure \ref{fig:radex_norm}, the velocity gradient $dv/dr$ was fixed at 1.0 $\rm km~s^{-1}~pc^{-1}$, and the abundances of CO and $^{13}$CO were set at $X$(CO) = $1 \times 10^{-4}$ and $X$($^{13}$CO) = $2 \times 10^{-6}$ as canonical values for Galactic disk clouds \citep[][and references therein]{2008ApJ...679..481P}. 
The abundance of HCO$^+$ was set at $X$(HCO$^+$) = $3 \times 10^{-9}$, which is the recommended value for diffuse gas in the Galactic disk \citep{2023ApJ...943..172L}. 
This value is consistent with the results for low-density ($<$$10^3$ cm$^{-3}$) regions in star-forming clouds \citep{2021A&A...648A.120R}. 

For the velocity components with relatively high CO brightness, we obtained $T_{\rm mb}$(CO)/$T_{\rm mb}$(HCO$^+$) $>$ 100, $T_{\rm mb}$(CO)/$T_{\rm mb}$($^{13}$CO) $\sim$ 8, and $T_{\rm mb}$(CO) $\sim$ 5 K. 
Typical molecular gas with $n$(H$_2$) $\sim 1 \times 10^3$ cm$^{-3}$ accounts for these parameters. 

If we half the $X$(CO) value to $5 \times 10^{-5}$ so that it is consistent with the CO/$^{13}$CO isotopologue ratio of 25 assumed for saturated CO absorbers, the contours go down so that their intercepts with the vertical axis are reduced by a factor of 1/1.4.  
This will not have a critical impact on the above-deduced values. 

For the nearby emission feature that corresponds to Component 17, we obtained $T_{\rm mb}$(CO)/$T_{\rm mb}$(HCO$^+$) $\sim$ 5, $T_{\rm mb}$(CO)/$T_{\rm mb}$($^{13}$CO) $\sim$ 50, and $T_{\rm mb}$(CO) $\sim$ 2.5 K. 
The extremely low $T_{\rm mb}$(CO)/$T_{\rm mb}$(HCO$^+$) value prefers high density, but there is no solution with the canonical values of $X$(CO), $X$($^{13}$CO), and $X$(HCO$^+$). 
The large $T_{\rm mb}$(CO)/$T_{\rm mb}$($^{13}$CO) close to the isotopologue ratio requires even CO emission to be optically thin. 
Because the reduction of the CO abundance will make the CO emission optically thinner, enhance the $T_{\rm mb}$(CO)/$T_{\rm mb}$(HCO$^+$) ratio, and reduce the $T_{\rm mb}$(CO), the CO-poor environment will explain the observed results. 
The \texttt{RADEX} calculation with 100 times poorer CO and $^{13}$CO abundances demonstrates a parameter space at $n$(H$_2$) $\sim  1 \times 10^3$ cm$^{-3}$ and $T_{\rm k}$ $\gtrsim$ 30 K that accounts for the observed parameters (Figure \ref{fig:radex_poor}). 
Note that Component 17 is one of the most prominent velocity components with notable HCO$^+$ absorption but without CO absorption (Figure \ref{fig:em_abs}) and the CO/HCO$^+$ abundance ratio was measured as small as $<$3 as shown in Tables \ref{table:col_co_abs} and \ref{table:col_hco+_hnc}. 
The abnormal line ratios observed in this adjacent area suggest that the CO-poor gas is extended. 

\section{Reliability of H$_2$ column density from $^{13}$CO $J$=1--0 emission studies}
\label{sec:reliability}

Here we examine the reliability of the H$_2$ column density of diffuse gas extracted from $^{13}$CO $J$=1--0 emission studies.

The brightness temperature $T_{\rm{b}}$ for the specific intensity $I_{\nu}$ is defined by 
\begin{equation}
\label{eq:d1}
T_{\rm{b}} \equiv \frac{c^2}{2 \nu^2 k_{\rm{B}}}I_{\nu}. 
\end{equation}
For black body radiation, $T_{\rm{b}}$ agrees with the temperature of the black body under the Rayleigh-Jeans approximation. 

If the background radiation with $T_{\rm{bg}}$ propagates through a uniform medium with its excitation temperature $T_{\rm{ex}}$ and opacity $\tau$, the radiative transfer equation gives 
\begin{equation}
\label{eq:d2}
T_{\rm{b}} = \left(\frac{h\nu/k_{\rm{B}}}{\exp({h\nu/k_{\rm{B}} T_{\rm{bg}})} - 1}\right)e^{-\tau} + \left(\frac{h\nu/k_{\rm{B}}}{\exp({h\nu/k_{\rm{B}} T_{\rm{ex}})} - 1}\right) (1-e^{-\tau}).
\end{equation}

If the medium occupies only a limited fraction $f$ of the telescope beam (beam-filling factor), 
\begin{equation}
\label{eq:d3}
\begin{split} 
T_{\rm{b}} & = f \left[\left(\frac{h\nu/k_{\rm{B}}}{\exp({h\nu/k_{\rm{B}} T_{\rm{bg}})} - 1}\right)e^{-\tau} + \left(\frac{h\nu/k_{\rm{B}}}{\exp({h\nu/k_{\rm{B}} T_{\rm{ex}})} - 1}\right) (1-e^{-\tau})\right] + (1-f)\left(\frac{h\nu/k_{\rm{B}}}{\exp({h\nu/k_{\rm{B}} T_{\rm{bg}})} - 1}\right) \\
& = f(1-e^{-\tau}) \left(\frac{h\nu/k_{\rm{B}}}{\exp({h\nu/k_{\rm{B}} T_{\rm{ex}})} - 1} - \frac{h\nu/k_{\rm{B}}}{\exp({h\nu/k_{\rm{B}} T_{\rm{bg}})} - 1}\right) + \left(\frac{h\nu/k_{\rm{B}}}{\exp({h\nu/k_{\rm{B}} T_{\rm{bg}})} - 1}\right).
\end{split} 
\end{equation}
In the position-switching observations, we usually measure the enhanced intensity over the uniform background, and the last term of Equation \ref{eq:d3} is eliminated to yield the main beam antenna temperature $T_{\rm{mb}}$ as
\begin{equation}
\label{eq:d4}
T_{\rm{mb}} = f(1-e^{-\tau}) \left(\frac{h\nu/k_{\rm{B}}}{\exp({h\nu/k_{\rm{B}} T_{\rm{ex}})} - 1} - \frac{h\nu/k_{\rm{B}}}{\exp({h\nu/k_{\rm{B}} T_{\rm{bg}})} - 1}\right) = f_{\rm{d}} \left(\frac{h\nu/k_{\rm{B}}}{\exp({h\nu/k_{\rm{B}} T_{\rm{ex}})} - 1} - \frac{h\nu/k_{\rm{B}}}{\exp({h\nu/k_{\rm{B}} T_{\rm{bg}})} - 1}\right).
\end{equation}
Here the dilution factor is defined as $f_{\rm{d}} \equiv f(1-e^{-\tau})$. 
In Equation \ref{eq:d4}, only $T_{\rm{mb}}$ and $\tau$ are considered as variables as functions of velocity for each velocity component. 

The excitation temperature $T_{\rm{ex}}$ is then related to $T_{\rm{mb}}$ and $f_{\rm{d}}$ as 
\begin{equation}
\label{eq:d5}
T_{\rm{ex}} = \frac{h \nu}{k_{\rm{B}}}\left[\ln \left(1 + \frac{1}{\frac{k_{\rm{B}}}{h \nu}\frac{T_{\rm{mb}}}{f_{\rm{d}}} +\frac{1}{\exp({h \nu / k_{\rm{B}} T_{\rm{bg}})-1}}}\right)\right]^{-1}. 
\end{equation}

For optically thick ($\tau \gg 1$) and beam-filling ($f = 1$) emission, $f_{\rm{d}}$ nearly equals to unity. 
However, if $f_{\rm{d}}$ is significantly smaller than unity as is expected for the $^{13}$CO $J$=1--0 emission and even for the CO $J$=1--0 emission toward diffuse gas, the observed $T_{\rm{mb}}$ shall be corrected for $f_{\rm{d}}$, and the analysis without this correction will severely underestimate $T_{\rm{ex}}$. 
%For $^{13}$CO $J$=1--0 ($h\nu/k_{\rm{B}}$ = 5.29 K), if $T_{\rm{mb}}$ = 1 K at the peak, $T_{\rm{ex}}$ deduced from the observed peak value of $T_{\rm{mb}}$ by assuming $f_{\rm{d}}$ = 1 is 4.0 K while it shall be 5.1 K for $f_{\rm{d}}$ = 0.5. 
%For CO $J$=1--0 ($h\nu/k_{\rm{B}}$ = 5.53 K), if $T_{\rm{mb}}$ = 5 K at the peak, $T_{\rm{ex}}$ deduced from the observed peak value of $T_{\rm{mb}}$ by assuming $f_{\rm{d}}$ = 1 is 8.3 K while it shall be 13.4 K for $f_{\rm{d}}$ = 0.5. 

Next, we deduce the column density $N$. 
For optically thin emission, $\tau \simeq 1-e^{-\tau}$, and Equation \ref{eq:d4} is transformed to, 
\begin{equation}
\label{eq:d6}
1-e^{-\tau} = \frac{k_{\rm{B}}}{h\nu} \frac{T_{\rm{mb}}}{f} \left(\frac{1}{\exp({h\nu/k_{\rm{B}} T_{\rm{ex}})} - 1} - \frac{1}{\exp({h\nu/k_{\rm{B}} T_{\rm{bg}})} - 1}\right)^{-1}.
\end{equation}
Then Equation \ref{eq:n_tot} is approximated by, 
\begin{equation}
\label{eq:d7}
\begin{split} 
N & \simeq \frac{3hQ(T_{\rm ex})}{8\pi^3\mu^2S_{\rm ul}} 
\frac{\exp(\frac{E_{\rm l}}{k_{\rm B} T_{\rm ex}})} {[1-\exp(\frac{-h\nu}{k_{\rm B}T_{\rm ex}})]} 
\int (1-e^{-\tau})\ dv \\ 
& = \frac{3hQ(T_{\rm ex})}{8\pi^3\mu^2S_{\rm ul}} 
\frac{\exp(\frac{E_{\rm l}}{k_{\rm B} T_{\rm ex}})} {[1-\exp(\frac{-h\nu}{k_{\rm B}T_{\rm ex}})]} 
\frac{k_{\rm B}}{h\nu} 
\left(\frac{1}{\exp({h\nu/k_{\rm{B}}T_{\rm{ex}})}-1} - \frac{1}{\exp({h\nu/k_{\rm{B}}T_{\rm{bg}})}-1}\right)^{-1} 
\int \frac{T_{\rm{mb}}}{f}\ dv, 
\end{split} 
\end{equation}

For pure rotation of a diatomic rotor like CO, the partition function $Q(T_{\rm ex})$ can be calculated as the sum of the populations of all energy levels $E_{\rm l}$ (= $\frac{1}{2}E_{J=1} J(J+1)$) degenerated by $g_{\rm u}$ (= $2J+1$), 
\begin{equation}
\label{eq:d8}
Q(T_{\rm ex}) = \sum_{J=0}^{\infty} g_{\rm u} \exp \left(-\frac{E_{\rm l}}{k_{\rm{B}}T_{\rm ex}}\right), 
\end{equation}
and we performed a direct sum over all energy levels obtained from CDMS for each $T_{\rm ex}$. 

For $^{13}$CO $J$=1--0 emission, $h\nu/k_{\rm{B}}$ = 5.29 K, $E_{\rm l}$ = 0 K. 
By assuming $T_{\rm{bg}}$ = $T_{\rm{CMB}}$ = 2.73 K, we derive the $^{13}$CO column density from its optically thin $J$=1--0 emission as, 
\begin{equation}
\label{eq:d9}
N({\rm ^{13}CO}) \simeq 
\frac{3.29 \times 10^{14}}{5.29\ {\rm K}}
\frac{Q(T_{\rm ex})}{[1-\exp(-5.29\ {\rm K}/T_{\rm ex})]} 
\left(\frac{1}{\exp(5.29\ {\rm K}/T_{\rm ex})-1}-0.168\right)^{-1}
\int \frac{T_{\rm{mb}}}{f}\ dv \ 
{\rm cm^{-2}~(km~s^{-1})^{-1}}.
\end{equation}

In the case of $k_{\rm{B}}T_{\rm ex}/E_{\rm l} \gg 1$, Equation \ref{eq:d8} can be approximated by, 
\begin{equation}
\label{eq:d10}
Q(T_{\rm ex})  \simeq \int_{0}^{\infty} g_{\rm u} \exp \left(-\frac{E_{\rm l}}{k_{\rm{B}}T_{\rm ex}}\right)dJ
= \int_{0}^{\infty} (2J+1) \exp \left(-\frac{E_{J=1} J(J+1)}{2k_{\rm{B}}T_{\rm ex}}\right)dJ
=\frac{2k_{\rm{B}}T_{\rm ex}}{E_{J=1}}=\frac{2k_{\rm{B}}T_{\rm ex}}{h\nu}.
\end{equation}

\begin{equation}
\label{eq:d11}
\begin{split} 
N({\rm ^{13}CO}) & \simeq 
\frac{3.29 \times 10^{14}}{5.29\ {\rm K}}
\frac{2 T_{\rm ex}/5.29\ {\rm K}}{[1-\exp(-5.29\ {\rm K}/T_{\rm ex})]} 
\left(\frac{\exp(-5.29\ {\rm K}/T_{\rm ex})}{1-\exp(-5.29\ {\rm K}/T_{\rm ex})}-0.168\right)^{-1}
\int \frac{T_{\rm{mb}}}{f}\ dv \ 
{\rm cm^{-2}~(km~s^{-1})^{-1}} \\
& \simeq 
\frac{6.58 \times 10^{14}}{5.29\ {\rm K}}
\frac{T_{\rm ex}/5.29\ {\rm K}}{5.29\ {\rm K}/T_{\rm ex}} 
\left(\frac{1}{5.29\ {\rm K}/T_{\rm ex}}-0.168\right)^{-1}
\int \frac{T_{\rm{mb}}}{f}\ dv \ 
{\rm cm^{-2}~(km~s^{-1})^{-1}} \\
& = 
1.24 \times 10^{14} 
\left(\frac{T_{\rm ex}}{5.29\ {\rm K}}\right)^2
\left(\frac{T_{\rm ex}}{5.29\ {\rm K}}-0.168\right)^{-1}
\int \frac{T_{\rm{mb}}}{f}\ dv \ 
{\rm cm^{-2}~(K~km~s^{-1})^{-1}}.
\end{split} 
\end{equation}

Now we need a good estimate of the $^{13}$CO excitation temperature $T_{\rm{ex,13}}$ and $f$ because both of them almost linearly affect the column density. 
However, it is hard to estimate $T_{\rm{ex,13}}$ from $T_{\rm{mb,13}}$ with Equation \ref{eq:d5} because $^{13}$CO $J$=1--0 emission is not optically thick for most of the molecular gas in the Galaxy. 
Instead, it is often assumed that $T_{\rm{ex,13}}$ is the same as $T_{\rm{ex,12}}$, and the CO $J$=1--0 emission is optically thick and beam-filling ($f_{\rm{d}} \simeq 1$). 
Under this assumption, Equation \ref{eq:d5} is approximated as, 
\begin{equation}
\label{eq:d12}
T_{\rm{ex,13}} \simeq T_{\rm{ex,12}} \simeq \frac{h\nu}{k_{\rm{B}}}\left[\ln \left(1 + \frac{1}{\frac{k_{\rm{B}}T_{\rm{mb,12}}}{h\nu} +\frac{1}{\exp({h\nu/k_{\rm{B}} T_{\rm{bg}})-1}}}\right)\right]^{-1}. 
\end{equation}

In reality, the above assumptions are never met at the same time for diffuse gas since we expect $T_{\rm{ex,12}}/T_{\rm{ex,13}} > 1$ because of the photon trapping effect that selectively enhances the excitation of optically thicker line. 
For diffuse and warm molecular gas with $T_{\rm{k}}$ = 50 K, $n$(H$_2$) = $3 \times 10^2$ cm$^{-3}$, $dv/dr$ = 1.0 $\rm km~s^{-1}~pc^{-1}$, and the canonical values of $X$(CO) = $1 \times 10^{-4}$ and $X$($^{13}$CO) = $2 \times 10^{-6}$ for Galactic disk clouds \citep[][and references therein]{2008ApJ...679..481P}, the \texttt{RADEX} calculation results in Figure \ref{fig:radex_Xco_abundance} show that $T_{\rm{ex,12}} \simeq$ 20 K and $T_{\rm{ex,13}} \simeq$ 7 K to yield $T_{\rm{ex,12}}/T_{\rm{ex,13}} \simeq 3$. 
Thus the assumption of $T_{\rm{ex,13}} \simeq T_{\rm{ex,12}}$ overestimates $N$ accordingly. 

For CO-poor components with the same physical conditions but with 100 times lower $X$(CO) and $X$($^{13}$CO), the photon-trapping effect for CO becomes less significant and $T_{\rm{ex,12}} \simeq$ 7.0 K and $T_{\rm{ex,13}} \simeq$ 6.2 K yield $T_{\rm{ex,12}}/T_{\rm{ex,13}} \simeq 1.1$. 
However, even the CO $J$=1--0 emission is not optically thick ($1-e^{-\tau_{12}} \simeq 0.2$), and $T_{\rm{ex,12}}$ from $T_{\rm{mb,12}}$ will be underestimated accordingly, resulting in underestimation of $N$($^{13}$CO). 
Since the H$_2$ column density is derived with $N({\rm H_2}) \equiv N({\rm ^{13}CO})/X({\rm ^{13}CO})$ by assuming $X$($^{13}$CO), which is $2 \times 10^{-6}$ as a canonical value for Galactic disk clouds \citep[][and references therein]{2008ApJ...679..481P} but may be smaller by 2--3 orders under diffuse environment as was shown in this study, the assumption of canonical $X$($^{13}$CO) will further underestimate $N({\rm H_2})$ by the same 2--3 orders. 

Estimation of the beam-filling factor $f$ is an additional challenge, particularly in cloud edges. 
If it is not unity, $T_{\rm{ex}}$ in Equations \ref{eq:d5} and \ref{eq:d11} is underestimated, and the assumption of $f = 1$ in Equation \ref{eq:d7} further underestimate the column density. 
In case we are interested only in the column density averaged over the telescope beam $fN$, the latter effect is canceled. 

Regarding the validity of the Rayleigh-Jeans approximation, the assumption of $T_{\rm{ex}} \gg h\nu/k_{\rm{B}}$ often fails because of the large $h\nu/k_{\rm{B}}$ for CO and low $T_{\rm{ex}}$ under diffuse environment. 
For $T_{\rm{ex}}$ = $h\nu/k_{\rm{B}}$ = 5.29 K and $T_{\rm{bg}}$ = $T_{\rm{CMB}}$ = 2.73 K, it gives $T_{\rm{mb}}/f_{\rm{d}} = (T_{\rm{ex}}-T_{\rm{bg}})$ = 2.56 K while it shall be 2.18 K. 
In other words, if we analyze the measured $T_{\rm{mb}}$ based on the Rayleigh-Jeans approximation, we will slightly underestimate $T_{\rm{ex}}$ accordingly. 

To conclude this section, $N$(H$_2$) of diffuse gas extracted from $^{13}$CO $J$=1--0 emission study is highly doubtful and is often severely underestimated, and even the order estimation may not be relevant. 
Care must be taken to set an upper limit to the diffuse molecular gas mass from such studies. 

\bibliography{sample631}{}
\bibliographystyle{aasjournal}



\end{document}